\documentclass{jpp}
\usepackage{amsmath, amssymb, mathabx, xspace, braket, nicefrac, color, longtable, setspace, MnSymbol}

\allowdisplaybreaks
\definecolor{darkred}{rgb}{0.75, 0, 0}
\usepackage[colorlinks=true, linkcolor=darkred, citecolor=darkred, hypertexnames=false]{hyperref}

\newcommand{\ie}{i.e.\@\xspace}
\newcommand{\perse}{per~se\@\xspace}

\DeclareMathAlphabet{\mathdutchcal}{U}{dutchcal}{m}{n}

\newcommand{\mc}[1]{\mathcal{#1}}

\newcommand{\mcc}[1]{\mathfrak{#1}}

\newcommand{\eq}[1]{(\ref{#1})}
\newcommand{\Eq}[1]{\eq{#1}}

\newcommand{\Fig}[1]{figure~\ref{#1}}

\newcommand{\Sec}[1]{section~\ref{#1}}
\newcommand{\Secs}[1]{sections~\ref{#1}}
\newcommand{\App}[1]{appendix~\ref{#1}}

\newcommand{\mt}[1]{\texorpdfstring{\({#1}\)}{}}
\newcommand{\m}[1]{$\smash{#1}$}
\newcommand{\mm}[1]{$#1$}

\DeclareMathOperator{\re}{re}
\DeclareMathOperator{\im}{im}
\DeclareMathOperator{\sgn}{sgn}
\DeclareMathOperator{\diag}{diag}
\DeclareMathOperator{\tr}{tr}
\DeclareMathOperator{\symb}{symb}
\DeclareMathOperator{\GAF}{\mcc{A}}
\DeclareMathOperator{\Gi}{Gi}

\newcommand{\oper}[1]{\hat{#1}}
\newcommand{\boper}[1]{\oper{\vec{#1}}}
\renewcommand{\vec}[1]{\boldsymbol{#1}}
\newcommand{\matr}[1]{\vec{#1}}
\newcommand{\T}{\intercal}

\newcommand{\inv}[1]{\bar{#1}}

\newcommand{\favr}[1]{\langle #1 \rangle}
\newcommand{\poisson}[1]{\lbrace #1 \rbrace}

\newcommand{\moyal}[1]{\lbrace \! \lbrace #1 \rbrace \! \rbrace}

\newcommand{\ldx}{\overset{{\scriptscriptstyle \leftarrow}}{\pd}_x}
\newcommand{\ldp}{\overset{{\scriptscriptstyle \leftarrow}}{\pd}_k}
\newcommand{\rdx}{\overset{{\scriptscriptstyle \rightarrow}}{\pd}_x}
\newcommand{\rdp}{\overset{{\scriptscriptstyle \rightarrow}}{\pd}_k}

\newcommand{\placeholder}{\sqbullet}
\newcommand{\pd}{\partial}
\newcommand{\dd}{\mathrm{d}}
\newcommand{\ii}{\mathrm{i}}
\newcommand{\ee}{\mathrm{e}}
\newcommand{\mapsfrom}{\leftmapsto}
\newcommand{\const}{\text{const}}
\newcommand{\bigO}{\mc{O}}

\newcommand{\ev}{\mcc{e}}
\newcommand{\kc}{\mc{K}}
\newcommand{\ki}{\varepsilon}
\newcommand{\MWf}{\mu}
\newcommand{\Eenv}{\tilde{\vec{e}}}
\newcommand{\spinhall}{U}
\newcommand{\dE}{\Upsilon}

\newcommand{\vg}{v_{\text{g}}}
\newcommand{\vvg}{\vec{v}_{\text{g}}}
\newcommand{\avg}{\bar{v}_{\text{g}}}
\newcommand{\avvg}{\bar{\vec{v}}_{\text{g}}}

\newcommand{\symp}[1]{\text{{\sf #1}}}
\newcommand{\sI}{\symp{I}}
\newcommand{\sJ}{\symp{J}}
\newcommand{\sS}{\symp{S}}
\newcommand{\sSi}{\inv{\sS}}
\newcommand{\sZ}{\symp{Z}}
\newcommand{\sg}{\symp{g}}
\newcommand{\sv}{\symp{v}}
\newcommand{\su}{\symp{u}}
\newcommand{\seta}{\symp{s}}
\newcommand{\sz}{\symp{z}}
\newcommand{\sy}{\symp{y}}
\newcommand{\yq}{q}
\newcommand{\yp}{p}
\newcommand{\oyq}{\oper{\yq}}
\newcommand{\oyp}{\oper{\yp}}
\newcommand{\sY}{\symp{Y}}
\newcommand{\sr}{\symp{r}}
\newcommand{\sa}{\symp{a}}

\newcommand{\Mwave}{\m{M}-wave\xspace}
\newcommand{\Mwaves}{\m{M}-waves\xspace}
\newcommand{\tMwaves}{\textit{M}-waves\xspace}

\begin{document}
\bibliographystyle{jpp}

\title{Geometrical optics in phase space}

\author{I.~Y.\ Dodin\aff{1,2}\corresp{\email{idodin@princeton.edu}},
N.~A.\ Lopez\aff{3}, Tingjing Xing\aff{2,4},\\
Rune Højlund Marholt\aff{5},
\and
Valerian~H.\ Hall-Chen\aff{4,6}
}
\affiliation{
\aff{1}Princeton Plasma Physics Laboratory, Princeton, NJ 08543, USA
\aff{2}Department of Astrophysical Sciences, Princeton University, Princeton, NJ 08544, USA
\aff{3}Rudolf Peierls Centre for Theoretical Physics, University of Oxford,
Oxford OX1 3PU, UK
\aff{4}Institute of High Performance Computing, Agency for Science, Technology and Research (A*STAR), Singapore 138632, Singapore
\aff{5}Section for Plasma Physics and Fusion Energy, Department of Physics, Technical University of Denmark, DK-2800 Kgs. Lyngby, Denmark
\aff{6}Future Energy Acceleration and Translation (FEAT), Strategic Research and Translational Thrust (SRTT), A*STAR, Singapore 138632, Singapore
}

\maketitle

\begin{abstract}
Geometrical optics (GO) is widely used for reduced modeling of waves in plasmas but fails near reflection points, where it predicts a spurious singularity of the wave amplitude. We show how to avoid this singularity by adopting a different representation of the wave equation. Instead of the physical space $x$ and the wavevector $k$, we use the ray time $\tau$ as the new canonical coordinate and the ray energy $h$ as the associated canonical momentum. To derive the envelope equation in the $\tau$-representation, we construct the Weyl symbol calculus on the $(\tau, h)$ space and show that the corresponding Weyl symbols are related to their $(x, k)$ counterparts by the Airy transform. This allows us to express the coefficients in the envelope equation through the known properties of the original dispersion operator. When necessary, solutions of this equation can be mapped to the $x$-space using a generalised metaplectic transform. But the field per se might not even be needed in practice. Instead, knowing the corresponding Wigner function usually suffices for linear and quasilinear calculations. As a Weyl symbol itself, the Wigner function can be mapped analytically, using the aforementioned Airy transform. We show that the standard Airy patterns that form in regions where conventional GO fails are successfully reproduced within MGO simply by remapping the field from the $\tau$-space to the $x$-space.  An extension to mode-converting waves is also presented. This formulation, which we call generalised metaplectic GO (MGO) offers a promising tool, for example, for reduced modeling of the O--X conversion in inhomogeneous plasma near the critical density, an effect that is important for fusion applications and also occurs in the ionosphere. Aside from better handling reflection, MGO is similar to GO and can replace it for any practical purposes.
\end{abstract}

\newpage
\setcounter{tocdepth}{2}
\tableofcontents

\newpage
\section{Introduction}

\subsection{Background}

The geometrical-optics (GO) approach is widely used for reduced modeling of waves in plasmas and other media \citep{book:kravtsov, book:tracy}. In this approach, instead of solving the wave equations \perse, one converts them into Hamilton's equations for wave rays and a first-order (in time or the ray path) differential equation for the wave envelope. Then, one calculates the ray trajectories that are relevant to the initial conditions (`reference rays') and solves the envelope equations on those trajectories. As an initial-value algorithm, this scheme is computationally inexpensive and has also yielded a number of spinoffs, such as beam tracing \citep{ref:pereverzev98, ref:poli2018, ref:chen22} and quasioptics, which can account for both transverse diffraction \citep{ref:balakin07} and mode conversion \citep{my:quasiop1, my:quasiop5}.\footnote{In its standard formulation, GO is also closely related to the Wentzel--Kramers--Brillouin (WKB) approximation, so we will use the terms `conventional GO' and WKB interchangeably.}

Still, the GO approach is limited in that it assumes that the characteristic scale \m{L} of the medium is much larger than the local wavelength \m{\lambda}. Even for high-frequency waves, such as those in the electron-cyclotron range, this assumption unavoidably breaks down near cutoffs (reflection points), where \m{\lambda/L} is infinite and the conventional GO predicts a spurious singularity of the field amplitude. For example, this is an issue in modeling the O--X mode conversion near the critical plasma density, which is a part of a promising scheme for heating overdense plasmas in fusion reactors \citep{ref:hansen85, ref:preinhaelter73} and also occurs in the ionosphere \citep{ref:mjolhus90, ref:eliasson16}. Since those waves experience many oscillations during the mode-conversion process, one can expect that \textit{some} GO-like model should be possible for them, even though the conversion occurs near a cutoff. This motivates a search for alternative formulations of GO that do not rely on the smallness of \m{\lambda/L} \perse. 

One such formulation, which has long been known, is based on Maslov's approach \citep{book:maslov, ref:ziolkowski84, ref:thomson85}. The idea of this approach is based on the fact that a wave experiencing reflection in the physical space satisfies the conditions of GO in the spectral space, so one can reinstate GO locally by applying the (spatial) Fourier transform. However, applying Maslov's method is cumbersome in that the exact moment for performing the Fourier transform is only loosely specified and thus codes have to be supervised. The more recent methods \citep{ref:littlejohn85, ref:littlejohn86, ref:kay94b, ref:zor96, ref:madhusoodanan98, ref:alonso97, ref:alonso99} remedy this issue by using continuous transformations of the wavefield along ray trajectories. However, these methods are restricted in the types of solutions sought and are usually applicable only to specific equations, such as the Schr\"odinger equation for quantum particles. Practical modeling of waves in plasmas, especially magnetised plasmas, requires a more general framework.

The first framework reasonably fit for plasma applications, called metaplectic geometrical optics (MGO), was developed in \citep{my:nimt, my:mgo, my:mgosing, my:mgounit, my:mgosteep}; for an overview, see \citep{my:mgoinv}. Its basic idea is as follows: instead of propagating the wavefield in the physical space, one propagates it, by integrating GO equations over a differentially small time, on a space that is locally tangent to the ray trajectory in the ray phase space. Then, one remaps the field to the space tangent to the reference ray at the next location on the ray and iterates. The remapping is done using metaplectic transforms (MT), a subject that we will introduce shortly (\Sec{sec:mtintro}). An advantage of this formulation is that it can be used for any linear wave equation, including integro-differential equations that emerge in kinetic wave theory. It also readily incorporates transverse diffraction and mode conversion and has been proven workable in calculations of the field caustics within various wave models \citep{my:mgosing}. That said, the need for constant remapping of the wavefield between neighboring tangent spaces remains an inconvenience in MGO. It would be better to have a differential equation that would describe the continuous evolution of the wave amplitude on reference rays themselves, while mapping the resulting solutions to the physical space would remain an optional diagnostic. The purpose of this paper is to propose such a formulation of MGO. (Below, the term `MGO' refers to the new formulation that we report here.)

\subsection{Toolbox}
\label{sec:mtintro}

Let us start by explaining what MTs are about. (A more detailed formulation is presented in \Sec{sec:mt}.) Suppose that one has some field \m{\psi_{\oper{x}}} on a coordinate space~\m{x}. (The reason for adding the index \m{\oper{x}} will become clear shortly.) Any function \m{\oper{U}\psi_{\oper{x}}}, where \m{\oper{U}} is a unitary transformation, contains just as much information as \m{\psi_{\oper{x}}} itself. Then, it makes sense to consider the set of all possible \m{\oper{U}\psi_{\oper{x}}} as a single object, a ket vector \m{\ket{\psi}}, and each \m{\oper{U}\psi_{\oper{x}}} as a representation of \m{\ket{\psi}}. For example, \m{\psi_{\oper{x}}(x)} itself can be understood as a projection of \m{\ket{\psi}} on the eigenvector of the position operator \m{\oper{x}} that corresponds to the eigenvalue \m{x}; \ie \m{\psi_{\oper{x}}(x) = \braket{\ev_{\oper{x}}(x)| \psi}}. Performing a `coordinate transformation', \ie choosing some \m{\oper{q}} as the new position operator instead of \m{\oper{x}}, induces a \m{q}-representation \m{\psi_{\oper{q}}(q) = \braket{\ev_{\oper{q}}(q)| \psi}}, where \m{\ev_{\oper{q}}(q)} is the eigenvector of \m{\oper{q}} corresponding to the eigenvalue \m{q}. It is connected to the original \m{x}-representation via a unitary integral operator, \m{\psi_{\oper{q}}(q) = \int M(q, x) \psi_{\oper{x}}(x)\,\dd x}, with kernel \m{M(q, x) = \braket{\ev_{\oper{q}}(q)| \ev_{\oper{x}}(x)}}. The mapping between \m{\psi_{\oper{x}}} and \m{\psi_{\oper{q}}} is an MT. In this paper, we introduce more than two position operators and, thus, multiple MTs as well.

One example of an MT is the Fourier transform, which maps functions in the \m{x}-representation to those in the wavevector, or \m{k}-representation, \ie corresponds to \m{\oper{q} = \oper{k}}. (Because the \m{x}-representation of the wavevector operator is \m{\oper{k}_{\oper{x}} = - \ii \pd_x}, this corresponds to \m{M(q, x) \propto \ee^{-\ii q x}}.) MTs are also used to introduce mixed representations, where \m{\oper{q} = A\oper{x} + B\oper{k}}, with constant \m{A} and \m{B}.\footnote{In the literature, the term `MT' is typically reserved just for this particular case, where \m{\oper{q}} is linear in both \m{\oper{x}} and \m{\oper{k}}. At the risk of irritating our readers, we will ignore this restrictive definition and use the term MT for the transformation of the field representation for any \m{\oper{q}(\oper{x}, \oper{k})}.} In this sense, MTs are a natural generalization of the Fourier transform and are well-studied as such \citep{ref:littlejohn86}. However, linear operator transformations (also used in the earlier formulation of MGO) are not enough for our purposes here. Since we aim to solve for the wavefield in the coordinate frame aligned with the ray, it will be necessary to consider the possibility of nonlinear dependence of \m{\oper{q}} on \m{\oper{x}} and \m{\oper{k}}. The corresponding MTs have been discussed in the quantum-physics literature to some extent (for example, see \cite{ref:mello75}), and their semiclassical limit has attracted interest in the area of atomic and molecular dynamics \citep{ref:miller74}. However, said literature is concerned with problems different from those that we are interested in here, so we basically start from scratch in this paper and will mention parallels with the existing literature only where they are relevant.

MTs are closely related to the Weyl symbol calculus and Wigner functions (\Sec{sec:primer}). For any given operator \m{\oper{A}} expressed through \m{\oper{x}} and \m{\oper{k}}, its (Weyl) symbol \m{A_\sz} is a function of \m{(x, k) \equiv \sz} that is typically close, in regimes that we are interested in, to the one that is obtained from \m{\oper{A}} by replacing each instance of \m{\oper{x}} and \m{\oper{k}} with \m{x} and \m{k}, respectively.\footnote{For example, the susceptibility \m{\chi(x, k)} of an inhomogeneous medium is, at least approximately, the Weyl symbol of the response operator \m{\oper{\chi}} that enters Maxwell's equations. (Depending on a problem, \m{x} can be a spacetime coordinate; then \m{k} also includes the frequency.)} (In the coordinate representation, this means replacing just \m{\pd_x} with \m{\ii k}.) In particular, the symbol of \m{\ket{\psi}\bra{\psi}}\footnote{Up to a constant coefficient that depends on a convention.} on the space \m{\sz} is called the Wigner function of \m{\psi_{\oper{x}}} and equals the Fourier image of \m{\psi_{\oper{x}}(x + s/2)\psi^*_{\oper{x}}(x - s/2)} with respect to \m{s}, assuming \m{\psi_{\oper{x}}} is a scalar. 

One can also use base operators other than \m{\oper{x}} and \m{\oper{k}}, say, \m{\oper{\tau}} and \m{\oper{h}}. Then, symbols, including Wigner functions, become functions of \m{(\tau, h) \equiv \sr}, respectively. If the relation between \m{(\oper{x}, \oper{k})} and \m{(\oper{\tau}, \oper{h})} is linear, the new symbols are related to the old ones via a simple formula \m{A_\sr(\sr) = A_\sz(\sz(\sr))}, \ie exhibit symplectic invariance; however, this is generally not the case for nonlinear transformations. A firm grasp on these facts is important to understanding the following section.

\subsection{Synopsis}
\label{sec:spoiler}

Developing MGO means answering the following questions: (i)~What is the natural position operator \m{\oper{\tau}} for MGO? (ii)~How does one write GO equations in the \m{\tau}-representation? (iii)~What is the small parameter that replaces the GO parameter \m{\lambda/L}? (iv)~How does one map a solution from the \m{\tau}-representation to the \m{x}-representation? Below, we summarise how we answer these questions in the rest of the paper.

Suppose a wave is governed by \m{\oper{H}\ket{\psi} = 0}, where \m{\oper{H}} is a Hermitian operator. (A~small anti-Hermitian addition to \m{\oper{H}} can be easily accommodated as a perturbation.) A systematic derivation of the conventional GO from this equation (\Sec{sec:go}) starts with introducing the Weyl symbol of \m{\oper{H}}, denoted \m{H_\sz}, where the index \m{\sz} indicates that the corresponding Weyl calculus is constructed on the original ray phase-space \m{(x, k) \equiv \sz}. Then, one Taylor-expands \m{H_\sz} around the ray with \m{\lambda/L} as a small parameter and converts this approximate symbol into a new operator, which serves as an approximation of \m{\oper{H}}. This leads to an approximate amplitude equation on a reference ray. Assuming that \m{H_\sz} is a scalar (vector waves are discussed below), the reference ray itself is governed by Hamilton's equations, with \m{H_\sz} serving as the Hamiltonian.

In MGO, we introduce a different phase space, \m{(\tau, h) \equiv \sr}, which is aligned with the ray (\Sec{sec:nat}). It turns out that the natural position operator \m{\oper{\tau}} in this case is the one that is associated with the momentum operator \m{\oper{h} = \oper{H}} via the canonical commutation relation \m{[\oper{\tau}, \oper{h}] = \ii};\footnote{Notably, this \m{\tau} is also basically what \citet{tex:thooft24} calls the ontological variable in a quantum-mechanical context.} \ie \m{\tau} is the ray time. We introduce the Weyl calculus on the space \m{\sr} and consider the corresponding symbol \m{H_\sr}. MGO envelope equations are derived much like in GO, via Taylor-expanding \m{H_\sr} (\Sec{sec:mgoscalar}). However, the relevant small parameter in this case is not \m{\lambda/L} but \m{\epsilon \doteq (\Delta x\,\Delta k)^{-1}}, where \m{\Delta x} and \m{\Delta k} are the characteristic scales of the original symbol \m{H_\sz} along the coordinate axis and the wavenumber axis, correspondingly.\footnote{If \m{\oper{\tau}} coincides with \m{\oper{x}}, when MGO is identical to GO, then \m{\Delta x \sim L} and \m{\Delta k \sim k \sim 1/\lambda}, so \m{\epsilon} becomes the usual GO parameter \m{\lambda/L}.} This parameter can be understood as the (squared) symplectic curvature of the ray trajectory, or at least a characteristic value thereof.

How does one find \m{H_\sr}? In practice, one generally knows \m{\oper{H}} in the \m{x}-representation. This means that one knows \m{H_\sz}. To obtain MGO equations explicitly, one needs to express \m{H_\sr} though \m{H_\sz}. Since the mapping between \m{\sz} and \m{\sr} is nonlinear, exact symplectic invariance is not to be expected, so we derive the corresponding relation from scratch. We show that the mapping between the symbols of a given operator in any two representations is determined by the Wigner function of the MT kernel \m{M}. We also show that, for the \m{\sz \mapsto \sr} transformation at small \m{\epsilon} in particular, this mapping can be approximated with the Airy transform. For smooth symbols like \m{H_\sz}, this gives \m{H_\sr(\sr) = H_\sz(\sz(\sr)) + \bigO(\epsilon^2)}. The absence of \m{\bigO(\epsilon)} corrections in this formula is important, because those would have affected the envelope equation (at least, for vector waves discussed below), while \m{\bigO(\epsilon^2)} is negligible within the assumed accuracy. In other words, one can use \m{H_\sr(\sr) \approx H_\sz(\sz(\sr))} notwithstanding the lack of exact symplectic invariance. This makes the MGO envelope equation similar to the GO envelope equation, yet singularities are avoided, as \m{H_\sz} is differentiated with respect to different variables (\Sec{sec:mgoscalar}).

We also propose ways to approximate the MT kernel itself, which then enables mapping solutions of the MGO envelope equation to the physical space \m{x}. Related problems have been traditionally attracting attention in the quantum-mechanics literature. However, they are not particularly important in the context of modeling waves in plasmas, for several reasons. First of all, possible errors caused by inaccuracy of the mapping \m{\psi_{\oper{q}} \mapsto \psi_{\oper{x}}} do not accumulate along the ray trajectory, so there is no need to make this mapping particularly precise. The wavefield is propagated in MGO in a single space, \m{\tau}, and mapping to the \m{x}-space is just an optional diagnostic. Second, although wave simulations are traditionally expected to yield spectacular illustrations, the field in the \m{x}-space does not actually matter much for practical purposes. One might not even need to leave the \m{\tau}-space except for initializing the field and also calculating the final state, and those mappings may be trivial to apply. Third, even when spatial profiles are actually needed, they are not of the field \perse but rather of quadratic functionals thereof. Such functionals can always be expressed through the Wigner function of the wavefield. (In fact, \textit{any} reasonably general quasilinear calculations are expressed more naturally  through Wigner functions than through fields \perse; see \cite{my:ql, my:qlrmpp}.) As a Weyl symbol, a Wigner function can be remapped from the \m{\sr}-space to the \m{\sz}-space \textit{analytically}, using the aforementioned Airy transform or its asymptotics. We show that this allows one to easily recover the well-known WKB results, as well as the also well-known Airy profiles of wavefields near cutoffs (\Sec{sec:mgoscalar}), which are usually considered as an epitome of GO failure. In a way, the Airy function is a cornerstone of MGO, because \textit{any} wavefield locally looks like an Airy field in a certain representation.

These results largely extend to vector waves as well (\Sec{sec:mgogeneral}). However, since \m{H_\sz} is a matrix in this case, reference rays and \m{\oper{h}} are constructed out of its relevant (small) eigenvalues rather than \m{H_\sz} \perse. We also derive equations that capture mode conversion when \m{H_\sz} is degenerate. Finally, we introduce the concept of a metaplectic resonance, which generalises the concept of the Cherenkov resonance to waves that are quasimonochromatic in their natural \m{\tau}-representation but not necessarily in the physical \m{x}-representation (\Sec{sec:wp}). Applications of this theory, as well as the (potentially straightforward) extension of MGO to quasioptical beams are left to future work.

\subsection{Outline}
\label{sec:outline}

This rest of this paper is organised as follows. In \Sec{sec:primer}, we introduce the key notation, the Weyl symbol calculus, and Wigner functions. In \Sec{sec:mt}, we formalise the concept of an MT, derive some properties of MTs, and present examples. In \Sec{sec:go}, we restate the conventional GO, which is used as an important reference point in the rest of the paper. We also discuss how the MT kernel \m{M} itself can be approximated using GO methods in some cases. In \Sec{sec:nat}, we derive \m{\oper{\tau}}, \m{\oper{h}}, \m{M}, and its Wigner function \m{\mu} for mapping symbols between the \m{\sz}-space and the \m{\sr}-space. In \Sec{sec:mgoscalar}, we derive the MGO envelope equation for scalar waves and discuss how to map its solutions to the \m{x}-space. In \Sec{sec:mgogeneral}, we generalise MGO to vector waves and discuss mode conversion in particular. In \Sec{sec:wp}, we introduce the concept of a metaplectic resonance. 

Auxiliary calculations are presented in appendices. In \App{app:req}, we rederive the ray equations, both for completeness and also to introduce some terminology. In \App{app:aac}, we discuss the possibility of introducing the local angle--action variables, and the associated operators, for unbounded GO trajectories. In \App{app:ai}, we discuss some properties of the auxiliary functions associated with the aforementioned Airy transform. In \App{app:qz}, we rederive the Einstein--Brillouin--Keller quantization within MGO, where it becomes particularly natural. In \App{app:P}, we derive an expression for the electromagnetic dissipation power per phase-space volume (as opposed to the spatial volume, as usual). Finally, in \App{app:notation}, we summarise our main notations.

\section{A math primer}
\label{sec:primer}

\subsection{Notation}
\label{sec:notation}

For any given Hermitian operator \m{\oper{A}} on a given Hilbert space \m{\mathbb{H}}, one can introduce a complete set of its orthogonal eigenvectors \m{\ket{\ev_{\oper{A}}(\lambda)} \in \mathbb{H}},
\begin{gather}
\oper{A}\ket{\ev_{\oper{A}}(\lambda)} 
= \lambda \ket{\ev_{\oper{A}}(\lambda)},
\qquad
\lambda = \lambda^*,
\end{gather}
where \m{\lambda} are the corresponding eigenvalues. (We use \m{^*} to denote complex conjugation, so the second equation states that the eigenvalues of a Hermitian operator are real.) Assuming that the eigenspace of \m{\oper{A}} is \m{\mathbb{R} \equiv (-\infty, \infty)},\footnote{A generalization to \m{\mathbb{R}^n} is straightforward \citep{my:ql}. Also, strictly speaking, one should introduce a rigged Hilbert space to allow for \eq{eq:norm}, but we do not strive for ultimate rigor here.} its eigenvectors can always be chosen such that
\begin{gather}\label{eq:norm}
\braket{\ev_{\oper{A}}(\lambda')|\ev_{\oper{A}}(\lambda'')} 
= \delta(\lambda' - \lambda''),
\end{gather}
where \m{\delta} is the Dirac delta function. Then, they satisfy the identity
\begin{gather}\label{eq:unit}
\int \dd\lambda\,
\ket{\ev_{\oper{A}}(\lambda)}\bra{\ev_{\oper{A}}(\lambda)} 
= \oper{1},
\end{gather}
where \m{\oper{1}} is a unit operator. (Integrals in this paper are taken over \m{\mathbb{R}} except where specified otherwise.) Accordingly, \m{\oper{A}} can be expressed as
\begin{align}\label{eq:operator}
\oper{A} 
& = \int \dd \lambda'\,
\ket{\ev_{\oper{A}}(\lambda')}
\bra{\ev_{\oper{A}}(\lambda')}\oper{A}
\int \dd \lambda''\,
\ket{\ev_{\oper{A}}(\lambda'')}
\bra{\ev_{\oper{A}}(\lambda'')}
\notag\\
& = \int \dd \lambda'\, \dd \lambda''
\ket{\ev_{\oper{A}}(\lambda')}
\lambda''\braket{\ev_{\oper{A}}(\lambda')|\ev_{\oper{A}}(\lambda'')}
\bra{\ev_{\oper{A}}(\lambda'')}
\notag\\
& = \int \dd \lambda
\ket{\ev_{\oper{A}}(\lambda)} \lambda \bra{\ev_{\oper{A}}(\lambda)}.
\end{align}

Among such operators on \m{\mathbb{H}}, let us assign some operator \m{\oper{x}} as the \textit{position operator}. For any given vector \m{\ket{\psi}} in \m{\mathbb{H}}, the eigenvectors of \m{\oper{x}} define the \m{x}-representation of \m{\ket{\psi}}, or a \textit{field} \m{\psi_{\oper{x}}(x)}, via
\begin{gather}
\psi_{\oper{x}}(x) \doteq 
\braket{\ev_{\oper{x}}(x)| \psi}.
\end{gather}
Correspondingly, for any given operator \m{\oper{O}}, the field associated with \m{\oper{O}\ket{\psi}} is as follows:
\begin{gather}
(\oper{O}_{\oper{x}}\psi_{\oper{x}})(x) = 
\braket{\ev_{\oper{x}}(x)| \oper{O} | \psi}.
\end{gather}
Using \Eq{eq:unit} for \m{\oper{A} = \oper{x}}, one can rewrite this as
\begin{align}
(\oper{O}_{\oper{x}}\psi_{\oper{x}})(x) & = 
\int \dd x'
\braket{\ev_{\oper{x}}(x)| \oper{O} | \ev_{\oper{x}}(x')}
\braket{\ev_{\oper{x}}(x') | \psi}
\notag\\
& = \int \dd x'\, O_{\oper{x}}(x, x')\, \psi_{\oper{x}}(x'),
\end{align}
where the kernel \m{O_{\oper{x}}(x, x') \doteq \braket{\ev_{\oper{x}}(x)| \oper{O} | \ev_{\oper{x}}(x')}} is the `matrix element' of \m{\oper{O}} in the \m{x}-representation, or simply the \m{x}-representation of \m{\oper{O}}. In the special case when \m{\oper{O}_{\oper{x}}} is a local or differential operator (that is, if \m{(\oper{O}_{\oper{x}}\psi_{\oper{x}})(x)} can be expressed through \m{\psi_{\oper{x}}} and its derivatives \m{\psi^{(m)}_{\oper{x}}} with finite \m{m}), then \m{O_{\oper{x}}(x, x')} can be expressed through the delta function, \m{\delta(x - x')}, and its corresponding derivatives, \m{\delta^{(m)}(x - x')}.

Using \Eq{eq:operator}, one can further use this function to express \m{\oper{O}} as follows:
\begin{gather}\label{eq:Oxx}
\oper{O} = \int \dd x'\,\dd x''
\ket{\ev_{\oper{x}}(x')}
O_{\oper{x}}(x', x'')
\bra{\ev_{\oper{x}}(x'')},
\end{gather}
and its Hermitian adjoint as
\begin{gather}\label{eq:Oxxd}
\oper{O}^\dag = \int \dd x'\,\dd x''
\ket{\ev_{\oper{x}}(x')}
O_{\oper{x}}^*(x'', x')
\bra{\ev_{\oper{x}}(x'')}.
\end{gather}
By definition,
\begin{gather}
x_{\oper{x}}(x, x') = \braket{\ev_{\oper{x}}(x)| \oper{x} | \ev_{\oper{x}}(x')}
= x' \braket{\ev_{\oper{x}}(x)| \ev_{\oper{x}}(x')}
= x' \delta(x - x')
= x \delta(x - x').
\end{gather}
Then, \m{(\oper{x}_{\oper{x}} \psi_{\oper{x}})(x) = x\psi_{\oper{x}}(x)}, so
\begin{gather}\label{eq:xop}
\oper{x}_{\oper{x}} = x.
\end{gather}

Let us also introduce the \textit{wavevector operator} \m{\oper{k}} as a generator of spatial translations:
\begin{gather}
\exp(-\ii a \oper{k})\ket{\ev_{\oper{x}}(x)}
= \ket{\ev_{\oper{x}}(x + a)}
\end{gather}
(here, \m{a} is any real constant), whence
\begin{align}
(\exp(-\ii a \oper{k}_{\oper{x}})\psi_{\oper{x}})(x)
& = \braket{\ev_{\oper{x}}(x) | \exp(-\ii a \oper{k}) | \psi}
\notag\\
& = \braket{\psi | \exp(\ii a \oper{k}) | \ev_{\oper{x}}(x) }^*
\notag\\
& = \braket{\psi | \ev_{\oper{x}}(x - a) }^*
\notag\\
& = \psi_{\oper{x}}(x - a).
\label{eq:shift}
\end{align}
At \m{a \to 0}, one has \m{\exp(-\ii a \oper{k}_{\oper{x}}) \to 1 - \ii a \oper{k}_{\oper{x}}} and \m{\psi_{\oper{x}}(x - a) \to \psi_{\oper{x}}(x) - a\,\pd_x\psi_{\oper{x}}(x)}, so \eq{eq:shift} leads to \m{-\ii \oper{k}_{\oper{x}} \psi_{\oper{x}}(x) \to - \pd_x\psi_{\oper{x}}(x)} for any \m{\psi_{\oper{x}}(x)}. Thus,
\begin{gather}\label{eq:deri}
\oper{k}_{\oper{x}} = -\ii\pd_{x}
\end{gather}
and  \m{k_{\oper{x}}(x, x') = -\ii \delta'(x - x')}. (The prime in \m{\delta'} denotes the derivative with respect to the argument.) Such \m{\oper{x}} and \m{\oper{k}} are called Schr\"odinger operators, and here we will also refer to them as base operators. Note that the base operators satisfy the canonical commutation relation
\begin{gather}\label{eq:commut}
[\oper{x}, \oper{k}] = \ii.
\end{gather}
The converse is also true: if given Hermitian operators \m{\oper{x}} and \m{\oper{k}} satisfy \eq{eq:commut}, then there exists a representation, called the \m{x}-representation, in which \eq{eq:xop} and \eq{eq:deri} are satisfied; see, \eg \citep[section~2.12]{book:galindo}. 

For \m{\xi_k(x) \doteq \braket{\ev_{\oper{x}}(x) | \ev_{\oper{k}}(k)}}, which is the \m{x}-representation of an eigenvector of \m{\oper{k}}, \eq{eq:deri} yields \m{-\ii\pd_{x}\xi_k = k \xi}. Assuming the same normalization as in \Eq{eq:norm}, one obtains\footnote{For \m{n}-dimensional \m{x}-space, here and further one should replace \m{2\upi} with \m{(2\upi)^n}.}
\begin{gather}\label{eq:xqme}
\braket{\ev_{\oper{x}}(x) | \ev_{\oper{k}}(k)}
 = \frac{1}{\sqrt{2\upi}}\,\ee^{\ii k x},
\end{gather}
where we chose the initial phase such that \m{\xi_k(0)} is real. This convention will also be assumed from now on for the eigenfunctions of \textit{all} other base operators.

Notably, \eq{eq:xqme} means that the \m{k}-representation of any given \m{\ket{\psi}} is the Fourier transform of the corresponding~\m{\psi_{\oper{x}}}:
\begin{gather}
\psi_{\oper{k}} 
\doteq \braket{\ev_{\oper{k}}(k) | \psi}
= \int \dd x
\braket{\ev_{\oper{k}}(k) | \ev_{\oper{x}}(x)}
\braket{\ev_{\oper{x}}(x) | \psi}
= \frac{1}{\sqrt{2\upi}} \int \dd x\,
\ee^{-\ii k x} \psi_{\oper{x}}(x).
\label{eq:ft}
\end{gather}
Obviously, the \m{k}-representation of \m{\oper{k}} is  \m{\oper{k}_{\oper{k}} = k}. The \m{k}-representation of \m{\oper{x}} is found from
\begin{align}
(\oper{x}_{\oper{k}} \psi_{\oper{k}})(k)
& = \int \dd k' \braket{\ev_{\oper{k}}(k) | \oper{x} | \ev_{\oper{k}}(k')}
\braket{\ev_{\oper{k}}(k') | \psi}
\notag\\
& = \int \dd x'\,\dd k' \braket{\ev_{\oper{k}}(k) | \oper{x} | \ev_{\oper{x}}(x')}
\braket{\ev_{\oper{x}}(x')| \ev_{\oper{k}}(k')}
\braket{\ev_{\oper{k}}(k') | \psi}
\notag\\
& = \int \dd x'\,\dd k'\, x' \braket{\ev_{\oper{x}}(x') | \ev_{\oper{k}}(k)}^*
\braket{\ev_{\oper{x}}(x')| \ev_{\oper{k}}(k')}
\braket{\ev_{\oper{k}}(k') | \psi}
\notag\\
& = \int \dd k' \braket{\ev_{\oper{k}}(k') | \psi} 
\frac{1}{2\upi}\int \dd x'\,x' \ee^{\ii(k' - k)x'}
\notag\\
& = \ii\pd_{k} \int \dd k' \braket{\ev_{\oper{k}}(k') | \psi} \delta(k' - k)
\notag\\
& = \ii\pd_{k} \psi_{\oper{k}}(k).
\end{align}
Then, \m{\oper{x}_{\oper{k}} = \ii \pd_k}, whence \m{-\oper{x}} can be understood as a generator of translations in the spectral space.

\subsection{Weyl symbols}
\label{sec:weyl}

Any given operator \m{\oper{O}} can be expressed as a linear superposition of non-commuting phase-space translations, \ie combinations of the base operators \citep{book:tracy, book:degosson06}. Here, we present only the formulas relevant to the discussion below. First of all, one has
\begin{subequations}\label{eq:wwo}
\begin{align}\label{eq:woperx}
\oper{O} & = \frac{1}{2\upi} \int \dd x \,\dd k\,\dd s\,
\ket{\ev_{\oper{x}}(x + s/2)}
O(x, k)\,\ee^{\ii k s}
\bra{\ev_{\oper{x}}(x - s/2)}
\\
& = \frac{1}{2\upi} \int \dd x \,\dd k\,\dd r\,
\ket{\ev_{\oper{k}}(k + r/2)}
O(x, k)\,\ee^{-\ii r x}
\bra{\ev_{\oper{k}}(k - r/2)},
\end{align}
\end{subequations}
where \m{O(x, k) \equiv \symb \oper{O}} is the Weyl symbol (or just `symbol') of \m{\oper{O}} given~by
\begin{subequations}
\begin{align}\label{eq:wsymbx}
O(x, k) 
& = \int \dd s
\braket{\ev_{\oper{x}}(x + s/2) | \oper{O} | \ev_{\oper{x}}(x - s/2)}
\ee^{- \ii k s}
\\
& = \int \dd r
\braket{\ev_{\oper{k}}(k + r/2) | \oper{O} | \ev_{\oper{k}}(k - r/2)}
\ee^{\ii r x}.
\label{eq:oxkr}
\end{align}
\end{subequations}
The symbol of an operator is related to its \m{x}-representation via the Fourier transform with respect to the argument difference \m{x - x'} at fixed \m{(x + x')/2}:
\begin{eqnarray}\label{eq:Mxx}
\braket{\ev_{\oper{x}}(x) | \oper{O} | \ev_{\oper{x}}(x')} 
= \frac{1}{2\upi}\int\dd k\,
\ee^{\ii k (x - x')} \,
O\left(\frac{x + x'}{2}, k\right),
\end{eqnarray}
and similarly for the \m{k}-representation. This also leads to the following notable properties of Weyl symbols:
\begin{subequations}
\begin{eqnarray}
\braket{\ev_{\oper{x}}(x)| \oper{O} | \ev_{\oper{x}}(x)} = 
\frac{1}{2\upi}\int \dd k\,O(x, k),
\\
\braket{\ev_{\oper{k}}(k)| \oper{O} | \ev_{\oper{k}}(k)} = 
\frac{1}{2\upi}\int \dd x\,O(x, k).
\end{eqnarray}
\end{subequations}

An operator unambiguously determines its symbol, and vice versa. We denote this isomorphism as \m{\oper{O} \Leftrightarrow O}. The mapping \m{\oper{O} \Rightarrow O} is called the Wigner transform, and \m{O \Rightarrow \oper{O}} is called the Weyl transform. For uniformity, we will refer to these as the direct and inverse Wigner--Weyl transform. The Wigner--Weyl isomorphism \m{\Leftrightarrow} is natural in that it has the following properties:
\begin{eqnarray}\label{eq:natural}
 \oper{1} \Leftrightarrow 1, 
 \quad 
 \oper{x} \Leftrightarrow x, 
 \quad 
 \oper{k} \Leftrightarrow k,
 \quad
 h(\oper{x}) \Leftrightarrow h(x),
 \quad  
 h(\oper{k}) \Leftrightarrow h(k),
 \quad
 \oper{O}^\dag \Leftrightarrow O^*,
\end{eqnarray}
where \m{h} is any function and \m{\oper{O}} is any operator. For multi-component operators, the Wigner--Weyl transform is applied element-by-element. In particular, if \m{\oper{O}} is a matrix operator, one finds that \m{\smash{\oper{O}}^\dag \Leftrightarrow O^\dag}, where \m{O} is the symbol matrix.

The product of two operators maps to the so-called Moyal product, or star product, of their symbols:
\begin{eqnarray}\label{eq:moyalx}
\oper{A}\oper{B}\,
\Leftrightarrow 
\,
A(x, k) \star B(x, k)
\doteq
A(x, k)\ee^{\ii\oper{\mc{L}}/2} B(x, k).
\end{eqnarray}
Here, \m{\oper{\mc{L}} \doteq \ldx \rdp - \ldp \rdx}, and the arrows indicate the directions in which the derivatives act. For example, \m{A \oper{\mc{L}} B} is just the canonical Poisson bracket:
\begin{equation}\label{eq:poissonx}
A \oper{\mc{L}} B 
= \poisson{A, B} \doteq
 \frac{\pd A}{\pd x}\frac{\pd B}{\pd k}
- \frac{\pd A}{\pd k}\frac{\pd B}{\pd x}
\equiv \frac{\pd A}{\pd \sz^\alpha}\,\sJ^{\alpha\beta}\,\frac{\pd B}{\pd \sz^\beta},
\end{equation}
where \m{\sz \doteq (x, k)^\T}\footnote{\label{ft:sp}We use \m{^\T} to denote transposition. Hence, when \m{\sz} is treated as a vector, \m{(x, k)^\T \equiv \sz} is understood as a column vector and \m{(x, k) \equiv \sz^\T} is understood as a row vector. (We ignored the difference between the two in the introduction for simplicity.) The upper and lower indices are not distinguished in this paper, but the  Einstein summation convention for repeated indices is assumed. For two vectors \m{\sz_1 = (a_1, b_1)^\T} and \m{\sz_2 = (a_2, b_2)^\T}, one has \m{\sz_1^\T \sz_2 = a_1 a_2 + b_1 b_2 \equiv \sz_1 \cdot \sz_2 = \sz_2 \cdot \sz_1}. Keep in mind, though, that \m{\cdot} is not a scalar product \perse and does not define a length. In particular, the expression \m{\sz \cdot \sz} is generally meaningless, because, by definition, \m{\sz \cdot \sz = x^2+ k^2} yet \m{x^2} and \m{k^2} generally have non-matching dimensions.} and \m{\sJ} is the canonical symplectic form, which satisfies \m{\sJ\sJ = - \sI}:
\begin{gather}\label{eq:JI}
\sJ \doteq \left(
\begin{array}{cc}
0 & 1\\
-1 & 0\\
\end{array}
\right),
\qquad
\sI= \left(
\begin{array}{cc}
1 & 0\\
0 & 1\\
\end{array}
\right).
\end{gather}
These formulas readily yield
\begin{eqnarray}\label{eq:hk}
h(\oper{x})\oper{k}
\Leftrightarrow 
k h(x) + \frac{\ii}{2}\,h'(x),
\qquad
\oper{k} h(\oper{x})
\Leftrightarrow 
k h(x) - \frac{\ii}{2}\,h'(x)
\end{eqnarray}
(the prime denotes the derivative with respect to the argument), so
\begin{gather}
k h(x) 
\Leftrightarrow \frac{1}{2}\,(h(\oper{x})\oper{k} + \oper{k}h(\oper{x}))
= h(\oper{x})\oper{k} + \frac{1}{2}\,[\oper{k}, h(\oper{x})]
= h(\oper{x})\oper{k} - \frac{\ii}{2}\, h'(\oper{x}),
\end{gather}
or, in the \m{x}-representation, 
\begin{gather}
k h(x) \Leftrightarrow -\ii h(x)\pd_x - \ii h'(x)/2.
\label{eq:khsym}
\end{gather}
Also \(h(\oper{k}) \ee^{\ii K \oper{x}} \Leftrightarrow h(k) \star \ee^{\ii K x} = h(k + K/2) \ee^{\ii K x}\), and so on. In particular, the symbol of a commutator \m{[\oper{A}, \oper{B}]} can be expressed through the so-called Moyal bracket \m{\moyal{A, B}}:
\begin{gather}\label{eq:mb0}
[\oper{A}, \oper{B}] \Leftrightarrow 
A \star B - B \star A
\equiv \ii \moyal{A, B}.
\end{gather}

The Moyal product and the Moyal bracket are particularly handy when \m{\pd_{x} \pd_{k} \sim \epsilon \ll 1}. (Sometimes, \m{\epsilon} is equated with the GO parameter \m{\lambda/L}, but note that \m{\star} is symplectically invariant and GO is not. One might notice, though, that this \m{\epsilon} is similar to the MGO parameter; cf.\ \Sec{sec:mgogen}.) Since \m{\oper{\mc{L}} = \bigO(\epsilon)}, one can express the Moyal product as an asymptotic series in powers of~\m{\epsilon}:
\begin{equation}
 \star = \oper{1} + \ii \oper{\mc{L}} /2 - \oper{\mc{L}}^2/8 + \bigO(\epsilon^3).
\end{equation}
Accordingly, the Moyal bracket can be approximated with the Poisson bracket:
\begin{equation}
 \moyal{A, B} = \poisson{A, B} + \bigO(\epsilon^3)
\end{equation}
(assuming that \m{A} and \m{B} are scalars or commuting matrices). Likewise, the symbol of any operator in the limit \m{\epsilon \to 0} can be obtained simply by replacing \m{\oper{x}} with \m{x} and \m{\oper{k}} with \m{k}.

\subsection{Wigner functions} 
\label{sec:wigner}

Of particular interest in the Weyl symbol calculus are `density operators'
\begin{gather}
\oper{W}^{(\psi)} \doteq (2\upi)^{-1} \ket{\psi}\bra{\psi}
\end{gather}
for various \m{\ket{\psi}}. (The factor \m{2\upi} is sometimes omitted, depending on a convention.) The symbol of such an operator, called the Wigner function, 
\begin{gather}
W^{(\psi)} \doteq \symb \oper{W}^{(\psi)},
\end{gather}
is a real function and can be expressed as follows:
\begin{subequations}
\begin{align}
W^{(\psi)}(x, k) 
& = \frac{1}{2\upi} \int \dd s\, 
\psi_{\oper{x}}(x + s/2)\, \psi_{\oper{x}}^*(x - s/2)\,\ee^{-\ii k s}
\\
& = \frac{1}{2\upi} \int \dd r\,
\psi_{\oper{k}}(k + r/2)\, \psi_{\oper{k}}^*(k - r/2)\,\ee^{\ii r x}.
\end{align}
\end{subequations}
(Like in the previous formulas, \m{2\upi} here must be replaced with \m{(2\upi)^n} if \m{x} is \m{n}-dimensional.) In particular, note the two special cases:
\begin{gather}\label{eq:Weig}
 W^{(\psi)}(x, k) = \frac{1}{2\upi} 
 \left\lbrace
 \begin{array}{cc}
 \delta(x - x_0), & \ket{\psi} = \ket{\ev_{\oper{x}}(x_0)},\\
 \delta(k - k_0), & \ket{\psi} = \ket{\ev_{\oper{k}}(k_0)}.
 \end{array}
 \right.
\end{gather}

Any function bilinear in~\m{\psi_{\oper{x}}} and~\m{\psi^*_{\oper{x}}} can be expressed through \m{W^{(\psi)}}. For example, for any operators~\m{\oper{L}} and~\m{\oper{R}}, one has
\begin{align}
(\oper{L}\psi_{\oper{x}}(x))(\oper{R}\psi_{\oper{x}}(x))^*
 = \int \dd k\,{L}(x, k) \star W^{(\psi)}(x, k) \star {R}^*(x, k),
\label{eq:LR1}
\end{align}
where \m{{L}} and \m{{R}} are the corresponding symbols. As a corollary, one has
\begin{gather}\label{eq:wigfprop1}
|\psi_{\oper{x}}(x)|^2 = \int \dd k\,W^{(\psi)}(x, k),
\qquad
|\psi_{\oper{k}}(k)|^2 = \int \dd x\,W^{(\psi)}(x, k).
\end{gather}
If, for a wavefield, \m{|\psi_{\oper{x}}(x)|^2} and \m{|\psi_{\oper{k}}(k)|^2} are interpreted as the densities of quanta in the \m{x}-space and the \m{k}-space, respectively, one can attribute \m{W^{(\psi)}} as a quasiprobability distribution of wave quanta in phase space. The prefix `quasi' is commonly added because \m{W^{(\psi)}} can be negative, which is not something that one expects from a probability distribution. That said, a Wigner function averaged over a sufficiently large phase-space volume \m{\Delta x\,\Delta k \gtrsim 1} is guaranteed to be nonnegative. Such a function can be understood as the spectrum of the two-point correlation function \m{\favr{\psi_{\oper{x}}(x + s/2) \psi_{\oper{x}}^*(x - s/2)}} over \m{s} at given \m{x} and represents a local property of the field. For further details, see, for example, \citep{my:ql}.

\section{Metaplectic transform}
\label{sec:mt}

Now that we have introduced the Weyl symbol calculus and Wigner functions, let us expand on the MT definition introduced in \Sec{sec:mtintro}. What follows is not intended as a comprehensive theory of MTs, and neither do we always adhere to the accepted mathematical terminology. Readers are encouraged to consider this section as a presentation of the tools needed specifically for our formulation of MGO. If it helps, any parallels with the already existing theory of MTs can be considered accidental.

\subsection{Definition} 
\label{sec:Mdef}

Consider a transformation of the base operators to some new base operators:
\begin{gather}\label{eq:trf}
\oper{\sz} \equiv \left(
\begin{array}{c}
\oper{x}\\ \oper{k}
\end{array}
\right)
\mapsto 
\left(
\begin{array}{c}
\oper{q}\\ \oper{p}
\end{array}
\right)
\equiv \oper{\sy}.
\end{gather}
Here, \m{\oper{q}} can be defined as any Hermitian operator whose eigenvalue space is \m{\mathbb{R}}, and \m{\oper{p}} is defined such that its \m{q}-representation is \m{\oper{p}_{\oper{q}} = -\ii \pd_q}. Accordingly, 
\begin{gather}\label{eq:qpcomm}
[\oper{q}, \oper{p}] = \ii,
\end{gather}
and thus the corresponding symbols \m{q(x, k)} and \m{p(x, k)} satisfy\footnote{Using the terminology to be introduced \Sec{sec:transform}, \m{\sy(\sz) \equiv (q(x, k), p(x, k))^\T} is a \m{\sz}-symbol of \m{\sy}, to be denoted \m{\sy_\sz}. Likewise, \m{\sz(\sy) \equiv (x(q, p), k(q, p))^\T} is a \m{\sy}-symbol of \m{\sz}, to be denoted \m{\sz_\sy}.}
\begin{gather}\label{eq:QP}
\moyal{{q, p}} = 1. 
\end{gather}
In the GO limit, this corresponds to a variable transformation
\begin{gather}\label{eq:cct}
\sz \equiv \left(
\begin{array}{c}
x\\k
\end{array}
\right)
\mapsto 
\left(
\begin{array}{c}
q\\p
\end{array}
\right)
\equiv \sy
\end{gather}
with the following Jacobian matrix:
\begin{gather}
\Xi \doteq \left(
\begin{array}{cc}
\pd_x q & \pd_k q\\
\pd_x p & \pd_k p\\
\end{array}
\right).
\end{gather}
Since  \m{\Xi \sJ \Xi^\T = \poisson{q, p} \sJ} and \m{\poisson{q, p} \to 1} by the GO limit of \Eq{eq:QP}, this transformation conserves the symplectic form and therefore is canonical. Thus, \eq{eq:trf} can be interpreted as a canonical transformation for operators.\footnote{Note that, beyond the GO limit, having \m{\poisson{q, p} = 1} does not guarantee \eq{eq:qpcomm} and vice versa.} Also, as a reminder, canonical transformations conserve the Poisson bracket for any \m{A} and \m{B}; \ie
\begin{equation}\label{eq:poisson}
\poisson{A, B} 
\equiv \frac{\pd A}{\pd \sz^\alpha}\,\sJ^{\alpha\beta}\,\frac{\pd B}{\pd \sz^\beta}
= \frac{\pd A}{\pd \sy^\alpha}\,\sJ^{\alpha\beta}\,\frac{\pd B}{\pd \sy^\beta}.
\end{equation}

The operator transformation \eq{eq:trf} induces the following transformation of fields:
\begin{gather}\label{eq:MT1}
\underbrace{\braket{\ev_{\oper{q}}(q)| \psi}}_{\psi_{\oper{q}}(q)}
= \int \dd x
\underbrace{\braket{\ev_{\oper{q}}(q) | \ev_{\oper{x}}(x)}}_{M_{\oper{q} \mapsfrom \oper{x}}(q, x)}
\underbrace{\braket{\ev_{\oper{x}}(x) | \psi}}_{\psi_{\oper{x}}(x)}.
\end{gather}
This can also be written as
\begin{gather}\label{eq:psiM}
\psi_{\oper{q}}(q) = \int \dd x\,M(q, x)\,\psi_{\oper{x}}(x),
\qquad
M(q, x) \equiv M_{\oper{q} \mapsfrom \oper{x}}(q, x).
\end{gather}
When \m{\oper{q} = \oper{k}} (and thus \m{\oper{p} = -\oper{x}}), the metaplectic transform is simply the Fourier transform, as seen by comparing \eq{eq:MT1} with \eq{eq:ft}. Since the function \m{M_{\oper{q} \mapsfrom \oper{x}}(q, x)} defines the corresponding operator \m{\oper{M}_{\oper{q} \mapsfrom \oper{x}}}, one can as well express \Eq{eq:psiM} as\footnote{\label{ft:MTQ}As a special case, \eq{eq:cct} can represent the evolution of a classical Hamiltonian system, with \m{(x, k) \equiv (x_0, k_0)} as the initial coordinates and \m{(q, p) \equiv (x_T, k_T)} as the coordinates at time \m{T}. Then, the MT is the propagator for the wavefunction of the corresponding quantum system.}
\begin{gather}\label{eq:pMp}
\psi_{\oper{q}} = \oper{M} \psi_{\oper{x}},
\qquad
\oper{M} \equiv \oper{M}_{\oper{q} \mapsfrom \oper{x}}.
\end{gather}
We call this operator the MT induced by the operator transformation \eq{eq:trf}.\footnote{Some authors use other approaches to defining MTs, particularly, MTs corresponding to specific variable transformations. It is not our goal to overview this subject here, but see, for example, \citep{ref:littlejohn86, ref:mello75} and the references therein.} Its inverse is the metaplectic operator \m{(\oper{M}_{\oper{q} \mapsfrom \oper{x}})^{-1} = \oper{M}_{\oper{x} \mapsfrom \oper{q}} \equiv \oper{\inv{M}}} induced by the inverse transformation \m{(\oper{q}, \oper{p}) \mapsto (\oper{x}, \oper{k})} and has the kernel 
\begin{gather}\label{eq:Minv}
M_{\oper{x} \mapsfrom \oper{q}}(x, q) \equiv \inv{M}(x, q)
= \braket{\ev_{\oper{x}}(x) | \ev_{\oper{q}}(q)}
= M_{\oper{q} \mapsfrom \oper{x}}^*(q, x).
\end{gather}
By \Eq{eq:Oxxd}, this means that \m{
(\oper{M}_{\oper{q} \mapsfrom \oper{x}})^{-1} = 
\oper{M}_{\oper{q} \mapsfrom \oper{x}}^\dag,
} \ie \m{\oper{M}} is unitary.

A sequence of metaplectic transformations (MTs) \m{\oper{M}_1\oper{M}_0} corresponding to the operator transformations \m{\oper{x}_0 \mapsto \oper{x}_1 \mapsto \oper{x}_2} is an MT \m{\oper{M}_2} corresponding to the operator transformation \m{\oper{x}_0 \mapsto \oper{x}_2}, because
\begin{align}
\psi_{\oper{x}_2}(x_2) 
& = \int \dd x_1\,M_1(x_2, x_1)\,\psi_{\oper{x}_1}(x_1)
\notag\\
& = \int \dd x_1\,\dd x_0\,M_1(x_2, x_1)\,M_0(x_1, x_0)\psi_{\oper{x}_0}(x_0)
\notag\\
& = \int \dd x_0\,\mc{M}(x_2, x_0)\psi_{\oper{x}_0}(x_0),
\end{align}
where 
\begin{align}
\mc{M}(x_2, x_0) 
& \doteq \int \dd x_1\,M_1(x_2, x_1)\,M_0(x_1, x_0)
\notag\\
& = \int \dd x_1
\braket{\ev_{\oper{x}_2}(x_2) | \ev_{\oper{x}_1}(x_1)}
\braket{\ev_{\oper{x}_1}(x_1) | \ev_{\oper{x}_0}(x_0)}
\notag\\
& =
\braket{\ev_{\oper{x}_2}(x_2) | \ev_{\oper{x}_0}(x_0)}
\vphantom{\int}
\notag\\
& = M_2(x_2, x_0).
\vphantom{\int}
\end{align}
Thus, MTs form a group.

\subsection{\tMwaves} 
\label{sec:mwaves}

The function \m{M(q, x) \equiv \braket{\ev_{\oper{q}}(q) | \ev_{\oper{x}}(x)}} can be understood as the \m{q}-representation of \m{\ket{\ev_{\oper{x}}(x)}}. Also, by definition, 
\begin{gather}
\oper{x}\ket{\ev_{\oper{x}}(x)} = x \ket{\ev_{\oper{x}}(x)}.
\end{gather}
This leads to
\begin{gather}
(\oper{x}_{\oper{q}} M)(q, x) = x M(q, x).
\end{gather}
Let us symbolically represent \m{\oper{x}_{\oper{q}}} through the new base operators \m{\oper{q}_{\oper{q}} = q} and \m{\oper{p}_{\oper{q}} = -\ii \pd_q} as \m{X(q, -\ii \pd_q)}. This leads to the following equation:
\begin{gather}\label{eq:xM}
X(q, -\ii \pd_q) M(q, x) = x M(q, x).
\end{gather}

Equation \eq{eq:xM} is a (pseudo)differential equation for \m{M} as a function of \m{q}, with \m{x} as a parameter. Thus, the solution of \Eq{eq:xM} is defined up only to an arbitrary function of \m{x} and another equation for \m{M} is needed to specify this function. To derive such an equation, note that \m{\inv{M}(x, q) \equiv \braket{\ev_{\oper{x}}(x) | \ev_{\oper{q}}(q)} \equiv M^*(q, x)} can be understood as the \m{x}-representation of \m{\ket{\ev_{\oper{q}}(q)}}. Hence, on the one hand,
\begin{gather}
(\oper{k}_{\oper{x}} \inv{M})(x, q) = -\ii \pd_x \inv{M}(x, q)
 = - \ii \pd_x M^*(q, x).
\end{gather}
On the other hand,
\begin{gather}
(\oper{k}_{\oper{x}} \inv{M})(x, q) 
= \braket{\ev_{\oper{x}}(x) | \oper{k} | \ev_{\oper{q}}(q)}
= \braket{\ev_{\oper{q}}(q) | \oper{k} | \ev_{\oper{x}}(x)}{}\!^*
= (\oper{k}_{\oper{q}} M)^*(q, x).
\end{gather}
Thus, \m{(\oper{k}_{\oper{q}} M)(q, x) = \ii \pd_x M(q, x)}, or, assuming the notation \m{\oper{k}_{\oper{q}} = K(q, -\ii \pd_q)},
\begin{gather}\label{eq:kM}
K(q, -\ii \pd_q)M(q, x) = \ii \pd_x M(q, x).
\end{gather}
Equation \eq{eq:kM} can be interpreted as a Schr\"odinger equation (SE) for \m{M}, with \m{x} serving as the time variable. In this sense, \m{M} is a \textit{wave}, which we call an \Mwave. Likewise, in retrospect, \eq{eq:xM} can be interpreted as an SE for a stationary wave with `energy' \m{x}.

Together, \eq{eq:xM} and \eq{eq:kM} define \m{M} up to a constant complex factor. From the normalization condition, one finds that
\begin{align}
\delta(x' - x'')
& = \braket{\ev_{\oper{x}}(x')|\ev_{\oper{x}}(x'')} 
\notag\\
& = \int \dd q 
\braket{\ev_{\oper{x}}(x')|\ev_{\oper{q}}(q)} 
\braket{\ev_{\oper{q}}(q)|\ev_{\oper{x}}(x'')} 
\notag\\
& = \int \dd q \,
M^*(q, x') M(q, x'').
\label{eq:Mdelta}
\end{align}
This means that, for any given \m{x}, \m{M} satisfies
\begin{gather}\label{eq:Mnorm}
\int \dd q \,\dd s\, M^*(q, x + s/2) M(q, x - s/2) = 1.
\end{gather}
(This is also seen from the definition of \m{M} as a representation of a normalised eigenvector.) This defines the normalization factor up to a constant phase, which can be anything depending how the eigenvectors of \m{\oper{q}} are chosen relative to those of \m{\oper{x}}. This choice constitutes a gauge freedom of the theory.

Similar equations apply to the inverse-MT kernel \m{\inv{M}(x, q)}. That is, assuming the notation \m{\oper{q}_{\oper{x}} = Q(x, -\ii \pd_x)} and \m{\oper{p}_{\oper{x}} = P(x, -\ii \pd_x)}, one has
\begin{align}
Q(x, -\ii \pd_x) \inv{M}(x, q) & = q \inv{M}(x, q),
\label{eq:xMi}
\\
P(x, -\ii \pd_x) \inv{M}(x, q) & = \ii \pd_q \inv{M}(x, q).
\label{eq:kMi}
\end{align}
Equation \eq{eq:xMi} can be interpreted as an SE for a wave \m{\inv{M}}, with \m{q} serving as the time variable. Because \m{\inv{M}(x, q) \equiv M^*(x, q)}, we will also call it an \Mwave. Also, \eq{eq:xMi} can be interpreted as an SE for a stationary wave with `energy' \m{q}. Like before, one also finds 
\begin{align}
\delta(q' - q'')
& = \braket{\ev_{\oper{q}}(q')|\ev_{\oper{q}}(q'')} 
\notag\\
& = \int \dd x 
\braket{\ev_{\oper{q}}(q')|\ev_{\oper{x}}(x)} 
\braket{\ev_{\oper{x}}(x)|\ev_{\oper{q}}(q'')} 
\notag\\
& = \int \dd x \,
\inv{M}^*(x, q') \inv{M}(x, q''),
\label{eq:Mbd}
\end{align}
whence
\begin{gather}\label{eq:Mnormi}
\int \dd x\,\dd s\,\inv{M}^*(x, q + s/2) \inv{M}(x, q - s/2) = 1.
\end{gather}

Let us summarise our intermediate results. \Mwaves are normalised solutions of SEs \eq{eq:kM} and \eq{eq:kMi}, which can also be written in the form \eq{eq:Dpsi}:
\begin{subequations}\label{eq:Sgt0}
\begin{alignat}{2}
& \oper{\mc{H}}_{M} M = 0, 
\qquad
&& \oper{\mc{H}}_{M} \doteq  K(q, -\ii \pd_q) - \ii \pd_x,
\\
& \oper{\mc{H}}_{\inv{M}} \inv{M} = 0,
\qquad
&& \oper{\mc{H}}_{\inv{M}} \doteq P(x, -\ii \pd_x) - \ii \pd_q.\label{eq:PinvM}
\end{alignat}
\end{subequations}
Their solutions can be formally expressed as follows:
\begin{gather}
M(q, x) = \exp(-\ii x\oper{k}_{\oper{q}}) M(q, 0),
\qquad
\inv{M}(x, q) = \exp(-\ii q\oper{p}_{\oper{x}}) \inv{M}(x, 0).
\end{gather}
It is generally impossible to find explicit analytic expressions for the propagators \m{\exp(-\ii x\oper{k}_{\oper{q}})} and \m{\exp(-\ii q\oper{p}_{\oper{x}})} for a given variable transformation. However, it is possible to do so for some important special cases and also asymptotically. This will be discussed later, in \Secs{eq:mtexamples}  and \ref{sec:mwavego}.

\subsection{Symplectic pseudo-measure} 

As a function of two variables, an \Mwave can be considered as a field on the two-dimensional `spacetime' \m{(x, q)}. The variable \m{x} serves as the time variable in the Schr\"odinger equation \eq{eq:kM}, so \m{\ii \pd_x} serves as the frequency operator and \m{-k} serves as the frequency variable. Hence, MGO naturally involves Wigner functions of \m{M} in the form 
\begin{gather}
2\upi W^{(M)}_\sz(x, q, -k, p) \equiv \MWf(\sy, \sz)
\end{gather}
(the utility of the added coefficient \m{2\upi} will become clear shortly), \ie
\begin{gather}\label{eq:Wdef}
\MWf(\sy, \sz) = \frac{1}{2\upi} \int \dd s\,\dd s'\,
M(q + s'/2, x + s/2) \, M^*(q - s'/2, x - s/2)\,\ee^{-\ii p s' + \ii k s}.
\end{gather}
For the inverse transformation, the corresponding function is
\begin{align}
\inv{\MWf}(\sz, \sy) 
& = \frac{1}{2\upi} \int \dd s\,\dd s'\,
\inv{M}(x + s/2, q + s'/2) \, \inv{M}^*(x - s/2, q - s'/2)\,\ee^{-\ii k s + \ii p s'}
\notag\\
& = \frac{1}{2\upi} \int \dd s\,\dd s'\,
M^*(q + s'/2, x + s/2) \, M(q - s'/2, x - s/2)\,\ee^{\ii p s' - \ii k s}
\notag\\
& = \MWf^*(\sy, \sz).\vphantom{\int}
\label{eq:aux112}
\end{align}
Like any Wigner function, \m{\MWf} is real, so one can also express \eq{eq:aux112} simply as
\begin{gather}\label{eq:muinv}
\inv{\MWf}(\sz, \sy)  = \MWf(\sy, \sz).
\end{gather}
Also note that
\begin{align}
\int \MWf(\sy, \sz)\,\dd \sz & = \int \dd x\,\dd s\,\dd s'\,
M(q + s'/2, x + s/2) \, M^*(q - s'/2, x - s/2)\,\delta(s)\,\ee^{-\ii p s'}
\notag\\
& = \int \dd s'\,\ee^{-\ii p s'}\int \dd x\,
M(q + s'/2, x) \, M^*(q - s'/2, x)
\notag\\
& = \int \dd s'\,\ee^{-\ii p s'}\,\delta(s')
\notag\\
& = 1
\label{eq:Wint}
\end{align}
(here we used \eq{eq:Mbd}) and, similarly, \m{\int \MWf(\sy, \sz)\,\dd\sy = 1} too. In summary then,
\begin{gather}\label{eq:dW}
\int \MWf(\sy, \sz)\,\dd \sz = \int \MWf(\sy, \sz)\,\dd \sy = 1.
\end{gather}
As to be seen below (\Sec{sec:transform}), these serve as measures on the \m{\sz}-space and the \m{\sy}-space, respectively (strictly speaking, pseudo-measures, because \m{\MWf} can be negative). Below, the arguments of \m{\MWf} will be occasionally omitted where they are obvious from the context.

\subsection{Transformation of operators}
\label{sec:transform}

A metaplectic transform induces a transformation of the operator representation \m{\oper{O}_{\oper{x}} \mapsto \oper{O}_{\oper{q}}}, which is found as follows. Note that
\begin{align}
(\oper{O}_{\oper{q}}\psi_{\oper{q}})(q) 
& = \braket{\ev_{\oper{q}}(q)|\oper{O}|\psi}
\notag\\
& = \int \dd q'
\braket{\ev_{\oper{q}}(q)| \oper{O} | \ev_{\oper{q}}(q')}
\braket{\ev_{\oper{q}}(q') | \psi}.
\end{align}
Using \Eq{eq:Oxx}, one can also represent this as follows:
\begin{align}
(\oper{O}_{\oper{q}}\psi_{\oper{q}})(q) 
& = \int \dd q'\,\dd x'\,\dd x''
\braket{\ev_{\oper{q}}(q)| \ev_{\oper{x}}(x')} 
O_{\oper{x}}(x', x'')
\braket{\ev_{\oper{x}}(x'')| \ev_{\oper{q}}(q')}
\psi_{\oper{q}}(q')
\notag\\
& = \int \dd q'\,\dd x'\,\dd x''\,
M(q, x')\, O_{\oper{x}}(x', x'')\, M^*(q', x'')\, \psi_{\oper{q}}(q')
\notag\\
& = \int \dd q'\,
(\oper{M}\oper{O}_{\oper{x}} \oper{M}^\dag)(q, q') \,\psi_{\oper{q}}(q'),
\end{align}
whence
\begin{gather}\label{eq:Oqx}
\oper{O}_{\oper{q}} = \oper{M}\oper{O}_{\oper{x}} \oper{M}^\dag.
\end{gather}

Let us use the following formula for the Weyl symbol of a given operator (cf.~\eq{eq:wsymbx}):
\begin{gather}
O_\sy(q, p) = \int \dd s'\,
\braket{
\ev_{\oper{q}}(q + s'/2)
| \oper{O} |
\ev_{\oper{q}}(q - s'/2)
}\ee^{-\ii p s'}.
\end{gather}
The subindex in the symbol's notation refers to the fact that this symbol is defined on the phase space \m{\sy \doteq (q, p)^\T}. We will call it the \m{\sy}-symbol of \m{\oper{O}}, and the symbol defined on the phase space \m{\sz \doteq (x, k)^\T} will be called the \m{\sz}-symbol of \m{\oper{O}}. Recall that
\begin{gather}
\oper{O} = \frac{1}{2\upi}
\int \dd x\,\dd k\,\dd s\,
\ket{\ev_{\oper{x}}(x + s/2)}
O_\sz(x, k)
\bra{\ev_{\oper{x}}(x - s/2)}
\ee^{\ii k s}.
\end{gather}
Then,
\begin{align}
O_\sy(q, p) 
& = \frac{1}{2\upi} \int \dd x\,\dd k\,\dd s\,\dd s'\,
\braket{\ev_{\oper{q}}(q + s'/2)|\ev_{\oper{x}}(x + s/2)}
\braket{\ev_{\oper{x}}(x - s/2) | \ev_{\oper{q}}(q - s'/2)}
\notag\\
&\hspace{5cm}\times O_\sz(x, k)\,\ee^{-\ii p s' + \ii k s}.
\end{align}
Expressing the brackets through \m{M}, one obtains the following expression for the \m{\sy}-symbol in terms of the \m{\sz}-symbol:
\begin{align}
O_\sy(q, p) & = \frac{1}{2\upi} \int \dd x\,\dd k\,\dd s\,\dd s'\,
M(q + s'/2, x + s/2) \, M^*(q - s'/2, x - s/2)
\notag\\
&\hspace{5cm}\times O_\sz(x, k)\,\ee^{-\ii p s' + \ii k s},
\label{eq:OO}
\end{align}
and the inverse transformation can be written similarly. Equivalently, this result can be expressed in the following compact form:
\begin{gather}\label{eq:OW}
O_\sy(\sy) = \int O_\sz(\sz)\,\MWf(\sy, \sz)\,\dd \sz,
\qquad
O_\sz(\sz) = \int O_\sy(\sy)\,\MWf(\sy, \sz)\,\dd \sy.
\end{gather}
(If \m{O_\sz \equiv 1}, then \eq{eq:OO} readily yields \m{O_\sy \equiv 1}, and vice versa, by \eq{eq:dW}.) In particular, these allow one to recalculate the Wigner functions from one representation to another.

\subsection{Examples}
\label{eq:mtexamples}

In this section, we present several examples of MTs that are used as building blocks of the theory presented in later sections.

\subsubsection{Shift}
\label{sec:pshift}

For example, let us consider a variable transformation that is a phase-space shift by a constant vector \m{\Delta \doteq (\Delta_q, \Delta_p)^\T}:
\begin{gather}
\oper{q} = \oper{x} + \Delta_q \oper{1},
\qquad
\oper{p} = \oper{k} + \Delta_p \oper{1}.
\end{gather}
This corresponds to
\begin{gather}
X(q, -\ii \pd_q) = q - \Delta_q,
\qquad
K(q, -\ii \pd_q) = -\ii \pd_q - \Delta_p.
\end{gather}
Then, \eq{eq:xM} requires that \m{M(q, x) = C(q)\delta(q - x  - \Delta_q)}, where \m{C} is some function. From \eq{eq:kM}, one finds that \m{(-\ii \pd_q - \ii \pd_x - \Delta_p)M(q, x) = 0}, whence \m{\pd_q C = \ii \Delta_p C}. With \eq{eq:Mnorm} for the normalization, this gives
\begin{gather}
M(q, p) = \ee^{\ii \Delta_p q}\,\delta(q - x  - \Delta_q)
\end{gather}
(up to an arbitrary constant phase factor, which we choose equal to one), and, therefore,
\begin{gather}
\psi_{\oper{q}}(q) = \ee^{\ii \Delta_p q}\psi_{\oper{x}}(q - \Delta_q).
\end{gather}
Notice that even at \m{\Delta_q = 0}, when \m{q} is the same as \m{x}, the function \m{\psi_{\oper{q}}} is not the same as \m{\psi_{\oper{x}}} due to nonzero \m{\Delta_p}. Also,
\begin{align}
\MWf & = \frac{1}{2\upi} \int \dd s\,\dd s'\,
\ee^{\ii \Delta_p s' - \ii p s' + \ii k s}\,\delta(q + s'/2 - x - s/2  - \Delta_q)\,\delta(q - s'/2 - x + s/2  - \Delta_q)
\notag\\
& = \frac{1}{2\upi} \int \dd s\,\dd s'\,
\ee^{\ii (\Delta_p - p) s' + \ii k s}\,\delta(q - x - \Delta_q)\,\delta(s - s')
\notag\\
& = \frac{1}{2\upi} \int \dd s\,
\ee^{\ii (k + \Delta_p - p) s}\,\delta(q - x - \Delta_q)
\notag\\
& = \delta(x - q + \Delta_q)\,\delta(k - p + \Delta_p).
\end{align}
Then, by \eq{eq:OW}, one obtains
\begin{gather}
O_\sy(q, p) = O_\sz(q - \Delta_q, p - \Delta_p) = O_\sz(x(q, p), k(q, p)).
\end{gather}
Simply put, this means that \m{O_\sy(\sy) = O_\sz(\sz)}; \ie the symbol of an operator is invariant with respect to phase-space shifts.

\subsubsection{Rescaling}
\label{sec:resc}

Let us consider a rescaling canonical transformation:
\begin{gather}
\oper{q} = \oper{x}/\alpha,
\qquad
\oper{p} = \alpha\oper{k},
\end{gather}
where \m{\alpha} is a nonzero real constant. (The same factor appears in both equations to keep the transformation canonical.) Clearly, \m{M(q, x) = C\delta(q - x/\alpha)}, where \m{C} is a constant. From \eq{eq:Mnorm}, one finds that \m{C = |\alpha|^{-1/2}} up to an arbitrary phase, so
\begin{gather}
M(q, x) = |\alpha|^{-1/2}\,\delta(q - x/\alpha),
\qquad
\psi_{\oper{q}}(q) = |\alpha|^{1/2} \psi_{\oper{x}}(\alpha q).
\end{gather}

\subsubsection{Linear symplectic transformation}
\label{sec:lst}

Let us also consider a linear symplectic transformation (LST), that is, 
\begin{gather}\label{eq:ls}
\left(
\begin{array}{c}
\oper{q} \\ \oper{p}
\end{array}
\right)
=
\underbrace{
\left(
\begin{array}{cc}
A & B\\
C & D
\end{array}
\right)
}_\sS
\left(
\begin{array}{c}
\oper{x} \\ \oper{k}
\end{array}
\right),
\qquad
\end{gather}
where the (constant) coefficients satisfy
\begin{gather}\label{eq:abcd}
AD - BC = 1, 
\end{gather}
\ie \m{\det \sS = 1}.\footnote{Additional constraints apply at \m{\dim x > 1}, when \m{A}, \m{B}, \m{C}, and \m{D} are matrices. See, for example, \citep{my:nimt}.} This subsumes relabeling coordinates and momenta; specifically, \m{A = D = 0} and \m{B = - C = \pm 1} corresponds to the transformation \m{(\oper{q}, \oper{p}) = (\pm \oper{k}, \mp \oper{x})}. (Note that \m{\oper{k}} and \m{\oper{x}} necessarily enter the latter formula with opposite signs to ensure symplecticity.) The rescaling transformation considered in \Sec{sec:resc} is also subsumed as a limit (\m{B \to 0}, \m{C \to 0}, and \m{A \to 1/D}).

Equivalently, \eq{eq:ls} can be represented in the following inverted form:
\begin{gather}
\left(
\begin{array}{c}
\oper{x} \\ \oper{k}
\end{array}
\right)
=
\underbrace{\left(
\begin{array}{cc}
\inv{A} & \inv{B}\\
\inv{C} & \inv{D}
\end{array}
\right)}_{\sSi}
\left(
\begin{array}{c}
\oper{q} \\ \oper{p}
\end{array}
\right),
\qquad
\end{gather}
where \m{\sSi = \sS^{-1}} is also symplectic; specifically,
\begin{gather}\label{eq:ssm}
\left(
\begin{array}{cc}
\inv{A} & \inv{B}\\
\inv{C} & \inv{D}
\end{array}
\right)
=
\left(
\begin{array}{cc}
D & -B\\
-C & A
\end{array}
\right),
\end{gather}
so \m{\det \sSi = \det \sS = 1}. Then, \eq{eq:xM} and \eq{eq:kM} yields
\begin{gather}
(\inv{A}q - \ii \inv{B} \pd_q - x) M(q, x) = 0,
\qquad
(\inv{C}q - \ii \inv{D} \pd_q - \ii \pd_x) M(q, x) = 0.
\end{gather}
With the normalization \eq{eq:Mnorm} taken into account, a straightforward calculation yields the well known result \citep{ref:littlejohn86}:
\begin{gather}\label{eq:MLST}
M(q, x) = (2\upi \ii \inv{B})^{-1/2}\, \exp\left(- \frac{\ii(\inv{A}q^2 - 2xq + \inv{D}x^2)}{2\inv{B}}\right).
\end{gather}
As usual, an arbitrary constant phase can be added to the overall expression, depending on a convention. The phase of the square root as a function of \m{B} must be chosen such that MTs remain a group. Loosely speaking, one can require that a phase-space rotation\footnote{Formally, a rotation matrix corresponds to \m{\inv{A} = \inv{D} = \cos\alpha} and \m{-\inv{C} = \inv{D} = \sin\alpha}; hence, the rotation angle is usually defined as \m{\alpha = \arg(\inv{A} + \ii \inv{B})}. Remember, though, that \m{\inv{A}} and \m{\inv{B}} generally have different units (or otherwise are normalised arbitrarily). Thus this expression for \m{\alpha} is questionable unless \m{\inv{A}} or \m{\inv{B}} is zero, \ie unless \m{\alpha} is \m{\upi/2} times an integer.} by \m{\pm 2\upi} causes the phase of \m{\inv{B}} to change by \m{\pm 2\upi} too; then, the phase of \m{\inv{B}^{1/2}} changes by \m{\pm \upi}. (For a more detailed discussion, see \citep{ref:littlejohn86}.) Then, the MT corresponding to a complete phase-space rotation in either direction is not an identity operator \m{\oper{1}} but \m{-\oper{1}}.

The corresponding transformation of symbols is calculated as follows. First, note that
\begin{align}
M(q + s'/2, x + s/2)\,& M^*(q - s'/2, x - s/2) \,\ee^{- \ii p s' + \ii k s}
\notag\\
& = \frac{1}{2\upi |\inv{B}|}\, 
\exp\left(\frac{\ii s(q - \inv{D}x + \inv{B}k) + \ii s'(x - \inv{A} q - \inv{B}p)}{\inv{B}}\right),
\end{align}
and
\begin{align}
\MWf & = \int 
\frac{\dd s'}{2\upi |\inv{B}|}\,\exp\left(\frac{\ii s'(x - \inv{A} q - \inv{B}p)}{\inv{B}}\right)
\int \frac{\dd s}{2\upi}\,
\exp\left(\ii s\left(\frac{q - \inv{D}x}{\inv{B}} + k\right)\right)
\notag\\
& = \delta(x - \inv{A} q - \inv{B}p)\, \delta\!\left(k - \frac{\inv{D} x - q}{\inv{B}}\right)
\notag\\
& = \delta(x - \inv{A} q - \inv{B}p)\, \delta\!\left(k - \frac{\inv{D}(\inv{A} q + \inv{B}p) - q}{\inv{B}}\right)
\notag\\
& = \delta(x - \inv{A} q - \inv{B}p)\, \delta(k - \inv{C}q - \inv{D}p),
\label{eq:WLST}
\end{align}
where we used \eq{eq:abcd}. Then, by \eq{eq:OW}, one obtains
\begin{gather}\label{eq:sinv}
O_\sy(q, p) = O_\sz(\inv{A}q + \inv{B}p, \inv{C}q + \inv{D}x)
= O_\sz(x(q, p), k(q, p)).
\end{gather}
This shows that the symbols of operators are invariant with respect to LSTs. In particular, Wigner functions are LST-invariant. Also, for \m{\oper{O} = \oper{\sy}}, \eq{eq:sinv} gives \m{\sz_\sy = \sSi \sy} and, similarly, \m{\sy_\sz = \sS \sz}; \ie the symbols of the base operators transform at LSTs exactly as the base operators themselves.

\subsubsection{Eikonal transform} 
\label{sec:eiktr}

Let us also consider the `eikonal transform' (this term  will become clear shortly):
\begin{gather}\label{eq:eikvt}
\oper{q} = \oper{x},
\qquad
\oper{p} = \oper{k} - \kc(\oper{x}) = \oper{k} - \kc(\oper{q}),
\end{gather}
which satisfies the commutation relation \eq{eq:qpcomm}. Like in \Sec{sec:pshift}, one finds that
\begin{gather}
M(q, p) = \ee^{-\ii \theta(q)}\,\delta(q - x),
\end{gather}
where \m{\theta \doteq \int \kc(q)\,\dd q = \int \kc(x)\,\dd x}. Then, \m{\psi_{\oper{q}}} is given by
\begin{gather}
\psi_{\oper{q}}(q) = \psi_{\oper{q}}(x) = \ee^{-\ii \theta(x)}\psi_{\oper{x}}(x),
\end{gather}
and
\begin{align}
\MWf & = \frac{1}{2\upi} \int \dd s\,\dd s'\,
\ee^{-\ii \theta(q + s'/2) + \ii \theta(q - s'/2) - \ii p s' + \ii k s}\,
\delta(q + s'/2 - x - s/2)\delta(q - s'/2 - x + s/2)
\notag\\
& = \frac{1}{2\upi} \int \dd s\,\dd s'\,
\ee^{-\ii \theta(q + s'/2) + \ii \theta(q - s'/2) - \ii p s' + \ii k s}\,
\delta(s' - s)\delta(q - x)
\notag\\
& = \frac{1}{2\upi} \int \dd s\,
\ee^{-\ii \theta(q + s/2) + \ii \theta(q - s/2) + \ii (k - p) s}\,
\delta(q - x).
\end{align}
Then, by \eq{eq:OW}, the operator symbols are transformed as follows: 
\begin{gather}\label{eq:auxenv}
O_\sy(q, p) = \frac{1}{2\upi} \int \dd k\,\dd s\,
 O_\sz(q, k)\,\ee^{\ii (k - p)s -\ii \theta(q + s/2) + \ii \theta(q - s/2)}.
\end{gather}
Using Taylor expansion, one has
\begin{gather}
\theta\left(q + \frac{s}{2}\right) - \theta\left(q - \frac{s}{2}\right) 
= \kc(q)\, s + \frac{1}{24}\frac{\pd^2 \kc(q)}{\pd q^2}\, s^3 + \ldots,
\end{gather}
where we used \m{\pd_q \theta(q) = \kc(q)}. Thus, 
\begin{align}
O_\sy(q, p) 
& = \frac{1}{2\upi} \int \dd k\,\dd s\,
 O_\sz(q, k)\,
 \left(1 - \frac{\ii}{24}\frac{\pd^2 \kc(q)}{\pd q^2}\,s^3 + \ldots\right)\ee^{\ii (k - p - \kc(q))s}
\notag\\
& = \frac{1}{2\upi} \left(1 - \frac{1}{24}\frac{\pd^2 \kc(q)}{\pd q^2}\,\frac{\pd^3}{\pd p^3} + \ldots\right)\int \dd k\,\dd s\,
 O_\sz(q, k)\,
 \ee^{\ii (k - p - \kc(q))s}
\notag\\
& = \frac{1}{2\upi} \int \dd k\,\dd s\,
 O_\sz(q, k)\,
 \ee^{\ii (k - p - \kc(q))s} + \bigO(\epsilon^2)
\notag\\
& = O_\sz(q, p + \kc(q)) + \bigO(\epsilon^2).
\vphantom{\int}
\label{eq:envs}
\end{align}

In particular, if \m{\theta(q)} is the eikonal and \m{\psi_{\oper{q}}} is the wave envelope, then \m{\kc} serves as the reference wavevector, and the symbol transformation reproduces the one derived, for example, in \citep{my:quasiop1}. The importance of \eq{eq:envs} in this case is in that, \textit{within GO, where \m{\bigO(\epsilon^2)} is negligible}, one may use
\begin{gather}\label{eq:eiksym}
O_\sy(q, p) 
\approx O_\sz(q, p + \kc(q))
= O_\sz(x(q, p), k(q, p)).
\end{gather}

\section{Geometrical optics through metaplectic transforms}
\label{sec:go}

\subsection{Basic equations}
\label{sec:trad}

The Weyl symbol calculus and MTs provide a natural framework for formulating GO equations \citep{book:tracy}. Here, we restate the conventional GO \citep{ref:mcdonald88, my:quasiop1}, which will be an important reference point for our further discussion. Let us start with a generic linear wave equation\footnote{Specific \m{\oper{D}_{\oper{x}}} that govern electromagnetic waves in dielectric media, plasma in particular, are discussed, for example, in \citep[section VI]{my:quasiop1}.}
\begin{gather}\label{eq:Dpsi}
\oper{D}_{\oper{x}}\psi_{\oper{x}} = 0.
\end{gather}
Here, \m{\psi_{\oper{x}}} is a field on some \m{x}-space (generally, spacetime), which we assume one-dimensional for simplicity, and \m{\oper{D}_{\oper{x}}} is some dispersion operator, which can be an integral operator or a differential operator as a special case. Suppose that \m{\psi_{\oper{x}}} is quasimonochromatic, \ie representable~as
\begin{gather}
\psi_{\oper{x}}(x) = \ee^{\ii\theta(x)}\Psi(x),
\end{gather}
where the (real) phase \m{\theta} is fast compared to its gradient \m{\kc \doteq \pd_x \theta} and to the envelope~\m{\Psi}. Our goal is to derive an approximate equation for this envelope, which can be written as
\begin{gather}\label{eq:enveq}
\oper{\mc{D}}_{\oper{x}}\Psi = 0,
\qquad
\oper{\mc{D}} \doteq \ee^{-\ii\theta(\oper{x})}\oper{D}\ee^{\ii\theta(\oper{x})}.
\end{gather}

As we have just discussed (\Sec{sec:eiktr}), the mapping from \m{\psi_{\oper{x}}} to \m{\Psi} can be considered as an MT that corresponds to the variable transformation \eq{eq:eikvt}, so\footnote{Here, \m{k(x)} denotes a given \textit{function} of \m{x}, while \m{k} without an argument denotes a \textit{variable}. This abuse of notation will be continued below and adopted for other variables as well.}
\begin{gather}\label{eq:Hpr}
\oper{\mc{D}}_{\oper{x}} = \oper{D}_{\oper{q}}, 
\qquad
\oper{x} = \oper{q},
\qquad
\oper{p} = \oper{k} - k(\oper{x}),
\end{gather}
\ie the envelope-evolution operator is just the \m{q}-representation of the original `full-wave' dispersion operator \m{\oper{D}}. Then, its symbol can be calculated using \eq{eq:eiksym}. Since  \m{\Psi} that \m{\oper{\mc{D}}_{\oper{x}}} acts upon is smooth, only small values of \m{p} matter in this case, so one can further approximate \eq{eq:eiksym} with its first-order Taylor expansion:\footnote{The second-order Taylor expansion is used for modeling transverse diffraction of quasioptical beams in multiple dimensions. Assuming that the transverse scale of a beam is much smaller than its longitudinal scale, \eq{eq:eiksym} remains applicable in that case \citep{my:quasiop1}.}
\begin{gather}
D_\sy(q, p) \approx D_\sz(q, k(q)) + p V(q),
\qquad
V(q) \doteq (\pd_p D_\sz(q, p))_{p = k(q)}.
\end{gather}
Loosely, the expansion parameter here is the ratio of the characteristic wavenumber \m{p} of the envelope and \m{k(q)}. Denoting the former as the inverse inhomogeneity scale, \m{1/L}, and introducing the wavelength \m{\lambda \sim 1/k(q)}, this small parameter can be roughly estimated as
\begin{gather}\label{eq:epsgo}
\epsilon \sim \lambda/L.
\end{gather}

Using the analog of \eq{eq:khsym} for base operators \m{\oper{q}} and \m{\oper{p}}, one immediately obtains
\begin{gather}
\oper{D}_{\oper{q}} \approx D_\sz(q, k(q)) - \ii V(q) \pd_q 
- \frac{\ii}{2}\,V'(q),
\end{gather}
where the prime denotes the derivative with respect to the argument, so \mm{V'(q) = (\pd^2_{qp} D_\sz(q, p) + p'(q)\,\pd^2_{pp} D_\sz(q, p))_{p = p(q)}}. Using \eq{eq:Hpr}, one can then rewrite \eq{eq:enveq} as follows:
\begin{gather}\label{eq:Leq}
(H_\sz(x) + \ii \Gamma_\sz(x))\Psi - \ii V(x) \pd_x \Psi - \frac{\ii}{2}\,V'(x)\Psi \approx 0,
\end{gather}
where we introduced \m{H_\sz(x) \doteq \re D_\sz(x, k(x))} and \m{\Gamma_\sz(x) \doteq \im D_\sz(x, k(x))}. (For vector waves, these would be the Hermitian and anti-Hermitian parts of \m{D_\sz(q, k(q))}, correspondingly.) It is common to assume that \m{\re D_\sz} is an order-one function, while \m{\im D_\sz} is order-\m{\epsilon}. Also, remember that \m{\pd_x \Psi} and \m{V'} are order-\m{\epsilon} as well, so one can redefine \m{V}~as
\begin{gather}
V(x) \doteq (\pd_k H_\sz(x, k))_{k = k(x)}
\end{gather}
without loss of accuracy. Then all coefficients in \eq{eq:Leq} are real, except \m{\ii} \perse.

Note that the function \m{\theta} has been unspecified so far. One can choose it such that\footnote{For vector waves, one might want to adopt a slightly different equation for \m{k(x)}. We revisit this subject in a broader context in \Sec{sec:mgogeneral}.}
\begin{gather}\label{eq:dc}
H_\sz(\sz) = 0
\end{gather}
(here, \m{\sz \equiv (x, k)^\T}, as usual), which can be considered as a local dispersion relation. For given initial conditions, this determines the `reference ray' trajectory \m{k(x)}, which can also be described by Hamilton's ray equations (\App{app:req})
\begin{gather}\label{eq:raysz}
\dot{\sz} = \sJ \pd_\sz H_\sz \equiv \poisson{\sz, H_\sz}.
\end{gather}
The dot in \m{\dot{\sz}} denotes a derivative with respect to the ray time (which can be different from the physical time \m{t}; see \App{app:req}) and \m{H_\sz} serves as the ray Hamiltonian. Accordingly, we henceforth address \m{\oper{H}}, \ie the Hermitian part of \m{\oper{D}}, as the wave Hamiltonian and \m{\oper{\Gamma}}, \ie the anti-Hermitian part of \m{\oper{D}}, as the dissipation operator. 

The reference ray determines the coefficients in \eq{eq:Leq}, which becomes
\begin{gather}\label{eq:act0}
V(x) \pd_x \Psi + \frac{1}{2}\,V'(x)\Psi = \Gamma_\sz(x) \Psi,
\qquad
\pd_x (V(x) |\Psi|^2) = 2 \Gamma_\sz(x) |\Psi|^2.
\end{gather}
(As a reminder, \m{V'(x)} in these equations is generally not just \m{\pd^2_{xk} H_\sz} but also includes a term \m{k'(x)\,\pd^2_{kk} H_\sz}.) Since the coefficients in \eq{eq:act0} are real, the phase of \m{\Psi} is conserved and one also readily finds \m{|\Psi|}:
\begin{gather}\label{eq:soln}
|\Psi|^2 = \frac{C}{V(x)} \exp\left(\int \frac{2\Gamma_\sz(x)}{V(x)}\,\dd x\right),
\end{gather}
where the constant \m{C} is determined by the initial conditions. In particular, if \m{\oper{D}} is Hermitian, \eq{eq:act0} becomes a conservation law:
\begin{gather}\label{eq:act}
\pd_x (V(x) |\Psi|^2) = 0.
\end{gather}
For stationary waves, this can be understood as energy conservation. More generally, when \m{x} is time or a multi-dimensional spacetime variable, \eq{eq:act} represents conservation of the wave action, which is conserved even when the energy is not. For details, see, for example, \citep[section 7]{my:ql}.  A generalization to vector waves is discussed in \citep{my:quasiop1} and, in the broader context of MGO, in \Sec{sec:mgogeneral}.

Although often convenient, this model significantly relies on the smallness of the GO parameter \eq{eq:epsgo}. Near cutoffs, where the local \m{\lambda} is large (infinite), \eq{eq:soln} predicts \m{|\Psi|^2 \sim 1/V(x) \to \infty}, which indicates limitations of said approximation. In this case, it is useful to promote \eq{eq:Dpsi} to an abstract vector equation,
\begin{gather}\label{eq:Dpsi2}
\oper{D} \ket{\psi} = 0,
\end{gather}
and seek an alternative representation of \eq{eq:Dpsi2} in which the corresponding wavelength changes with the new coordinate slowly or not at all. We will discuss this in \Sec{sec:nat}, after we have introduced some more analytical tools in the next sections.

\subsection{Geometrical optics of \tMwaves: special case}
\label{sec:mwavegoS}

It is instructive to apply the above approach to \Mwaves in particular, which can be studied using GO methods like any other waves. In this section, we consider an important special case where the new variables \m{(q, p)} are linked to \m{H_\sz} as follows. (Related transformations will be considered in \Secs{sec:aac2} and \ref{sec:mgogen}.) 

Like in \Sec{sec:trad}, suppose that the reference-ray trajectory is given by some \m{k = k(x)}. Then, near the reference ray, one has\footnote{Basically, this is Hayes's representation \citep{ref:hayes73, my:qponder} of \m{H_\sz}.}
\begin{gather}\label{eq:hayesH}
H_\sz(\sz) \approx V(x)(k - k(x)),
\end{gather}
where \m{V(x) \doteq (\pd_k H_\sz)_{k = k(x)}} can be interpreted as the group velocity on the reference ray. Let us choose the new variables \m{(q, p)} such that \m{p = H_\sz}, whence \m{k} can be readily expressed as a function of \m{(x, p)}:
\begin{gather}\label{eq:ph}
k(x, p) = k(x) + p/V(x).
\end{gather}
To find the corresponding canonical coordinate \m{q}, we look for a generating function in the type-2 form, \m{F = F(x, p)}, such that \citep{book:goldstein}
\begin{gather}\label{eq:pt}
k(x, p) = \pd_x F(x, p), 
\qquad
q(x, p) = \pd_p F(x, p).
\end{gather}
The former gives 
\begin{gather}
F(x, p) = \int^{x}_0 k(\tilde{x}, p)\,\dd \tilde{x},
\label{eq:varthetaF}
\end{gather}
where we set the integration constant to zero by choice. (The tilde is used to distinguish the dummy integration variable \m{\tilde{x}} from the actual canonical coordinate \m{x}.) Then,
\begin{gather}\label{eq:tauqh}
q(x, p) = \int^{x}_0 \frac{\pd k(\tilde{x}, p)}{\pd p}\,\dd \tilde{x} 
= \int^{x}_0 \frac{\dd \tilde{x}}{V(\tilde{x})} \equiv q(x).
\end{gather}

Now we can derive the corresponding approximation for \Mwaves. One way to do this is to perform an eikonal MT on \m{M} and directly apply the results of \Sec{sec:trad}. But since we have already introduced the expansion \eq{eq:hayesH}, one might as well notice that the Weyl transform of \eq{eq:hayesH} readily yields the following approximation of \m{\oper{H}_{\oper{x}}}:
\begin{gather}
\oper{H}_{\oper{x}} \approx -\ii V(x) \pd_x - \frac{\ii}{2}\,V'(x) - V(x)k(x).
\end{gather}
Then, \eq{eq:kMi} leads to the following equations for \m{\inv{M} \equiv \inv{M}(x, q)}:
\begin{gather}
\ii \pd_q \inv{M} = \left(-\ii V(x) \pd_x - \frac{\ii}{2}\,V'(x) - V(x)k(x)\right)\inv{M}.
\end{gather}
Also, \eq{eq:xMi} gives \m{(q - q(x))\inv{M}(x, q) = 0}, whence \m{\inv{M}(x, q)\, \propto\, \delta(q - q(x))}. Thus,
\begin{gather}\label{eq:bMGO}
\inv{M}(x, q) = \frac{\ee^{\ii \vartheta(x)}}{\sqrt{V(x)}}\,\delta(q - q(x)),
\qquad
\vartheta(x) \doteq \int^{x}_0 k(\tilde{x})\,\dd \tilde{x},
\end{gather}
where the normalization is chosen such that \eq{eq:Mbd} is satisfied:
\begin{align}
\int \dd x \,\inv{M}^*(x, q') \inv{M}(x, q'')
& = \int \frac{\dd x}{|V(x)|} \,\delta(q' - q(x))\,\delta(q'' - q(x))
\notag\\
& = \delta(q' - q'') \int \frac{\dd x}{|V(x)|} \,\delta(q' - q(x))
\notag\\
& = \delta(q' - q'') \int \dd q\,\delta(q' - q)
\notag\\
& = \delta(q' - q'').
\end{align}

Note that \eq{eq:bMGO} readily provides a general GO solution to \eq{eq:Dpsi}. Indeed, according to \eq{eq:Dpsi2}, a wave \m{\ket{\psi}} is a eigenstate of \m{\oper{H} = \oper{p}} corresponding to the eigenvalue \m{p = 0}. Then, the \m{q}-representation of this field is
\begin{gather}\label{eq:psia}
\psi_{\oper{q}}(q) = a \braket{\ev_{\oper{q}}(q) | \ev_{\oper{p}}(0)}
= \frac{a}{\sqrt{2\upi}} = \const,
\end{gather}
where \m{a} is a constant amplitude and we have used \eq{eq:xqme}. Accordingly, 
\begin{gather}\label{eq:psixqW}
\psi_{\oper{x}}(x) = \int \dd q\,\inv{M}(x, q)\,\psi_{\oper{q}}(q)
= \psi_{\oper{q}} \int \dd q\,\inv{M}(x, q)
= \psi_{\oper{q}}\,\frac{\ee^{\ii \vartheta(x)}}{\sqrt{V(x)}}.
\end{gather}
This coincides with the well-known WKB solution for eikonal waves, with \m{\psi_{\oper{q}}} serving as a constant factor determined by the initial conditions.

Of course, the same result can as well be obtained by directly applying GO to \m{\psi_{\oper{x}}}. However, the advantage of using \Mwaves is that it separates the problem of calculating the evolution of \m{\ket{\psi}} from calculating its representation. Errors that may emerge from the inaccuracy of the mapping \m{\psi_{\oper{q}} \mapsto \psi_{\oper{x}}} \textit{do not accumulate} (except, possibly, in the phase). For example, even though \eq{eq:psixqW} fails when \m{V} turns to zero, its validity gets reinstated as soon as \m{V} becomes large enough. In other words, one does not need to integrate the amplitude equation \textit{through} singularity. One needs to solve continuously only for \m{\psi_{\oper{q}}} (which is constant here but can be more complicated in general) while mapping it to \m{\psi_{\oper{x}}} only occasionally, when it is easy to do with a desired accuracy.

\subsection{Geometrical optics of \tMwaves: generic case}
\label{sec:mwavego}

Let us now consider a generic case when GO \Mwaves are not delta-shaped but quasimonochromatic in both \m{x} and \m{q}.\footnote{The relation between this case and the case considered in \Sec{sec:mwavegoS} is explained at the end of this section.} Then, one can search for a solution of \eq{eq:Sgt0} in the eikonal form, as usual:
\begin{gather}
\inv{M}(x, q) \equiv M^*(q, x) = \ee^{\ii \Theta(x, q)}\mcc{M}(x, q),
\end{gather}
where \m{\Theta} is a real phase (eikonal) and \m{\mcc{M}} is a real amplitude. Such \Mwaves are well studied in quantum mechanics; see, for example, \citep[section~II]{ref:miller74}. Here, we offer a concise derivation of those known results from an alternative perspective.

Assuming that \m{\Theta} is fast compared to \m{\mcc{M}}, it satisfies the two `local dispersion relations', or Hamilton--Jacobi equations, that flow from \eq{eq:Sgt0}:
\begin{subequations}\label{eq:Sgt}
\begin{align}
K(q, -\pd_q\Theta) & = \pd_x\Theta,\\
P(x, \pd_x\Theta)  & = -\pd_q\Theta.
\end{align}
\end{subequations}
Since \m{K(q, -\pd_q\Theta) = K(q, p) = k} and \m{P(x, \pd_x\Theta) = P(x, k) = p}, \eq{eq:Sgt} can be expressed~as
\begin{gather}\label{eq:kp}
k = \pd_x\Theta(x, q),
\qquad
p = -\pd_q\Theta(x, q).
\end{gather}
As a function that satisfies \eq{eq:kp}, \m{\Theta(x, q)} can be recognised as the type-1 generating function of the canonical transformation \eq{eq:cct}. Generating functions of other types emerge if one re-attributes the old or (and) new coordinates as the momenta and vice versa, with the appropriate sign changes to preserve symplecticity (\Sec{sec:lst}). For example, in canonical variables \m{(q', p') \doteq (p, -q)}, \eq{eq:kp} becomes
\begin{gather}
k = \pd_x\Theta(x, p'),
\qquad
q' = \pd_{p'}\Theta(x, p'),
\end{gather}
which makes \m{\Theta} a type-2 generating function. Likewise, using \m{(x', k') \doteq (k, -x)}, one has
\begin{gather}
x' = -\pd_{k'}\Theta(k', q),
\qquad
p = -\pd_q\Theta(k', q),
\end{gather}
so \m{\Theta} becomes a type-3 generating function. Finally, using both \m{(q', p')} and \m{(x', k')}, one can write
\begin{gather}
x' = -\pd_{k'}\Theta(k', p'),
\qquad
q' = \pd_{p'}\Theta(k', p'),
\end{gather}
in which case \m{\Theta} serves as a type-4 generating function. A reader interested in brushing up on this topic is referred to \citep{book:goldstein}.

Since \m{K(q, -\ii \pd_q)} serves as the Hamiltonian for the dynamics in `time' \m{x}, the symbol \m{K(q, p)} serves as the ray Hamiltonian. Since this Hamiltonian does not depend on \m{x} explicitly, it is conserved on rays, \m{\dd_x K = 0}, where \m{\dd_x} is the convective derivative associated with the `group velocity' \m{K_p(x, q) \doteq (\pd_p K)(q, p(x, q))} as a field on the `spacetime' \m{(q, x)}. Similarly, \m{P(x, -\ii \pd_x)} serves as the Hamiltonian for the dynamics in `time' \m{q}. Since this Hamiltonian does not depend on \m{q} explicitly, it is conserved on rays, \m{\dd_q P = 0}, where \m{\dd_q} is now the convective derivative associated with the `group velocity' \mm{P_k(x, q) \doteq (\pd_k P)(x, k(x, q))} as a field on the `spacetime' \m{(x, q)}. From
\begin{subequations}
\begin{align}
& 0 = \dd_x K = \pd_x k(x, q) + K_p \pd_q k(x, q),
\\
& 0 = \dd_q P = \pd_q p(x, q) + P_k \pd_x p(x, q),
\end{align}
\end{subequations}
one obtains the following equations that will be useful in the next section:
\begin{subequations}\label{eq:vgKP}
\begin{gather}
K_p(x, q) = - \frac{\pd_x k(x, q)}{\pd_q k(x, q)} 
= - \frac{\Theta_{xx}}{\Theta_{xq}},
\\
P_k(x, q) = - \frac{\pd_q p(x, q)}{\pd_x p(x, q)} 
= - \frac{\Theta_{qq}}{\Theta_{xq}}.
\end{gather}
\end{subequations}
The lower indices in \m{\Theta \equiv \Theta(x, q)} denote the corresponding partial derivatives, and the argument \m{(x, q)} will now be occasionally omitted for brevity.

To calculate the envelope \m{\mcc{M}}, let us start with writing the `action conservation theorems' \eq{eq:act} for \Mwaves in the form
\begin{subequations}\label{eq:Mmcc}
\begin{gather}
\pd_x \mcc{M}^2 + (\pd_q K_p) \mcc{M}^2 + K_p \pd_q \mcc{M}^2 = 0,
\\
\pd_q \mcc{M}^2 + (\pd_x P_k) \mcc{M}^2 + P_k \pd_x \mcc{M}^2 = 0.
\end{gather}
\end{subequations}
Substituting \m{\pd_q \mcc{M}^2} from the second equation into the first one yields
\begin{gather}\label{eq:lnM}
\ln \mcc{M}^2 = \int \dd x\,\frac{K_p(\pd_x P_k) - \pd_q K_p}{1 - P_k K_p}
= \ln |\Theta_{xq}| + \const,
\end{gather}
where the latter equality is obtained by direct calculation using \eq{eq:vgKP}. The integration constant is independent of \m{x}. Since one can equally arrive at \eq{eq:lnM} by excluding \m{\pd_x \mcc{M}^2} instead of \m{\pd_q \mcc{M}^2} and integrating over \m{q} instead of \m{x}, this constant cannot depend on \m{q} either. Thus, the GO solution for \Mwaves is
\begin{gather}\label{eq:Mampl}
\inv{M}(x, q) = C \sqrt{\Theta_{xq}}\,\ee^{\ii\Theta}.
\end{gather}
The factor \m{C} is determined by the normalization condition \eq{eq:Mnormi}:
\begin{align}
1 \approx |C|^2 \int \dd x\,\dd s |\Theta_{xq}(x, q)|\,\ee^{-\ii \Theta_{q}(x, q) s}
= |C|^2 \int \dd \xi\,\dd s\,\ee^{-\ii \xi s}
= 2\upi |C|^2,
\end{align}
so \m{|C| = (2\upi)^{-1/2}}. (Here, we introduced a variable transformation \m{x \mapsto \xi \doteq \Theta_q(x, q)} at fixed \m{q} and used that \m{\dd\xi = \Theta_{qx}\dd x}.) Then, up to a piecewise-constant phase, one has
\begin{gather}\label{eq:Mampl2}
\inv{M}(x, q) = \sqrt{\frac{\Theta_{xq}}{2\upi}}\,\ee^{\ii\Theta},
\end{gather}
and, as a reminder, \m{\Theta_{xq} = \pd_q k = - \pd_x p}.\footnote{For a transformation \m{(x, k) \mapsto (q, p)} that represents the evolution of a classical Hamiltonian system (see the footnote on p.~\pageref{ft:MTQ}), equation \eq{eq:Mampl2} is equivalent to the Van Vleck approximation for the propagator of the corresponding Schr\"odinger equation; for example, see \citep{ref:blair22}.} The initial phase is determined by the preferred gauge (\Sec{sec:mwaves}). Once it is set, the remaining phase is determined by the branch of the square root and changes by \m{\pm \upi/2} when \m{\Theta_{xq}} goes through zero. This is the standard Maslov-index problem, which we will not revisit here, but see \citep[p.~85-86]{ref:miller74}. 

One might wonder why \eq{eq:Mampl2} is not delta-shaped unlike the result in \Sec{sec:mwavegoS}, which is based on similar approximations. The explanation is as follows. In \Sec{sec:mwavegoS}, we used a type-2 generating function \m{F(x, p)} to obtain \m{q = q(x, p)} and \m{k = k(x, p)}. Normally, one can invert the former to obtain \m{p = p(x, q)} and \m{k = k(x, p(x, q))}; then one can reformulate this as a type-1 transformation, and the results of the present section apply. However, the specific transformation in \Sec{sec:mwavegoS} is singular in that, by \eq{eq:tauqh}, our specific \m{q(x, p)} happens to be independent of \m{p}. This makes the inversion is impossible and leads to a delta-shaped \Mwave \eq{eq:bMGO}.

An example illustrating \eq{eq:Mampl2} will be discussed in \Sec{sec:mjphi}, and this formula is also useful for deriving the quantization condition for closed orbits (\Sec{app:qz}). However, keep in mind that the \eq{eq:Mampl2} is constructed only as the leading-order approximation and we will also need alternative approximations of \Mwaves for our purposes (\Sec{sec:mgogen}).

\section{Natural coordinates}
\label{sec:nat}

Now let us discuss how to use MTs for reduced wave modeling beyond the realm of conventional GO. This section is focused on introducing the key concepts. Specifics, such as envelope equations, and examples will be presented in \Sec{sec:mgogen}.

\subsection{Preliminaries}

\subsubsection{Tangent space}
\label{sec:tang}

Our goal here is to construct a phase-space coordinate system aligned with the reference ray. We will assume that the ray is governed by the Hamiltonian \m{H_\sz} and thus satisfies \m{H_\sz(\sz) = 0}, like in \Sec{sec:trad}. (However, other Hamiltonians can be advantageous or even necessary in some cases, as will be discussed in \Sec{sec:mgogeneral}.) Consider a narrow region of a fixed phase-space point \m{\sz_0} that satisfies \m{H_\sz(\sz_0) = 0}. Assuming that \m{H_\sz} is sufficiently smooth (the quantitative condition for this will be discussed in \Sec{sec:mgo}), let us consider its Taylor expansion in \m{\sz - \sz_0}:
\begin{gather}\label{eq:doplit}
H_\sz(\sz) = (\sz - \sz_0) \cdot \underbrace{\pd_\sz H_\sz(\sz_0)}_{-\sJ \sv} +\, \dE,
\qquad
\sv \doteq \sJ \pd_\sz H_\sz(\sz_0),
\end{gather}
where we used \m{\sJ\sJ = -\sI}. Here, \m{\sv} is understood as the phase-space velocity at \m{\sz_0}, so we will also write it as \m{\sv = \dot{\sz}_0}. (The index 0 denotes that the corresponding quantity is evaluated at \m{\sz_0}.) The term \m{\dE} subsumes the second- and higher-order terms of the Taylor series. Albeit nonnegligible, it is small at small \m{\sz - \sz_0}, so GO applicability is determined entirely by the first term, \m{-(\sz - \sz_0) \cdot \sJ \sv}. Because this term is linear in \m{\sz}, a linear variable transformation is generally enough to reinstate GO at least locally. This is done as follows.

Assuming the same notation as earlier, let us adopt the operator transformation \eq{eq:trf} in the form
\begin{gather}\label{eq:yzS}
\oper{\sy} = \sS (\oper{\sz} - \sz_0),
\end{gather}
where \m{\sS} is a symplectic matrix. As discussed in \Sec{sec:lst}, this corresponds to the phase-space transformation \m{\sy(\sz) \equiv \sy_\sz = \sS (\sz - \sz_0)}, or, equivalently, \m{\sz(\sy) \equiv \sz_\sy = \sz_0 + \sSi\sy}, where \m{\sSi} is a symplectic matrix that is the inverse of \m{\sS}. Let us represent \m{\sSi} through its columns \m{\seta_q} and \m{\seta_p},
\begin{gather}\label{eq:sinve}
\sSi = \left(
\begin{array}{cc}
\seta_q & \seta_p
\end{array}
\right),
\end{gather}
and assume the notation \m{\sy = (q, p)^\T}, as before. Then, \m{q} and \m{p} can be understood as the components of \m{\sz - \sz_0} in the basis formed by the vectors \m{\seta_q} and \m{\seta_p}:
\begin{gather}
\sz - \sz_0 = \seta_q q + \seta_p p.
\end{gather}
Also, the transformation \eq{eq:yzS} can be expressed through \eq{eq:sinve} using symplecticity of \m{\sS}; that is, inverting \m{\sS \sJ \sS^\T = \sJ} yields \m{\sS = - \sJ\sSi^\T\sJ} (see also \eq{eq:ssm}).

Let us choose \m{\seta_q} such that it is parallel to \m{\sv}, \ie the \m{q}-axis is tangent to the ray,
\begin{gather}\label{eq:sqvdef}
\seta_q = \sv/v.
\end{gather}
Because of this, we will call \m{\sy} a tangent phase space. The nonzero scalar \m{v} can be anything for now (but this freedom will be removed in \Sec{sec:mgo} for nonlinear coordinate transformations, where we also comment on the physical meaning of \m{v}). It may be tempting to choose \m{\seta_p} such that it is perpendicular to \m{\seta_q}; however, angles are not well defined on a symplectic space for the lack of a metric.\footnote{Since \m{x} and \m{k} have different units, angles on the \m{\sz} plane are unit-dependent, and the linear term in \eq{eq:doplit} has no natural units. Natural units will be introduced in \Sec{sec:mgo} by studying~\m{\dE}.} Instead, the only natural constraint on \m{\seta_p} is that \m{\sSi} must be symplectic, \ie \m{\sJ = \sSi^\T \sJ \sSi}:
\begin{gather}
\sJ
= 
\left(
\begin{array}{c}
\seta_q^\T 
\\
\seta_p^\T
\end{array}
\right)
\sJ
\left(
\begin{array}{cc}
\seta_q & \seta_p
\end{array}
\right)
=
\left(
\begin{array}{cc}
\seta_q \cdot \sJ \seta_q & \seta_q \cdot \sJ \seta_p\\
\seta_p \cdot \sJ \seta_q & \seta_p \cdot \sJ \seta_p\\
\end{array}
\right)
= (\seta_q \wedge \seta_p) \sJ,
\end{gather}
where \m{\symp{a} \wedge \symp{b} \doteq \symp{a} \cdot \sJ \symp{b} = - \symp{b} \wedge \symp{a}}. Hence, the requirement that \m{\sSi} must be symplectic is equivalent to the requirement that \m{\seta_q \wedge \seta_p = 1}, whence
\begin{gather}\label{eq:etap}
\seta_p = \su/u,
\qquad
u \doteq \sv \wedge \su/v.
\end{gather}
We choose \m{\su} such that \m{u} is nonzero. Assuming the notation \m{\sv = (a_1, a_2)^\T} and \m{\su = (b_1, b_2)^\T}, one has \m{\sv \wedge \su = a_1 b_2 - a_2 b_1}, so having nonzero \m{u} is equivalent to having \m{\su} linearly independent from \m{\sv}, \ie having \eq{eq:sinve} nondegenerate. Other than that, \m{\su} can be chosen arbitrarily (but see also \Sec{sec:mgo}). 

Because the transformation \eq{eq:sinve} is linear, it preserves the symbol of \m{\oper{H}}, so \m{H_\sy(\sy) = H_\sz(\sz) = \dE - (\sSi \sy)^\T \sJ \sv}. Notice that 
\begin{align}
- \sy^\T
\left(
\begin{array}{c}
\seta_q^\T\\
\seta_p^\T
\end{array}
\right) 
\sJ \sv 
= - \sy^\T
\left(
\begin{array}{c}
\sv \wedge \sv/v\\
\su \wedge \sv/u
\end{array}
\right)
=
 \left(
\begin{array}{cc}
q & p
\end{array}
\right) 
\left(
\begin{array}{c}
0\\
v
\end{array}
\right)
= vp,
\end{align}
so, in summary, \m{H_\sy(\sy) = \dE + vp}. The corresponding ray equations are
\begin{gather}\label{eq:raysy}
\dot{\sy} = \sJ \pd_\sy H_\sy \equiv \poisson{\sy, H_\sy},
\end{gather}
or, explicitly,
\begin{gather}
\dot{q} = v + \pd_p \dE,
\qquad
\dot{p} = - \pd_q \dE.
\end{gather}
Since \m{\dE = \bigO((\sz - \sz_0)^2)}, these lead to \m{\dot{q}_0 = v} and \m{\dot{p}_0 = 0}. In this sense, \m{v} can be understood as the `symplectic speed' at \m{\sz_0}. However, \m{v} must not be confused with the length of the phase-space-velocity vector \m{\sv}, because the vector length is generally undefined on a symplectic space (see the footnote on \pageref{ft:sp}). 

Also note that, since \m{\dE} is small, the wavelength \m{2\upi/p} evolves slowly. This justifies the GO approximation and \eq{eq:raysy} \textit{a~posteriori}. As one can also see easily, \eq{eq:raysy} are equivalent to the ray equations in the \m{x}-representation, \eq{eq:raysz}. In other words, \textit{the ray equations \eq{eq:raysz} remain applicable even near cutoffs}, where GO is inapplicable in the \m{x}-representation. In practice, this fact is often taken for granted, but it is nontrivial and requires justification. As seen from the above argument, such `unreasonable effectiveness' of the ray equations is due to the existence of a symplectic transformation \m{\sz \mapsto \sy} that reinstates GO locally for any smooth \m{H_\sz} while conserving the Poisson bracket and the dispersion-operator symbol.

With GO reinstated in the \m{q}-representation, one can also search for the wavefield in the eikonal form \m{\psi_{\oper{q}} = \ee^{\ii\theta(q)}\Psi(q)} and derive the envelope operator \m{\oper{\mc{H}}_\sy} for the envelope \m{\Psi} in the \m{q}-space the same way the envelope equation was derived in \Sec{sec:trad} in the \m{x}-space. If \m{\dE} is small \textit{globally}, \ie if \m{H_\sz} is mostly linear in \m{\sz} at all \m{\sz}, this envelope equation will remain sufficient indefinitely. However, such dispersion operators are rarely found in practice. A typical \m{H_\sz} is a nonlinear functions of phase-space variables, so the choice of the \m{q}-representation that is convenient at some \m{\sz} is likely to become inconvenient at other \m{\sz}. Then a single linear transformation is not enough.

\begin{figure*}
\centering
\includegraphics[width=.9\textwidth]{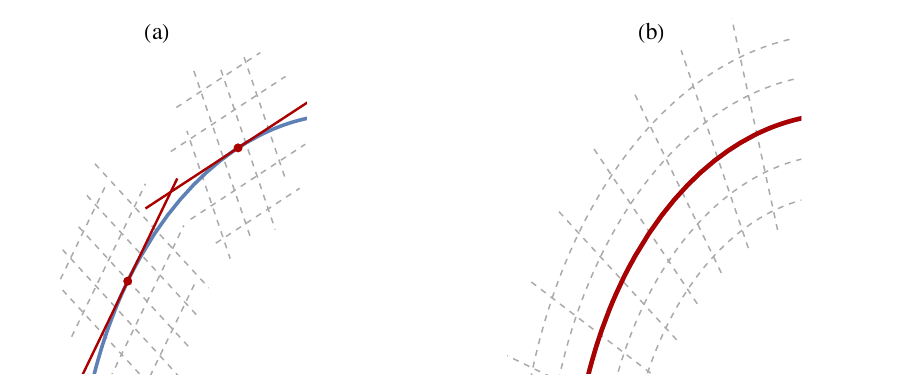}
\caption{\label{fig:spaces}
The two ways of constructing the \m{\sy}-coordinate grids around a reference ray (blue; coincides with the dispersion surface) in the \m{\sz}-space: (a)~The grids are constructed in patches near predefined locations (red dots) on the ray via linear variable transformations. The \m{q}-axes (red) are tangent to the ray. (b)~A single grid is constructed via a nonlinear variable transformation. The coordinate axis (red) coincides with the ray indefinitely. 
}
\end{figure*}

There are two ways to deal with this problem. One is to define different linear transformations at different points \m{\sz_0^{(i)}} along the ray and use MTs to map the wavefield from the \m{q}-space at \m{\sz_0^{(i)}} to the \m{q}-space at \m{\sz_0^{(i + 1)}} (\Fig{fig:spaces}(a)). For example, one can choose \m{\sz_0^{(i + 1)}} to be infinitesimally close to \m{\sz_0^{(i)}}; then the MT for remapping the field will be a near-identity transformation, which can be convenient. This approach is summarised in \citep{my:mgoinv}. The alternative is to use just one but \textit{nonlinear} transformation such that the new coordinate space is aligned with the ray indefinitely (\Fig{fig:spaces}(b)). No field remapping is needed in this case other than to initialise the field in the new coordinate space space and to output the final results back to the \m{x}-space. This is the approach that we discuss below.

\subsubsection{Ray path as a coordinate}
\label{sec:mgo}

Let us consider a generic ray that is curved, \ie has \m{\ddot{\sz}} linearly independent from \m{\dot{\sz}} in the region of interest. (Straight rays can be handled as a limit; see below.) Then, at any given \m{\sz_0} on the ray, the vectors 
\begin{gather}\label{eq:uv2}
\sv = \dot{\sz}_0, 
\qquad
\su = \ddot{\sz}_0
\end{gather}
form a symplectic basis\footnote{A similar approach is used in \citep{ref:kamran09}. Also note that here we assume that \m{\sz} is two-dimensional for simplicity. In case of a higher-dimensional phase space \m{\sZ}, one can think of our \m{\sz}-space as the osculating plane spanned by \m{\dot{\sZ}_0} and~\m{\ddot{\sZ}_0}.}
\begin{gather}\label{eq:ss}
\seta_q = \sv/v,
\qquad
\seta_p = \su/u,
\qquad
\seta_q  \wedge \seta_p = 1,
\end{gather}
where \m{u \doteq \sv \wedge \su/v} and the scalar \m{v} remains to be defined. This basis can be used to construct new ray-aligned coordinates \m{\sr = (\tau, h)} (here, `\m{\sr}' stands for `ray') such that the ray itself would be an isoline of the new momentum \m{h}. Then, one can derive the envelope equation as a PDE in the new-coordinate representation. The envelope remains slow in this representation because the reference wavevector satisfies the dispersion relation indefinitely and there are no cutoffs in this representation by construction of \m{\sr }.

Since we are interested only in the local dynamics on the ray, it will be sufficient for us to find \m{\sr(\sz)} and said PDE near a predefined generic \m{\sz_0} on the ray. Then, it is convenient to search for the mapping \m{\sz \mapsto \sr} as a superposition of two transformations:
\begin{gather}\label{eq:split}
\underbrace{(x, k)^\T}_{\sz} \mapsto 
\underbrace{(\yq, \yp)^\T}_{\sy}  \mapsto 
\underbrace{(\tau, h)^\T}_\sr.
\end{gather}
The first one, \m{\sy = \sSi (\sz - \sz_0)}, is the same as in \Sec{sec:tang}, except now \m{\su} is specified by \eq{eq:uv2}; \ie the \m{\yq} axis is tangent to the ray at \m{\sz = \sz_0}. The second transformation, \m{\sy \mapsto \sr}, is nonlinear and modifies the coordinate grid such that the new coordinate axis coincides with the ray. (The momentum axis is also modified accordingly, such that the canonical commutation relation \eq{eq:qpcomm} is preserved.) The advantage of the splitting \eq{eq:split} is that the mapping \m{\sz \mapsto \sy} is easy to construct exactly and, for smooth rays, \m{\sy \mapsto \sr} can also be found analytically, at least asymptotically. Below, we show how to do this step by step.

\subsection{Osculating harmonic oscillator}
\label{sec:iy}

Let us start with a simplified model, when \m{H_\sz} can be adequately approximated (locally) with its second-order Taylor expansion in \m{\sz - \sz_0}:\footnote{Limitations of this approximation are discussed at the end of this section. Also, alternative expansions will be introduced in \Secs{sec:parop} and \ref{sec:mgogen}.}
\begin{gather}\label{eq:dopnlit}
H_\sz(\sz) = - (\sz - \sz_0) \cdot \sJ \sv 
+ \frac{1}{2}\,(\sz - \sz_0) \cdot \sg(\sz - \sz_0).
\end{gather}
Here, \m{\sv \doteq \sJ \pd_\sz H_\sz(\sz_0) \equiv \dot{\sz}_0}, as earlier, and \m{\sg} is a symmetric matrix given by \m{\sg \doteq (\pd^2_\sz H)_0}. A related description of this model can also be found in \citep[Section 5.3.3]{book:tracy} for a specific form of \m{H_\sz}. Here, we allow for a general \m{H_\sz} in order to keep the theory symplectically invariant, at least in its final form discussed in \Sec{sec:mgogen}. 

Equation \eq{eq:dopnlit} leads to the following Hamilton's equations:
\begin{gather}\label{eq:dzh}
\dot{\sz} = \sv + \sJ \sg(\sz - \sz_0),
\qquad
\ddot{\sz} = \sJ \sg\dot{\sz}, 
\end{gather}
so \m{\ddot{\sz}_0 = (\sJ \sg\dot{\sz})_0 = \sJ \sg\sv}. Then, \m{\su = \sJ \sg\sv} and \m{u = - \ki/v}. Here, \m{v} is a normalization factor that was introduced in \eq{eq:sqvdef} and remains to be specified. Also,
\begin{gather}\label{eq:kidef}
\ki \doteq (\ddot{\sz} \wedge \dot{\sz})_0 
=
(\sJ \sg \sv)^\T \sJ \sv
= \sv \cdot \sg\sv,
\end{gather}
where we used \m{\sJ^\T\sJ = \sI} and \m{\sg^\T = \sg}. Also note that 
\begin{gather}
\sv \cdot \sg\su
= \sv \cdot \sg \sJ \sg\sv
= (\sg\sv) \wedge (\sg\sv) = 0,
\\
\su \cdot \sg \su 
= (\sJ \sg\sv)^\T \sg (\sJ \sg\sv) 
= \sv^\T \sg^\T \sJ^\T \sg \sJ \sg \sv 
= \sv \cdot \sg (-\sJ \sg \sJ \sg) \sv 
= g \ki,
\end{gather}
where we used \m{\sJ^\T = -\sJ}, and, for \m{2 \times 2} matrices, \m{-\sJ \sg \sJ \sg = g\sI}, where \m{g \doteq \det \sg}. Similarly, the time derivative of \m{\ddot{\sz} \wedge \dot{\sz}} is
\begin{gather}
\dddot{\sz} \wedge \dot{\sz}
= (\sJ \sg \ddot{\sz}) \wedge \dot{\sz}
= \ddot{\sz}^\T \sg^\T \sJ^\T \sJ \dot{\sz}
= \ddot{\sz}^\T \sg \dot{\sz}
= \ddot{\sz}^\T \sJ^\T (\sJ \sg \dot{\sz})
= -\ddot{\sz} \wedge \ddot{\sz}
= 0,
\end{gather}
so \m{\ddot{\sz} \wedge \dot{\sz} = (\ddot{\sz} \wedge \dot{\sz})_0}. This means that, to the extent that the model \eq{eq:dopnlit} is applicable, one can replace the definition \eq{eq:kidef} with \m{\ki = \ddot{\sz} \wedge \dot{\sz}}.

Let us use these results to see how \eq{eq:dopnlit} is transformed by the mapping \m{\sz \mapsto \sy}. The first term in \eq{eq:dopnlit} becomes \m{\yp v}, like before. The second term becomes
\begin{align}
\frac{1}{2}\,(\sz - \sz_0) \cdot \sg(\sz - \sz_0)
& = \frac{1}{2}\,\sy^\T 
\left(
\begin{array}{c}
\sv^\T/v\\
\su^\T/u
\end{array}
\right)
\sg
\left(
\begin{array}{cc}
\sv/v & \su/u
\end{array}
\right)
\sy
\notag\\
& = \frac{1}{2}\,\sy \cdot
\left(
\begin{array}{cc}
\sv \cdot \sg \sv/v^2 & \sv \cdot \sg \su/vu\\
\su \cdot \sg \sv/vu & \su \cdot \sg \su/u^2
\end{array}
\right)
\sy
\notag\\
& = \frac{1}{2}\,\sy \cdot
\left(
\begin{array}{cc}
\ki/v^2 & 0\\
0 & gv^2/\ki
\end{array}
\right)
\sy.
\end{align}
Assuming the notation \m{\sigma_f \doteq \sgn f} for any \m{f}, let us introduce \m{\Omega \doteq \sigma_\ki\sqrt{|g|}} and choose \m{v = (\ki^2/|g|)^{1/4}}, so that the absolute values of the diagonal elements in the above matrix be the same. Then, \m{\ki/v^2 = |g|v^2/\ki = \Omega}, so \m{v = |\ki/\Omega|^{1/2}}, \m{u = - \sigma_\ki |\ki \Omega|^{1/2}}, and
\begin{gather}
\left(
\begin{array}{cc}
\ki/v^2 & 0\\
0 & gv^2/\ki
\end{array}
\right)
= \Omega\left(
\begin{array}{cc}
1 & 0\\
0 & \sigma_g
\end{array}
\right).
\end{gather}
Then,
\begin{gather}\label{eq:rayH}
H_{\sy}(\sy) = H_\sz(\sz)
 = \frac{\Omega}{2}\,\big(\yq^2 + \sigma_g (\yp + R)^2 - \sigma_g R^2\big),
\end{gather}
where \m{R \doteq \sigma_g v/\Omega}, so \m{|R| = |v/\Omega| = |\ki|^{1/2}|g|^{-3/4}}.\footnote{Notably, \m{|R|^{2/3} = |\sz'' \wedge \sz'|^{1/3} \equiv |\mcc{s}'|}, where the primes denote derivatives with respect to the phase \m{\Omega t} and \m{\mcc{s}} is the `symplectic arc length' introduced in \citep{ref:kamran09}.} One can recognise this procedure as the Williamson diagonalization of \m{\sg} (at least, for \m{g > 0}), with \m{\Omega} being the symplectic eigenvalue of \m{\sg} \citep{ref:nicacio21}. Note that \m{\Omega} and \m{R} are symplectic invariants by definition, and so are \m{\ki}, \m{g}, and \m{v}. Also, it is easily seen that \m{R} is dimensionless and the other  quantities have the following units (the square brackets here stand for `units of'):
\begin{gather}
[\ki] = [H_\sz]^3, 
\qquad
[g] = [H_\sz]^2,
\qquad
[\Omega] = [H_\sz],
\qquad
[v] = [H_\sz].
\end{gather}
Also useful for us will be the following formula:
\begin{gather}\label{eq:RR}
|\ki|^{1/3} = |\Omega| |R|^{2/3} = v/|R|^{1/3}.
\end{gather}

\begin{figure*}
\centering
\includegraphics[width=.95\textwidth]{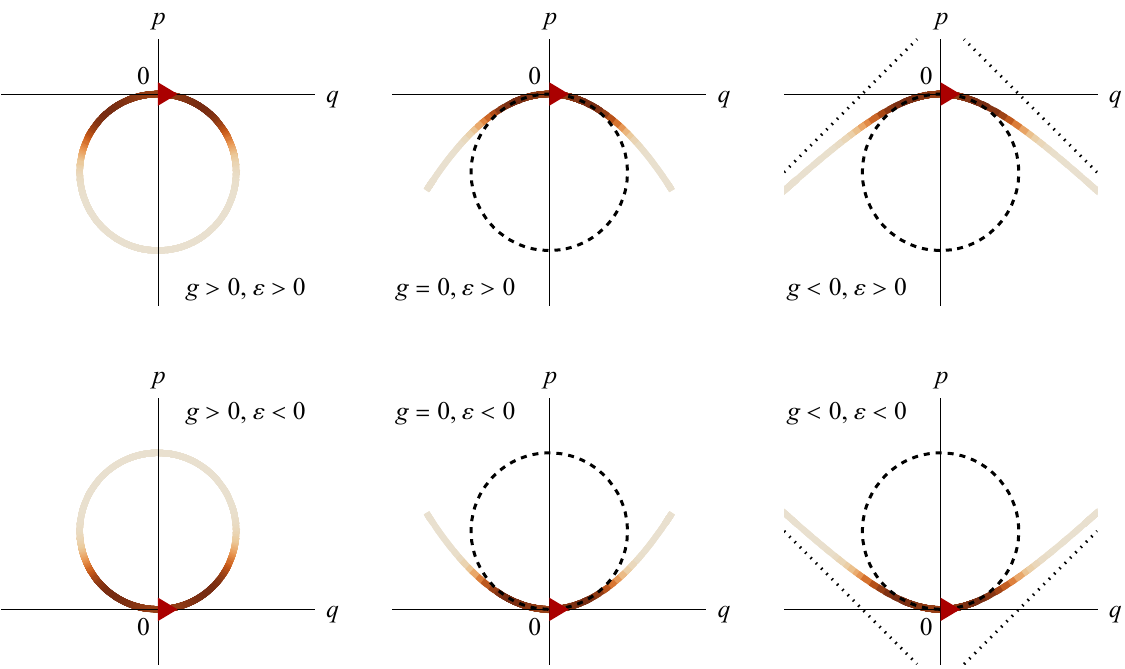}
\caption{\label{fig:signs}
Ray trajectories \m{H_{\sy}(\sy) = 0} for various signs of~\m{g} and~\m{\ki}. Darker colors mark, loosely, the areas of where the assumed model is adequate. Lighter colors mark the areas that are beyond the validity domain of the model. The arrows mark the direction of the ray propagation, which is always towards positive~\m{\yq}. In the right columns, dashed are the osculating circles. The radius of each circle is \m{|R| = |v/\Omega|}. Dotted are the asymptotes of the hyperbolas.
}
\end{figure*}

The phase-space variables \m{(q, p)^\T \equiv \sy} can be understood as a local approximation to the natural canonical variables of \m{\oper{H}}. The matrix \m{\sg} serves as a natural (pseudo)metric induced by \m{\oper{H}} on the symplectic space; then, \m{\ki = ||\sv||_\sg^2} can be understood as the square of \m{\sv} in this metric. The ray trajectories in the \m{\sy}-space satisfy \m{H_{\sy}(\sy) = 0}, so, within the approximation \eq{eq:rayH}, they are conic sections: circles at \m{g > 0}, parabolas at \m{g = 0}, and hyperbolas at \m{g < 0}. However, we are interested only in small \m{\yq} and \m{\yp}, in which case \m{\sigma_g} has little effect on the ray trajectories (\Fig{fig:signs}).\footnote{Strictly speaking, at \m{g < 0}, the dispersion curve has two branches, potentially allowing for mode conversion, \ie energy tunneling from one dispersion surface to another (\Sec{sec:mc}). However, this is largely irrelevant for us here, as our goal in this section is to construct phase-space variables around a given reference ray rather than solve a wave equation.} Hence, locally, the ray dynamics can be thought of as the dynamics of a harmonic oscillator with the Hamiltonian
\begin{gather}\label{eq:oscosc}
H_{\sy}(\sy) = \frac{\Omega}{2}\,\big(\yq^2 + (\yp + R)^2 - R^2\big).
\end{gather}
We call it the `osculating' harmonic oscillator (OHO). The radius of the phase-space orbit of the OHO is \m{|R|}. This defines the local characteristic scale of the ray orbit. Since our approximation \eq{eq:dopnlit} is generally adequate only on scales small compared to this scale, our model is applicable only when both \m{q} and \m{p} are much less than \m{R}.

The parameter \m{\Omega} can be understood as OHO's angular frequency. At \m{\Omega > 0}, or \m{\ki > 0}, the phase-space trajectory is oriented clockwise, as usual. (Note that the canonical phase of the oscillator normally grows when the polar angle on the \m{(\yq, \yp)} plane decreases.) At \m{\Omega < 0}, or \m{\ki < 0}, the trajectory is counter-clockwise. At \m{\Omega = 0}, which corresponds to \m{\ki = 0}, the trajectory is flat, \ie \m{R = \infty}. Hence, it is convenient to classify the ray trajectories in terms of their (dimensionless) `symplectic curvature'
\begin{gather}\label{eq:kapep}
\kappa \doteq \frac{1}{R} = \frac{\Omega}{v} = \sigma_\ki\,\frac{|g|^{3/4}}{|\ki|^{1/2}},
\end{gather}
and \m{R} will be called the symplectic radius, accordingly. (Here and further, \mm{\sgn R \equiv \sgn \kappa = \sigma_\ki}.) Note that the symplectic curvature can be very different from the regular curvature of a trajectory in a metric space. In particular, one can easily show that
\begin{gather}\label{eq:epsdef}
\ki = - \frac{\dd^2 x}{\dd k^2}\left(\frac{\pd H_\sz}{\pd x}\right)^3 
= - \frac{\dd^2 k}{\dd x^2}\left(\frac{\pd H_\sz}{\pd k}\right)^3,
\end{gather}
where the full derivatives are taken along the ray trajectory, \m{H_{\sz}(\sz) = 0}. Thus, at inflection points (where \m{\ki \to 0}), \m{\kappa} is infinite, whereas the regular curvature would be zero.

Finally, notice the following. The OHO approximation \eq{eq:oscosc} for \m{H_\sz} corresponds to the following approximation of \m{\oper{H}} itself: \m{\oper{H} \approx (\Omega/2)(\oper{\yq}^2 + (\oper{\yp} + R)^2 - R^2)}. This means that, in the absence of dissipation, the wave equation \eq{eq:Dpsi2} becomes the equation of a quantum Harmonic oscillator (QHO) with \m{\hbar = 1}. On the ray that is a circle with radius \m{|R|}, the (`quantum') mode number \m{n} of this oscillator, \ie the action in the natural units, is 
\begin{gather}
n = \frac{R^2}{2} - \frac{1}{2}.
\end{gather}
This readily determines applicability conditions for GO in the \m{q}-representation in the region \m{q \ll R} that we are interested in. Indeed, as commonly known, one can adequately approximate the dynamics of such QHO using GO at \m{n \gg 1}, \ie \m{R \gg 1}. This makes \m{\kappa} a natural small parameter for MGO. One might be surprised that this parameter becomes infinite at inflection points, where rays become straight and thus, supposedly, MGO should work only better. But notice that, near inflection points, \m{R} evolves faster than the ray moves along the OHO trajectory \perse and thus the very concept of the orbit radius becomes irrelevant. In other words, the quadratic approximation \eq{eq:dopnlit} is inapplicable in such regimes and one should either include cubic terms or simply ignore the effects produced by the quadratic terms, in which case MGO is reinstated. What matters then is not the local \m{R} \perse but the characteristic \m{R}. We will return to this subject in \Sec{sec:mgogen}.

\subsection{Parabolic approximation}
\label{sec:aac2}

There is more than one way to define the transformation \mt{\sy \mapsto \sr} near a given \m{\sz_0}. One set of obvious variables to try are the angle--action variables of the OHO. However, as discussed in \App{app:aac}, this approach is not optimal, so let us try a different one. 

\subsubsection{Operator transformation}
\label{sec:parop}

As pointed out in \Sec{sec:iy}, the OHO approximation is valid only at \m{q \ll R}, where \m{p \sim q^2/R \ll q}. Because of this, one can neglect \m{p^2} compared to \m{q^2} in \eq{eq:rayH} and obtain the following `parabolic' approximation of the ray Hamiltonian:\footnote{One can cast \eq{eq:rayH2} in an even simpler, one-parameter form,  \m{H_{\sY} = P + \ki Q^2/2}, by introducing yet another set of canonical variables, \m{\sY \equiv (Q, P)^\T \doteq (\yq/v, pv)^\T}.}
\begin{gather}
H_{\sy}(\sy) = H_\sz(\sz) 
\approx pv + \frac{\Omega}{2}\,\yq^2.
\label{eq:rayH2}
\end{gather}
(Note that this holds for any signs of \m{g} and \m{\ki}, so the discontinuous function \m{\sgn} never even appears in this formulation.) Let us now introduce new canonical variables \m{(\tau, h)^\T \equiv \sr} such that \m{h = pv + \Omega \yq^2/2}. To find a suitable \m{\tau}, let us search for the corresponding generating function \m{F} in the type-2 form, \m{F = F(q, h)} \citep{book:goldstein}:
\begin{gather}
p(q, h) = \pd_q F(q, h), 
\qquad
\tau(q, h) = \pd_h F(q, h).
\end{gather}
Since \m{p(q, h) = (h - \Omega q^2/2)/v}, one has \m{F(q, h) = (h q - \Omega q^3/6)/v} (up to an arbitrary function of \m{h}, which we choose to be zero), so
\begin{gather}
\tau = q/v, 
\qquad
h = pv + \Omega \yq^2/2,
\end{gather}
where we have repeated the expression for \m{h} for completeness. Since \m{v} and \m{w} are defined as values at a fixed \m{\sz_0}, the corresponding operators are given simply by
\begin{gather}\label{eq:QPop}
\oper{\tau} = \oper{q}/v, 
\qquad
\oper{h} = \oper{p}v + \Omega \oper{q}^2/2.
\end{gather}
(Basically, \eq{eq:QPop} is just a rescaled and truncated small-\m{p} approximation of \eq{eq:phij}.) One can readily see that they exactly satisfy the canonical commutation relation \m{[\oper{\tau}, \oper{h}] = \ii}.

\subsubsection{\tMwaves}

Besides avoiding infinite series, the above model is advantageous in that it allows one to explicitly calculate \Mwaves in a closed form. Indeed, \eq{eq:kMi} for \m{\inv{M} \doteq M_{\oper{q} \mapsfrom \oper{\tau}}} becomes an equation solvable analytically:
\begin{gather}\label{eq:Mp}
\ii \pd_\tau \inv{M} = -\ii v \pd_q \inv{M} + \frac{1}{2}\,w q^2 \inv{M}
\end{gather}
(and \eq{eq:kM} is just the complex conjugate thereof). Since \m{\oper{\tau} = \oper{q}/v}, one can adopt \m{\ket{\ev_{\oper{\tau}}(\tau)} = \sqrt{v}\ket{\ev_{\oper{q}}(v\tau)}}; then the initial condition for \eq{eq:Mp} is
\begin{gather}
\inv{M}(q, 0) 
\equiv \braket{\ev_{\oper{q}}(q) | \ev_{\oper{\tau}}(0)}
= \sqrt{v}\braket{\ev_{\oper{q}}(q) |  \ev_{\oper{q}}(0)}
= \sqrt{v}\,\delta(q).
\end{gather}
The corresponding solution of \eq{eq:Mp} is
\begin{gather}\label{eq:Mpar}
\inv{M}(q, \tau) = \frac{1}{\sqrt{v}}\,\delta\left(\tau - \frac{q}{v}\right)
\exp\left(-\frac{\ii \Omega q^3}{6v}\right).
\end{gather}
Accordingly, the pseudo-measure \eq{eq:Wdef}, with \m{M \equiv \inv{M}^*}, is given by
\begin{align}
\MWf 
& = \frac{1}{2\upi} \int \dd s'\,\dd s\,
M(\tau + s/2, q + s'/2) \, M^*(\tau - s/2, q - s'/2)\,\ee^{-\ii h s + \ii p s'}
\notag\\
& = \frac{1}{2\upi v} \int \dd s\,\dd s'\,
\delta\left(\tau - \frac{q}{v} + \frac{s - s'/v}{2}\right)
\delta\left(\tau - \frac{q}{v} - \frac{s - s'/v}{2}\right)
\notag\\
& \qquad \qquad\times
\exp\left(
\frac{\ii \Omega}{6v}\left(q + \frac{s'}{2}\right)^3
- \frac{\ii \Omega}{6v}\left(q - \frac{s'}{2}\right)^3
-\ii h s + \ii p s'
\right)
\notag\\
& = \frac{1}{2\upi v} \int \dd s\,\dd s'\,
\delta\left(\tau - \frac{q}{v}\right)
\delta\left(s - \frac{s'}{v}\right)
\exp\left(
\frac{\ii \Omega}{6v}\left(q + \frac{s'}{2}\right)^3
- \frac{\ii \Omega}{6v}\left(q - \frac{s'}{2}\right)^3
-\ii h s + \ii p s'
\right)
\notag\\
& = \frac{1}{2\upi} \int \dd s\,
\delta\left(\tau - \frac{q}{v}\right)
\exp\left(
\frac{\ii \Omega}{6v}\left(q + \frac{s v}{2}\right)^3
- \frac{\ii \Omega}{6v}\left(q - \frac{s v}{2}\right)^3
+ \ii s (p v - h)
\right)
\notag\\
& = \frac{1}{2\upi} \int \dd s\,
\delta\left(\tau - \frac{q}{v}\right)
\exp\left(
\frac{\ii \ki s^3}{24} + \ii s (H_\sy(\sy) - h)
\right),
\label{eq:aux3}
\end{align}
where we used \m{\Omega v^2 = \ki}. This result can also be expressed as
\begin{gather}
\MWf(\sr, \sy) = \Ai_\ki(H_\sy(\sy) - h)\,\delta(\tau - q/v),
\label{eq:auxWht}
\end{gather}
where we have introduced the following (real) function, with \m{\gamma} as a real parameter:
\begin{gather}
\Ai_\gamma(z) 
\doteq \frac{1}{2\upi}\int_{-\infty}^{\infty} 
\dd t\,\ee^{\ii z t + \ii \gamma t^3/24}
= \frac{2}{|\gamma|^{1/3}}\,\Ai\left(\frac{2}{|\gamma|^{1/3}}\,\sigma_\gamma z\right).
\label{eq:airyr}
\end{gather}
Here, \m{\sigma_\gamma \doteq \sgn \gamma}, and \m{\Ai} is the Airy function of the first kind. Importantly (\App{app:ai}),
\begin{gather}\label{eq:aidelta}
\lim_{\gamma \to 0} \Ai_\gamma(z) = \delta(z).
\end{gather}
As to be seen, the function \m{\Ai_\gamma}, which we will call a rescaled Airy function (RAF), plays a fundamental role in MGO and emerges in multiple contexts.

\subsection{General case}
\label{sec:mgogen}

Let us now present a formulation that is not restricted to the quadratic approximation \eq{eq:dopnlit} and also subsumes the standard WKB approximation \eq{eq:psixqW} as a special case.

\subsubsection{Operator transformation}
\label{sec:opt}

Let us start with assuming a generic \m{H_\sz} in the form
\begin{gather}\label{eq:HmcH}
H_\sz(\sz) = \mc{H}(x/\Delta x, k/\Delta k),
\end{gather}
where \m{\Delta x} and \m{\Delta k} are the characteristic scales of the system in \m{x} and \m{k}, respectively, and \m{\mc{H}} is some function with order-one scales. Consider new dimensionless variables
\begin{gather}
X \doteq R x/\Delta x,
\qquad
K \doteq R k/\Delta k,
\end{gather}
where \m{R} is a dimensionless constant. We require that the transformation \m{\sz \mapsto \sZ \equiv (X, K)^\T} be symplectic; then, \m{H_\sZ(\sZ) = H_\sz(\sz)} and 
\begin{gather}\label{eq:R2}
R = \sqrt{\Delta x \Delta k}.
\end{gather}
We will call this quantity a symplectic scale. To the extent that the quadratic approximation \eq{eq:dopnlit} is adequate at least locally, \m{R} coincides with the absolute value of the symplectic radius that was introduced in \Sec{sec:iy}. (For this reason, below, we occasionally mix the terms `symplectic scale' and `symplectic radius' and also refer to the inverse symplectic scale as symplectic curvature.) 

Much like in \Sec{sec:trad}, our goal now is to derive field equations that are accurate to the first (not zeroth) order in the `MGO parameter'
\begin{gather}\label{eq:mgopar}
\epsilon \doteq R^{-2} \equiv (\Delta x \Delta k)^{-1} \ll 1.
\end{gather}
It is similar to the GO parameter introduced in \Sec{sec:trad}, but the absolute value of \m{k} is now replaced with the spectral-inhomogeneity scale \m{\Delta k}; \ie the smallness of the local wavelength \perse is irrelevant. To the extent that the OHO model is applicable, \m{\epsilon} coincides with \m{\kappa^2}, where \m{\kappa} is the symplectic curvature introduced in \Sec{sec:iy}. 

Let us perform a linear canonical transformation \m{\sZ \mapsto \sy \equiv (q, p)^\T}, as in \Sec{sec:tang}, so that the \m{q} axis is along \m{\sv} that is tangent to the ray. Then, \m{H_\sy(\sy) = \mc{H}(q/R, p/R)}, so the ray trajectory satisfies \m{\mc{H}(q/R, p/R) = 0}. Its solution (for a given branch) can be written in the form \m{p = R\mc{P}(q/R) \equiv p(q)}, where \m{\mc{P}} is an order-one function that has order-one scales. (If \m{H_\sz} is linear in \m{\sz}, then \m{\mc{P} = 0} by construction. If \m{H_\sz} is quadratic, then \m{p(q) = -R + \sqrt{R^2 - q^2}}, so \m{p(q) \approx q^2/2R} at \m{q \ll R}, as in \Sec{sec:aac2}.) By Taylor-expanding \m{H_\sy(\sy)} in \m{p - p(q)}, one obtains
\begin{gather}\label{eq:hayes}
H_\sy(\sy) \approx V(q)(p - p(q)),
\end{gather}
where \m{V(q) \doteq (\pd_p H_\sy)_{p = p(q)}}, so \m{V(0) = v}. Like \eq{eq:hayesH}, \eq{eq:hayes} can be understood as Hayes's representation of \m{H_\sy}. For GO, this representation is precise in that it leads to the correct ray equations and correctly captures the leading-order amplitude dynamics, because those are entirely determined by the first-order derivatives of the symbol. The MT that follows from \eq{eq:hayes} will not be exact, because \m{\bigO([p - p(q)]^2)} terms have been neglected. Still, assuming \m{p - p(q) = \bigO(q^2)}, \eq{eq:hayes} differs from the exact \m{H_\sy} by at most \m{\bigO(q^4)}, which is assumed negligible (at not-too-large \m{q} that we are interested in locally).

Let us perform a canonical transformation \m{\sy \mapsto \sr \equiv (\tau, h)^\T} such that \m{h = H_\sy}, \ie
\begin{gather}
p(q, h) = p(q) + h/V(q).
\end{gather}
(As a reminder: \m{p(q)} is a prescribed function determined by the shape of the ray trajectory, while \m{p(q, h)} is the new momentum variable \m{p} expressed as a function of \m{q} and \m{h}.) Like in \Sec{sec:aac2}, to find what the new coordinate \m{\tau} should be to keep the transformation canonical, we look for a generating function \m{F} in the type-2 form:
\begin{gather}
p(q, h) = \pd_q F(q, h), 
\qquad
\tau(q, h) = \pd_h F(q, h).
\end{gather}
The former gives 
\begin{gather}
F(q, h) = \int^{q}_0 p(\tilde{q}, h)\,\dd \tilde{q},
\label{eq:varthetaF2}
\end{gather}
where we set the integration constant to zero by choice. (The tilde is used to distinguish the dummy integration variable \m{\tilde{q}} from the actual canonical coordinate \m{q}.) Then,
\begin{gather}\label{eq:tauqh2}
\tau(q, h) = \int^{q}_0 \frac{\pd p(\tilde{q}, h)}{\pd h}\,\dd \tilde{q} 
= \int^{q}_0 \frac{\dd \tilde{q}}{V(\tilde{q})} \equiv \tau(q).
\end{gather}
In other words, \m{\tau} and \m{h} are the ray time and the ray energy. (By the ray energy, we mean the value of the ray Hamiltonian, not the wave energy that depends on the wave amplitude.) The corresponding operators are
\begin{gather}
\oper{\tau} = \tau(\oper{q}),
\qquad
\oper{h} = \frac{1}{2}\,(\oper{V} \oper{p} + \oper{p} \oper{V})
- \oper{V} p(\oper{q}),
\end{gather}
where \m{\oper{V} \doteq V(\oper{q})} (note, though, that \m{\oper{p}} is not the same as \m{p(\oper{q})}), so \m{[\oper{\tau}, \oper{V}] = 0} and
\begin{align}
[\oper{\tau}, \oper{h}] 
& = [\oper{\tau}, \oper{V} \oper{p} + \oper{p} \oper{V}]/2
\notag\\
& = (\oper{\tau}\oper{V} \oper{p} - \oper{V} \oper{p}\oper{\tau}
+ \oper{\tau} \oper{p} \oper{V} - \oper{p} \oper{V} \oper{\tau})/2
\notag\\
& = (\oper{\tau}\oper{V} \oper{p} 
- \oper{V} [\oper{p}, \oper{\tau}] - \oper{V}\oper{\tau} \oper{p}
+ \oper{\tau} [\oper{p}, \oper{V}] + \oper{\tau} \oper{V} \oper{p}
- \oper{p} \oper{V} \oper{\tau})/2
\notag\\
& = ([\oper{\tau}\oper{V}, \oper{p}]
- \oper{V} [\oper{p}, \oper{\tau}]
+ \oper{\tau} [\oper{p}, \oper{V}])/2
\notag\\
& = \ii((\tau(q) V(q))'
+ V(q) \tau'(q)
- \tau(q) V'(q))/2
\notag\\
& = \ii V(q)\tau'(q)
\notag\\
& = \ii.
\end{align}

\subsubsection{\tMwaves}
\label{sec:gaf}

The approximation \eq{eq:hayes} corresponds to the following approximation of the operator \m{\oper{H}_{\oper{q}}} (cf.\ \Sec{sec:trad}):
\begin{gather}
\oper{H}_{\oper{q}} \approx -\ii V(q) \pd_q - \frac{\ii}{2}\,V'(q) - V(q)p(q).
\end{gather}
Then, \eq{eq:xMi} and \eq{eq:kMi} lead to the following equations for \m{\inv{M} \equiv \inv{M}(q, \tau)}:
\begin{gather}
(\tau - \tau(q))\inv{M} = 0,\\
\ii \pd_\tau \inv{M} = \left(-\ii V(q) \pd_q - \frac{\ii}{2}\,V'(q) - V(q)p(q)\right)\inv{M}.
\end{gather}
These can be readily solved:
\begin{gather}
\inv{M}(q, \tau) = \frac{\ee^{\ii \vartheta(q)}}{\sqrt{V(q)}}\,\delta(\tau - \tau(q)),
\qquad
\vartheta(q) \doteq \int^{q}_0 p(\tilde{q})\,\dd \tilde{q},
\end{gather}
which generalises \eq{eq:Mpar} to non-constant \m{V(q)} and arbitrary \m{p(q)}. To the extent that \eq{eq:hayes} is valid, this result is exact. One can also notice parallels with \eq{eq:bMGO}, which is the same formula applied to different coordinates.

To calculate the generalization of the pseudo-measure \eq{eq:auxWht}, notice that
\begin{gather}
\tau\left(q \pm \frac{s'}{2}\right)
\approx 
\tau(q) \pm \frac{s'}{2V} - \frac{V'}{8V^2}\,s'^2,
\end{gather}
where \m{V \equiv V(q)} and \m{V'(q) \equiv \dd V(q)/\dd q}, while the prime in \m{s'} is only a notation that distinguishes said variable from \m{s}.  Then,
\begin{align}
& \delta\left(
\tau + \frac{s}{2} - \tau\left(q + \frac{s'}{2}\right)
\right)
\delta\left(
\tau - \frac{s}{2} - \tau\left(q - \frac{s'}{2}\right)
\right)
\notag\\
& \qquad \approx 
\delta\left(
\tau -  \tau(q) + \frac{s - s'/V}{2} + \frac{V's'^2}{8V^2}
\right)
\delta\left(
\tau -  \tau(q) - \frac{s - s'/V}{2} + \frac{V's'^2}{8V^2}
\right)
\notag\\
& \qquad = 
\delta\left(\tau -  \tau(q) + \frac{V's'^2}{8V^2}\right)
\delta\left(s - \frac{s'}{V}\right)
\notag\\
& \qquad =
\delta\left(\tau -  \tau(q) + \frac{V's^2}{8}\right)
\delta\left(s - \frac{s'}{V}\right)
\notag\\
& \qquad \approx
\left(1 + \frac{V's^2}{8}\,\frac{\pd}{\pd \tau}\right)
\delta(\tau -  \tau(q))
\delta\left(s - \frac{s'}{V}\right),
\end{align}
where the last equality is based on the first-order Taylor expansion. (Admittedly, Taylor-expanding a delta function is a questionable procedure, but it makes sense as a shorthand for Taylor-expanding what actually matters, namely, integrals of such functions.) Then,
\begin{gather}\label{eq:mugen}
\MWf = \big[1 - (V'/8)\,\pd^3_{\tau h h}\big]\mu_0,
\qquad
\mu_0 = \GAF \delta(\tau(q) -  \tau),
\end{gather}
where \m{\GAF} is a generalization of \eq{eq:airyr}:
\begin{gather}\label{eq:gaf2}
\GAF \doteq \frac{1}{2\upi} \int \dd s\,\ee^{\ii\Phi},
\qquad
\Phi \doteq (p V - h) s - \int_{q - s V/2}^{q + s V/2} p(\tilde{q})\,\dd\tilde{q}.
\end{gather}
Remember that \m{p} is an independent variable, \m{V \equiv V(q)}, and \m{p(q)} is a function introduced in \Sec{sec:opt}.

\subsubsection{Symplectically invariant model for \mt{\mu}}
\label{sec:syns}

Since we are already neglecting terms of order \m{V''}, the factor \m{V'} in \eq{eq:mugen} can be considered commuting with the derivatives. Then, using \eq{eq:OW} and integrating by parts, one obtains the following formula for remapping symbols between the \m{\sy} and \m{\sr} representations:
\begin{gather}\label{eq:OOmu0}
O_\sy(\sy) 
= \int \left(O_\sr(\sr) + \frac{V'}{8} \,\pd^3_{\tau h h} O_\sr(\sr) \right)\MWf_0 \,\dd\tau\,\dd h.
\end{gather}
Below, we will be interested in \m{O_\sr} that are either \m{\tau}-independent or smooth, namely, have symplectic scales of order \m{R} or larger. In either case, the second term in \eq{eq:OOmu0} is \m{\bigO(R^{-4})}, \ie \m{\bigO(\epsilon^2)}, which is considered negligible. Thus, below we adopt \m{\mu \approx \mu_0}, \ie
\begin{gather}\label{eq:mugena}
\MWf \approx \GAF \delta(\tau(q) -  \tau).
\end{gather}

One can also show (\App{app:ai}) that, at \m{R \gg 1}, one can approximate \m{\GAF} as
\begin{gather}\label{eq:gafap}
\GAF \approx \Ai_\ki(H_\sy(\sy) - h)
\end{gather}
with \m{\ki = - V^3 p''}. In fact, even when \eq{eq:mgopar} is not satisfied, yet all higher-order derivatives of \m{p(q)} are small enough, \m{\GAF} can still be approximated with a delta function, \m{\GAF \approx \delta(H_\sy(\sy) - h)}, which may be sufficiently accurate for practical purposes. In this case, \m{\GAF} can be safely replaced with \m{\Ai_\ki} because the latter can be approximated with a delta function as well. Hence, we will assume \eq{eq:gafap} either literally, when \eq{eq:mgopar} is satisfied, or in the sense that both \m{\GAF} and  \m{\Ai_\ki} are close enough to \m{\delta(H_\sy(\sy) - h)}. Then, \eq{eq:mugena} becomes
\begin{gather}\label{eq:mugenab}
\MWf \approx \Ai_\ki(H_\sy(\sy) - h)\, \delta(\tau(q) -  \tau).
\end{gather}

Because \m{\sy} and \m{\sz} are connected by a linear transformation, which preserves \m{\mu}, one can also consider \eq{eq:mugenab} as the symplectic measure for the whole transformation \m{\sr \leftrightarrow \sz}:
\begin{gather}
\mu = \Ai_\ki(H_\sz(\sz) - h)\,\delta(\tau(\sz) - \tau).
\label{eq:aux4}
\end{gather}
At \m{\ki \to 0}, when the transformation \m{\sz \mapsto \sr} is linear,\footnote{By \eq{eq:epsdef}, zero \m{\ki} corresponds to inflection points on the dispersion curve, \ie locally straight rays. Thus, the ray-aligned coordinates \m{\sr} are locally identical to \m{\sy}.} \eq{eq:aux4} yields \mm{\MWf(\sr, \sz) = \delta(H_\sz(\sz) - h)\,\delta(\tau(\sz) - \tau)}, in agreement with \eq{eq:WLST}. Application of \eq{eq:aux4} in the general case requires that one expresses \m{\ki} as a function of \m{\sz} (as opposed to \m{q}). Since our calculation is valid only for small \m{q}, multiple coordinate charts would have to be introduced and properly merged for this, which is inconvenient. However, as long as \m{\ki} is smooth (or indistinguishable from zero), one can use \eq{eq:kidef} to extrapolate it to all \m{\sz} via
\begin{gather}
\ki = (\sJ \pd_\sz H_\sz) \cdot (\pd_\sz^2 H_\sz) (\sJ \pd_\sz H_\sz).
\label{eq:kidef2}
\end{gather}
This ensures that \m{\ki} is close to the desired value \m{\ki = - V^3 p''} in the ray vicinity, while far from the ray our result is not expected to be applicable in any case. Then, one can readily transform the symbols of operators with
\begin{gather}\label{eq:aitr}
O_\sz(\sz) 
\approx \int \Ai_\ki(H_\sz(\sz) - \tilde{h})\,O_\sr(\tau(\sz), \tilde{h})\, \dd \tilde{h},
\end{gather}
\ie \m{O_\sz(\sz)} is the Airy transform of \m{O_\sr} \citep{ref:widder79}. Equation \eq{eq:aitr} is one of the key results of this paper.

Assuming the symplectic scale of \m{O_\sr} is of order \m{R}, one can take this integral using \eq{eq:ainorm6}. This leads to \m{O_\sz(\sz)  \approx O_\sr(\tau(\sz), H_\sz(\sz)) + \bigO(\ki \pd_h^3 O_\sr)}. Based on \eq{eq:kidef2}, the second term on the right-hand side can be estimated as \m{\ki \pd_h^3 O_\sr \sim O_\sr/R^4}. Thus,
\begin{gather}\label{eq:OOap}
O_\sz(\sz) = O_\sr(\sr(\sz)) + \bigO(\epsilon^2),
\end{gather}
and \m{\bigO(\epsilon^2)} will be negligible for our purposes (\Sec{sec:mgoscalar}). This is another key result of this paper. Keep in mind, though, that \eq{eq:OOap} is invalid for small-scale symbols, such as the delta-shaped symbols discussed in \Sec{sec:mapping}. In that case, the more general \eq{eq:aitr} should be used instead.

\section{MGO for scalar waves}
\label{sec:mgoscalar}

\subsection{Dynamics in the \mt{\tau}-space}
\label{sec:envscal}

Armed with \eq{eq:OOap}, we are now ready to formulate the MGO envelope equation. Let us return to our original wave equation \eq{eq:Dpsi2}, which accounts for dissipation, and rewrite it in the \m{\tau}-representation:
\begin{gather}\label{eq:hg}
(\oper{H}_{\oper{\tau}} + \ii\oper{\Gamma}_{\oper{\tau}})\psi_{\oper{\tau}} = 0.
\end{gather}
(As earlier in this paper, we assume here that \m{\psi_{\oper{\tau}}} is a scalar function, but see \Sec{sec:vwaves} for a more general case.) Since \m{\oper{H}_{\oper{\tau}} = \oper{h}_{\oper{\tau}} = -\ii \pd_\tau}, this leads to 
\begin{gather}\label{eq:Dpsit}
(- \pd_\tau + \oper{\Gamma}_{\oper{\tau}})\psi_{\oper{\tau}} = 0.
\end{gather}
Assuming that \m{\Gamma_\sr = \bigO(\epsilon)}, the operator \m{\oper{\Gamma}_{\oper{\tau}}} can be approximated simply by its symbol on the reference ray, \m{\Gamma_\sr(\tau)}. Then, \eq{eq:Dpsit} becomes
\begin{gather}\label{eq:Dpsit2}
\pd_\tau \psi_{\oper{\tau}} = \Gamma_\sr(\tau) \psi_{\oper{\tau}}.
\end{gather}
Assuming that \m{\Gamma_\sr} is sufficiently smooth, \eq{eq:OOap} predicts that \m{\Gamma_\sr(\sr) \approx \Gamma_\sz(\sz(\sr))}, which is a known function when the dispersion operator is given. Then, \eq{eq:Dpsit2} can be readily integrated, 
\begin{gather}\label{eq:psisolng}
\psi_{\oper{\tau}} = a \exp \left(\int \dd\tau \Gamma_\sr(\tau)\right),
\qquad
a = \const,
\end{gather}
and, unlike in the GO equation \eq{eq:act0}, this solution exhibits no singularities. Also notably, the ray equations in the \m{\sr}-representation are particularly simple:
\begin{gather}\label{eq:thre}
\dot{\tau} = 1,
\qquad
\dot{h} = 0.
\end{gather}

A generalization to vector waves is discussed in \Sec{sec:mgogeneral}. One can also generalise the MGO envelope equation to quasioptical beams that experience transverse diffraction. This can be done similarly to how quasioptical equations were derived in \citep{my:quasiop1, my:quasiop2, my:quasiop3} for conventional GO. However, presenting these calculations in detail would require introducing additional cumbersome machinery that is not specific to MGO, so we postpone discussing this subject until further publications.

\subsection{Mapping solutions to the \mt{x}-space}
\label{sec:mapping}

Mapping solutions to the \mt{x}-space is generally the most challenging part in MGO applications \citep{my:mgosteep, ref:mgocode, ref:mgocode2}. However, knowing the field at \textit{all} \m{x} is redundant.\footnote{Likewise, in quantum simulations of linear waves \citep{my:qc1, my:qc3, my:qcdissip}, the cost of outputting the computation results can be a significant fraction of the total computation cost. Because of this, one usually aims to outputs `expectation values' rather than the field profiles \perse. We recommend adhering to this principle in MGO as well.} One might not even need to leave the \m{\tau}-space except for initializing the field and also calculating the final state, and those mappings may be trivial to do. In other cases, one can proceed as follows.

\subsubsection{General case}
\label{sec:gm}

To calculate how the wave evolves in the original \m{x}-space, one needs to remap the \m{\tau}-representation to the \m{x}-representation:
\begin{gather}\label{eq:psixt}
\psi_{\oper{x}}(x)
= \int \dd\tau M_{\oper{x} \mapsfrom \oper{\tau}}(x, \tau)\,
\psi_{\oper{\tau}}(\tau).
\end{gather}
Since \m{\oper{M}_{\oper{x} \mapsfrom \oper{\tau}}
= \oper{M}_{\oper{x} \mapsfrom \oper{q}}
\oper{M}_{\oper{q} \mapsfrom \oper{\tau}}}, this leads to
\begin{gather}
\psi_{\oper{x}}(x) = 
\int \dd q\,\frac{\ee^{\ii \vartheta(q)}}{\sqrt{V(q)}}\,
M_{\oper{x} \mapsfrom \oper{q}}(x, q)\,\psi_{\oper{\tau}}(\tau(q)),
\label{eq:Mxt}
\end{gather}
where \m{\vartheta(q) = \int^{q}_0 p(\tilde{q})\,\dd \tilde{q}}. Since \m{M_{\oper{x} \mapsfrom \oper{q}}} consists of a phase-space shift (\Sec{sec:pshift}) and an LST (\Sec{sec:lst}), the integrand in \eq{eq:Mxt} is easy to write explicitly in any given~case. 

How to approximate such an integral depends on the problem and can be done using various techniques, with accuracy requirements determined by specific applications. For example, a stationary-phase or steepest-descent methods might suffice, or one might be able to reduce \eq{eq:Mxt} to a known representation of a special function. In particular, to the extent that \eq{eq:Mxt} can be expressed through the Airy function (or, equivalently, Bessel functions of order \m{1/3}), it is similar to the results from \citep{ref:langer37, ref:chester57}. Similar integrals have been studied also in the context of semiclassical calculations in atomic and molecular physics; for example, see \citep{ref:kay94}.

Remember also that the above result is derived under the assumption that only small values of \m{q} contribute to this integral. This allows one to derive local approximations \m{\psi_{\oper{x}}} using different definitions of \m{(q, p)} for different regions of interest. A global representation is not necessarily desirable for practical applications. When it is, though, it can be constructed in one of two ways. 

One way to do this is to numerically find the type-1 generating function \m{\Theta(x, \tau)} of the transformation \m{(x, k) \mapsto (\tau, h)} using
\begin{gather}\label{eq:ktTh}
k = \pd_x\Theta(x, \tau),
\qquad
h = -\pd_\tau\Theta(x, \tau),
\qquad
\dd\Theta = k\,\dd x - h\,\dd\tau,
\end{gather}
and then invoke \eq{eq:psixt} in combination with the approximate formula \eq{eq:Mampl2}. For the specific coordinates that we assume here, the latter takes the form
\begin{gather}\label{eq:semiMxt}
M_{\oper{x} \mapsfrom \oper{\tau}}(x, \tau) 
= \sqrt{\frac{\Theta_{x\tau}}{2\upi}}\,\ee^{\ii\Theta}.
\end{gather}
This formula is also useful for determining the quantization condition for periodic orbits, as discussed in \App{app:qz}.

Alternatively, one may notice that \m{\psi_{\oper{x}}} \perse is rarely needed in applications. Instead, of interest are bilinear functionals of \m{\psi_{\oper{x}}} and \m{\psi_{\oper{x}}^*}, such as the energy density or the dissipated power. Any such functional can be expressed through the field's Wigner function.\footnote{See \citep[Appendix B.1]{my:ql} for the general formula. In \App{app:P} of the present paper, we explicitly show how to derive this form for the dissipation power in particular. In fact, \textit{any} reasonably general quasilinear calculations are expressed more naturally through Wigner functions than through fields \perse \citep{my:ql, my:qlrmpp}. This is similar to the situation in quantum mechanics, where expectation values, as opposed to wavefunctions, are all that matter.} By \eq{eq:Weig} and \eq{eq:psisolng}, the Wigner function of \m{\psi_{\oper{\tau}}} in the \m{\sr}-representation is
\begin{gather}
W_\sr(\tau, h) \approx W_0(\tau) \delta(h),
\qquad
W_0(\tau) = |a|^2\exp \left(2 \int \dd\tau \Gamma_\sr(\tau)\right).
\label{eq:wigaux}
\end{gather}
By \eq{eq:OW}, its \m{\sz}-representation is \m{W_\sz(\sz) 
= \int \dd \tau\,\dd h\,W_\sr(\tau, h)\,\MWf(\sr, \sz)}. This leads to a remarkably concise expression for \m{W_\sz}:
\begin{gather}
W_\sz(\sz) 
= W_0(\tau(\sz))\, \Ai_\ki(H_\sz(\sz)).
\label{eq:WAi1}
\end{gather}

\subsubsection{Eigenwaves}
\label{sec:eigb}

Let us now consider an important special case when dissipation is negligible, at least locally (\m{\oper{\Gamma} = 0}), and a wave satisfies \m{\oper{H}_{\oper{\tau}}\psi_{\oper{\tau}} \approx 0}. Such a wave can be understood as an eigenstate of \m{\oper{H}} corresponding to the zero eigenvalue, \m{h = 0}; \ie \m{\ket{\psi} = a_h\ket{\ev_{\oper{h}}(0)}}, where \m{a_h \equiv \sqrt{2\upi}a} is some constant amplitude. Then, using \eq{eq:xqme}, one obtains
\begin{gather}\label{eq:psitau}
\psi_{\oper{\tau}}(\tau) = a_h\braket{\ev_{\oper{\tau}}(\tau) | \ev_{\oper{h}}(0)}
= a = \const,
\end{gather}
which, of course, can also be obtained directly from \eq{eq:Dpsit2}. For \m{\psi_{\oper{x}}}, \eq{eq:Mxt} readily yields
\begin{gather}\label{eq:psies}
\psi_{\oper{x}} 
= \psi_{\oper{\tau}}
\int \dd q\,\frac{\ee^{\ii \vartheta(q)}}{\sqrt{V(q)}}\,
M_{\oper{x} \mapsfrom \oper{q}}(x, q).
\end{gather}
In the special case when \m{q \equiv x}, one has \m{M_{\oper{x} \mapsfrom \oper{q}}(x, q) = \delta(x - q)}, so the standard WKB solution is obtained:
\begin{gather}
\psi_{\oper{x}}(x) = \psi_{\oper{\tau}}\,\frac{\ee^{\ii \vartheta(x)}}{\sqrt{V(x)}}.
\end{gather}
More generally, \eq{eq:psies} represents, expectedly, the same WKB solution in the tangent space, which is then mapped to the \m{x}-space using a linear MT. One can approach the integral \eq{eq:psies} the same way as discussed in \Sec{sec:gm}.

\begin{figure*}
\centering
\includegraphics[width=.45\textwidth]{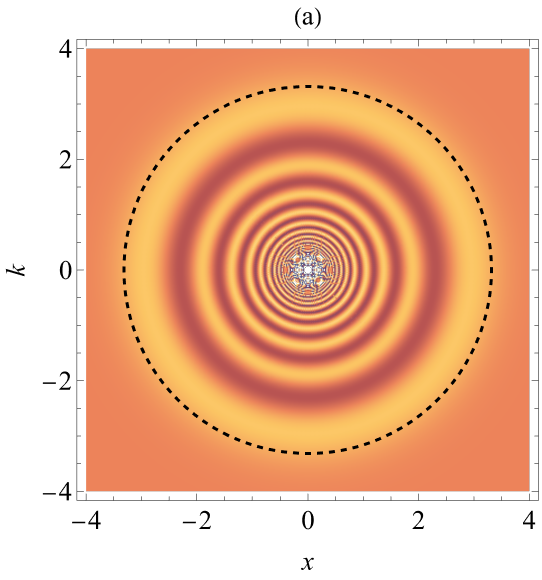}
\qquad
\includegraphics[width=.45\textwidth]{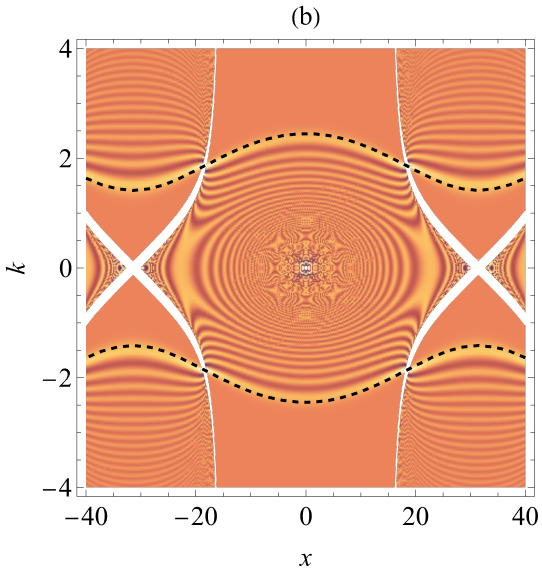}
\caption{\label{fig:airy}
Approximations \eq{eq:WAi}, with \eq{eq:kidef2} for \m{\ki}, for the Wigner functions \m{W_\sz} of fields satisfying \eq{eq:Dpsi2}: (a) \m{H_\sz(x, k) = (x^2 + k^2 - 11)/2}, which corresponds to a QHO with \m{n = 5}; (b) \m{H_\sz(x, k) = - 2 - \cos(x/10) + k^2/2}. The color intensity denotes the magnitude of \m{W_\sz} (arbitrary units). The approximations are accurate near the ray trajectories given by \m{H_\sz(x, k) = 0} (dashed). The white areas in~(b) correspond to \m{\ki \approx 0}. 
}
\end{figure*}

\begin{figure*}
\centering
\includegraphics[width=\textwidth]{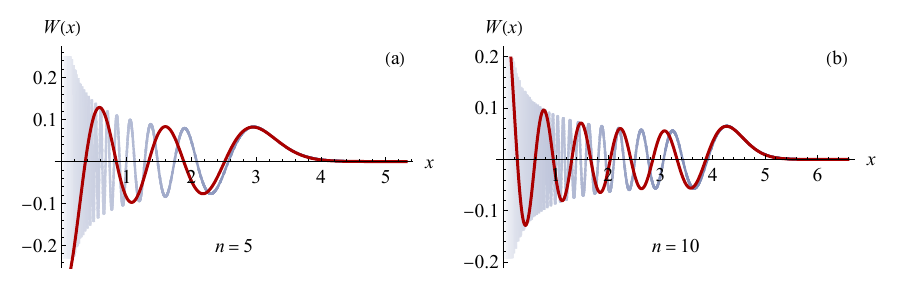}
\caption{\label{fig:airyc}
The Wigner functions \m{W(x) \equiv W_\sz(x, k = 0)} of a QHO, \mm{H_\sz = (x^2 + k^2)/2 - (n + 1/2)}: (a) \m{n = 5}, (b) \m{n = 10}. Red -- exact analytical result \eq{eq:laguerre}, blue -- approximation \eq{eq:WAi}, with \m{W_0 = (2\upi)^{-1}}. The latter is valid only close to the ray (dashed curve in \Fig{fig:airy}(a)), which, at \m{k = 0} considered here, corresponds to the reflection points \m{x = \sqrt{2n + 1}}.
}
\end{figure*}

Alternatively, one can use \eq{eq:wigaux} and \eq{eq:WAi1}, which, in this case, become
\begin{gather}
W_\sr(\tau, h) = W_0 \delta(h),
\label{eq:Wzeta0}
\\
W_\sz(\sz) \approx W_0 \Ai_\ki(H_\sz(\sz)),
\label{eq:WAi}
\end{gather}
with \m{W_0 = |a|^2 = \const}. Examples of such \m{W_\sz(\sz)} are presented in \Fig{fig:airy} for rays in a quadratic potential, which corresponds to a QHO, and a sinusoidal potential. Similar patterns are also seen in direct wave simulations; for example, see \citep[figure~3(a2)]{my:wkeadv} or \citep[figure~9]{ref:weinbub18}. Note that these complicated patterns are merely a result of mapping the Wigner function to the original coordinates \m{\sz}, as opposed to the natural coordinates \m{\sr}, where the Wigner functions have a simple form \eq{eq:Wzeta0}. Also note that \eq{eq:Wzeta0} is more precise than \eq{eq:WAi} in that it does not rely on approximations of the MT and is valid at all \m{\sz}. In contrast, \eq{eq:WAi} is accurate only in the vicinity of the ray trajectories, where \m{H_\sz} is small. This is illustrated in \Fig{fig:airyc}, which shows a comparison of \eq{eq:WAi} with the exact analytical result for a QHO, \m{H_\sz(x, k) = (x^2 + k^2)/2 - (n + 1/2)} \citep{tex:mostwoski21}:
\begin{gather}\label{eq:laguerre}
W_\sz(x, k) = \frac{(-1)^n}{\upi}\, e^{-\xi^2} L_n(2\xi),
\qquad
\xi \doteq x^2 + k^2.
\end{gather}
Here, \m{n} is a nonnegative integer that corresponds to the state number, \m{L_n} is \m{n}th Laguerre polynomial, and the standard normalization \m{a_h = 1} is assumed, which corresponds to \m{W_0 = (2\upi)^{-1}}.

To the extent that \m{\Ai_\ki} can be replaced with the delta function, \eq{eq:WAi} yields 
\begin{gather}\label{eq:psixw}
|\psi_{\oper{x}} (x)|^2 = \int \dd k\, W_\sz(x, k) 
\approx W_0 \int \dd k\,\delta(H_\sz(x, k))
= \frac{W_0}{|v_\text{g}|},
\end{gather}
where \m{v_\text{g} \doteq \pd_k H_\sz} and the derivative is evaluated on the dispersion curve at a given \m{x} and the corresponding \m{k} on the dispersion curve. This coincides with the result that flows from the action conservation for stationary waves (see, \eg \citep[Section 7.3.2]{my:ql}), 
\begin{gather}
\pd_x(v_\text{g} |\psi_{\oper{x}}|^2) = 0,
\end{gather}
and \m{|\psi_{\oper{x}}|^2} depends on \m{x} only due to inhomogeneity of \m{v_\text{g}}, \ie slowly. If a wave experiences reflection, one might want to account for the multiple branches, \ie two or more values of \m{k} at which the delta-function argument in \eq{eq:psixw} turns to zero at given \m{x}.\footnote{\label{ft:ign}Whether to account for one or multiple branches depends on the problem that one seeks to solve. For example, when resonant interactions with particles or other waves are considered, the resonance conditions are typically satisfied only for one branch. Then, the other branches are irrelevant and can be ignored.} Then, 
\begin{gather}\label{eq:psixw2}
|\psi_{\oper{x}} (x)|^2 = \sum_i \frac{W_0}{|v_\text{g}|_i},
\end{gather}
where the summation is taken over said branches. Like in the previous case, the dependence on \m{x} (due to inhomogeneity of \m{|v_\text{g}|_i}) is slow in this case. This is because the assumed delta approximation for \m{W_\sz} ignores the interference between multiple branches; \ie \eq{eq:psixw2} can be understood as \m{|\psi_{\oper{x}} (x)|^2} averaged over their relative phases.

\begin{figure*}
\centering
\includegraphics[width=\textwidth]{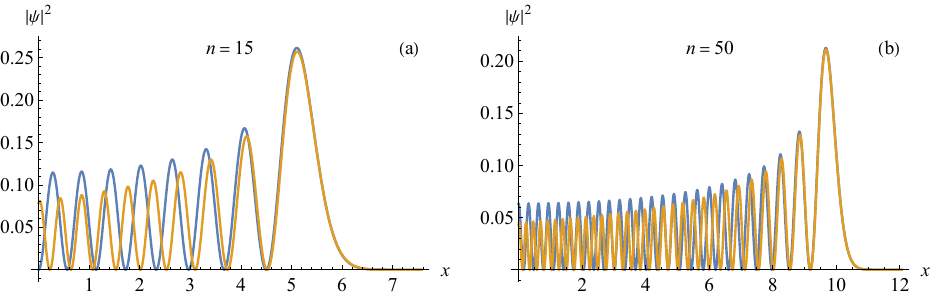}
\caption{\label{fig:airypsi}
The normalised eigenstates \m{|\psi|^2 \equiv |\psi_{\oper{x}}(x)|^2} of a QHO for two sample states: (a) \m{n = 15} and (b) \m{n = 50}. Blue -- exact analytical result; orange -- approximation \eq{eq:psiAiry}, with \m{W_0 = (2\upi)^{-1}} and \m{R = \sqrt{2n + 1}}. Unlike in \eq{eq:psixw3}-\eq{eq:aux8}, which assume \m{\sz_0 = 0} for simplicity, the \m{x}-axis origin here is at the center of the potential well, as usual; \ie, to reproduce these plots, one should replace \m{x} with \m{x - R} in \eq{eq:psiAiry}.
}
\end{figure*}

One can also integrate the actual Wigner function \eq{eq:WAi} without resorting to the delta approximation, at least as long as the parabolic approximation \eq{eq:rayH2} holds. For example, in the simplest case when \m{(q, p) \equiv (x, k)}, one readily obtains, that
\begin{gather}\label{eq:psixw3}
|\psi_{\oper{x}} (x)|^2 
= W_0 \int \dd k\,\Ai_\ki\left(v k + \frac{v x^2}{2R}\right)
= \frac{W_0}{v},
\end{gather}
which is in agreement with \eq{eq:psixw}. More generally, when \m{(q, p)} are connected with \m{(x, k)} via \eq{eq:ls} (we assume \m{\sz_0 = 0} for brevity), one can write 
\begin{gather}\label{eq:Hxkquad}
H_\sz(\sz) = \underbrace{- \frac{v}{B}\left(x + \frac{R D^2}{2B}\right)}_{U(x)}
+ \frac{v B^2}{2R}\bigg(\underbrace{\vphantom{\bigg(}k + \frac{A x}{B} + \frac{R D}{B^2}}_K\bigg)^2,
\end{gather}
where we used \eq{eq:abcd}. Then, using \citep{ref:valle97}
\begin{gather}
\int_{-\infty}^\infty \Ai(X + P^2)\,\dd P =  2^{2/3} \upi\,  \Ai^2(X/2^{2/3})
\end{gather}
and \eq{eq:RR}, one finds that
\begin{align}
|\psi_{\oper{x}} (x)|^2  
& = W_0 \int \dd K\, \Ai_\ki\left(U(x) + \frac{v B^2}{2R}\,K^2\right) 
\notag\\
& = W_0\,\frac{\upi v}{2^{1/3}|B|}\,
\Ai_\ki^2\left(\frac{U(x)}{2^{2/3}}\right).
\label{eq:aux7}
\end{align}
In particular, the case when \m{B = 1} and \m{D = 0} corresponds to having a reflection point at \m{x = 0}, when the \m{\sy}-space is rotated by \m{\upi/2} relative to the \m{\sz}-space. Then, one obtains
\begin{gather}\label{eq:psiAiry}
|\psi_{\oper{x}} (x)|^2 = 2\upi W_0\,\frac{2^{2/3}}{|R|^{1/3}}\,
\Ai^2\big(-\sigma_\ki |2R|^{1/3}x\big),
\end{gather}
which corresponds to approximating \eq{eq:Dpsi2} with the (appropriately rescaled) Airy equation near the reflection point. Although this is a commonly known result, a comparison of \eq{eq:psiAiry} with the exact solutions for a QHO is shown in \Fig{fig:airypsi} for completeness.

Let us also consider the opposite limit, that is, \m{B \to 0} and \m{D \to 1}. One might expect to recover \eq{eq:psixw3} in this case, yet \eq{eq:aux7} leads to a different result. Assuming the notation \m{\xi \doteq |R|^{2/3}/(2^{1/3}|B|) \gg 1}, it can be written as follows:
\begin{gather}\label{eq:aux8}
|\psi_{\oper{x}} (x)|^2 \to \frac{2W_0}{v}\,2\upi \xi\,
\Ai^2(-\xi^2) \approx \frac{2W_0}{v}\left(1 + \sin \frac{4\xi^{3}}{3} \right).
\end{gather}
This is understood as follows. Notice that now the problem is posed such that \m{|\psi_{\oper{x}} (x)|^2} automatically includes the contributions from \textit{two} branches, that is, the two roots \m{k(x)} of the parabolic Hamiltonian \eq{eq:Hxkquad}. At \m{B \to 0}, \m{D \to 1}, and \m{x \to 0}, one of these roots approaches zero, while the other one tends to infinity, \m{k \approx -2R/B^2 \to -\infty}. To the extent that the interference with this second branch can be ignored, one can average out the second term in \eq{eq:aux8} to zero and obtain a result consistent with \eq{eq:psixw2}, \m{|\psi_{\oper{x}} (x)|^2 \to 2W_0/v}. Likewise, if the second branch is ignored entirely,\footnote{See the previous footnote.} then
\begin{gather}\label{eq:singlebranch}
|\psi_{\oper{x}} (x)|^2 \to W_0/v,
\end{gather}
as expected. This can equivalently be done by replacing the function \m{\Ai^2} in \eq{eq:aux8} with \m{|\Ai + \ii\,\Gi|^2/4}, where \m{\Gi} is the Scorer function, to isolate the contribution from a single branch \citep{ref:lopez24}. The asymptotic form of \m{|\Ai + \ii\,\Gi|} for large \m{\xi} then gives \eq{eq:singlebranch}.

\section{MGO for vector waves}
\label{sec:mgogeneral}

\subsection{Basic equations}

Let us now consider the case when \m{\vec{\psi}_{\oper{\tau}}} is an \m{N}-dimensional field,\footnote{Note that the dimension of \m{\vec{\psi}_{\oper{\tau}}} is not related to the dimension of the coordinate space.} and, accordingly, \m{\boper{H}_{\oper{\tau}}} and \m{\boper{\Gamma}_{\oper{\tau}}} are \m{N \times N} matrices of operators (hence the bold font),
\begin{gather}\label{eq:hg2}
(\boper{H}_{\oper{\tau}} + \ii\boper{\Gamma}_{\oper{\tau}})\ket{\vec{\psi}} = 0.
\end{gather}
In this case, one cannot adopt \m{\oper{h}_{\oper{\tau}}} equal to \m{\boper{H}_{\oper{\tau}}}, because \m{\oper{h}_{\oper{\tau}} = - \ii \pd_\tau} must be a one-dimensional operator. (One can convert any Hermitian combination of the elements of \m{\boper{H}_{\oper{\tau}}} into \m{-\ii \pd_\tau} but not all of them at the same time.) Instead, we proceed as follows. Let us Weyl-expand \eq{eq:hg2} by analogy with \eq{eq:Leq}, using the smallness of \m{h} near the reference~ray:
\begin{gather}\label{eq:Leqm}
(\matr{H}_\sr(\tau) + \ii \matr{\Gamma}_\sr(\tau))\vec{\psi}_{\oper{\tau}} 
- \ii \matr{V}(\tau) \pd_\tau \vec{\psi}_{\oper{\tau}} 
- \frac{\ii}{2}\,\matr{V}'(\tau) \vec{\psi}_{\oper{\tau}} \approx 0.
\end{gather}
Here, the reference ray is yet to be specified, and so is \m{\oper{h}}. Also, the prime denotes a derivative with respect to the argument, \m{\matr{H}_\sr(\tau)} is the symbol \m{\matr{H}_\sr} on the reference ray,\footnote{As a reminder, the symbol \m{\matr{A} \doteq \symb\boper{A}} of a given matrix operator \m{\boper{A}} is simply the matrix composed of the symbols of the original matrix elements, \m{\symb \oper{A}_{ab}}.} \m{\matr{\Gamma}_\sr(\tau)} is the symbol \m{\matr{\Gamma}_\sr} on the reference ray, and 
\begin{gather}
\matr{V}(\tau) \doteq (\pd_h \matr{H}_\sr(\tau, h))_{h = 0}.
\end{gather}
As we have already mentioned in \Sec{sec:envscal}, this approach also allows including transverse diffraction, following the same procedure as used in \citep{my:quasiop1, my:quasiop2, my:quasiop3} for conventional GO, but we leave this generalization to future work.

\subsection{Single mode}
\label{sec:vwaves}

Since \m{\matr{\Gamma}_\sr} and \m{\pd_\tau} are order-\m{\epsilon} in \eq{eq:Leqm}, the term \m{\matr{H}_\sr\vec{\psi}_{\oper{\tau}}} must be order-\m{\epsilon} as well. This means that \m{\vec{\psi}_{\oper{\tau}}} is close to a local normalised eigenvector \m{\vec{\eta}} of \m{\matr{H}_\sr} that corresponds to some small eigenvalue \m{\Lambda} (which is real, because \m{\matr{H}_\sr} is Hermitian); \ie
\begin{gather}
\matr{H}_\sr \vec{\eta} = \Lambda \vec{\eta},
\qquad
\vec{\eta}^\dag\vec{\eta} = 1,
\qquad
\Lambda = \bigO(\epsilon).
\end{gather}
Let us assume, until \Sec{sec:mc}, that \m{\matr{H}_\sz} is non-degenerate, so that \m{\Lambda} is its \textit{only} small eigenvalue. Then, such \m{\vec{\psi}_{\oper{\tau}}} can be represented as 
\begin{gather}
\vec{\psi}_{\oper{\tau}} = \vec{\eta} a + \bar{\vec{\Psi}},
\qquad
\bar{\vec{\Psi}} = \bigO(\epsilon),
\qquad
\vec{\eta}^\dag \bar{\vec{\Psi}} = 0,
\end{gather}
where \m{a = a(\tau)} is a scalar amplitude and \m{\bar{\vec{\Psi}}} is the (small) component of \m{\vec{\psi}_{\oper{\tau}}} transverse to \m{\vec{\eta} = \vec{\eta}(\tau)}. This means that the wave consists mostly of a single mode\footnote{By modes, we mean monochromatic waves that would have been exact solutions at~\m{\pd_\tau = 0}.} with polarization \m{\vec{\eta}}, and the nonzero contribution \m{\bar{\vec{\Psi}}} from the other modes is only due to the fact that the wave evolves along the ray and therefore is not entirely monochromatic in \m{\tau}.

Let us multiply \eq{eq:Leqm} by \m{\vec{\eta}^\dag} from the left. After omitting \m{\bigO(\epsilon^2)} terms, this gives\footnote{As usual, the prime denotes a derivative with respect to the argument, \ie \m{\tau}. Also, \m{\pd_\tau \equiv a'} here, but would be a partial derivative for quasioptical beams in a multi-dimensional space.}
\begin{gather}
(\Lambda + \ii \underbrace{\vec{\eta}^\dag\matr{\Gamma}_\sr\vec{\eta}}_{\Gamma})a
- \ii \vec{\eta}^\dag \vec{V} \vec{\eta}' a 
- \frac{\ii}{2}\,\vec{\eta}^\dag \vec{V}' \vec{\eta} a
- \ii \underbrace{\vec{\eta}^\dag \vec{V} \vec{\eta}}_{\mc{V}} \pd_\tau a = 0,
\end{gather}
or, equivalently,
\begin{gather}\label{eq:pen2}
(\Lambda - \spinhall + \ii \Gamma)a
- \ii \mc{V} \pd_\tau a - \frac{\ii}{2}\,\mc{V}' a = 0,
\qquad
\spinhall \doteq \im ((\vec{\eta}^\dag)' \vec{V} \vec{\eta}),
\end{gather}
where we used 
\begin{align}
- \ii \vec{\eta}^\dag \vec{V} \vec{\eta}' - \frac{\ii}{2}\,\vec{\eta}^\dag \vec{V}' \vec{\eta}
& = - \ii \vec{\eta}^\dag \vec{V} \vec{\eta}' - \frac{\ii}{2}\,\pd_\tau(\vec{\eta}^\dag \vec{V} \vec{\eta}) 
+ \frac{\ii}{2}\,(\vec{\eta}^\dag)' \vec{V} \vec{\eta}
+ \frac{\ii}{2}\,\vec{\eta}^\dag \vec{V} \vec{\eta}'
\notag\\
& = - \frac{\ii}{2}\,\mc{V}' - \frac{\ii}{2}(\vec{\eta}^\dag \vec{V} \vec{\eta}' - (\vec{\eta}^\dag)' \vec{V} \vec{\eta})
\notag\\
& = - \frac{\ii}{2}\,\mc{V}' - \spinhall.
\end{align}
Also note that \m{\mc{V} \equiv \vec{\eta}^\dag (\pd_h \matr{H}_\sr) \vec{\eta}} can be written as
\begin{align}
\mc{V} 
& = \frac{\pd}{\pd h}\,(\vec{\eta}^\dag \matr{H}_\sr \vec{\eta}) 
- \frac{\pd \vec{\eta}^\dag}{\pd h}\,\matr{H}_\sr \vec{\eta}
- \vec{\eta}^\dag \matr{H}_\sr\, \frac{\pd \vec{\eta}}{\pd h}
\notag\\
& = \frac{\pd \Lambda}{\pd h} 
- \frac{\pd \vec{\eta}^\dag}{\pd h}\,\Lambda\vec{\eta}
- \vec{\eta}^\dag \Lambda\,\frac{\pd \vec{\eta}}{\pd h} 
\notag\\
& = \frac{\pd \Lambda}{\pd h}
- \frac{\pd (\vec{\eta}^\dag \vec{\eta})}{\pd h}\,\Lambda
\notag\\
& = \frac{\pd \Lambda}{\pd h},
\label{eq:mcV}
\end{align}
because \m{\vec{\eta}^\dag \vec{\eta} \equiv 1}. Then, it is convenient to choose \m{\oper{h}} such that \m{h = \Lambda}, so \eq{eq:mcV} yields \m{\mc{V} = 1}.

Now, let us finally specify the reference ray. One option is to choose it such that \m{\Lambda = 0} on this ray, as most commonly done in GO \citep{book:stix}. Then, \eq{eq:pen2} becomes
\begin{gather}\label{eq:Pvw}
\pd_\tau a = (\Gamma + \ii \spinhall)a,
\qquad
\pd_\tau |a|^2 = 2\Gamma |a|^2,
\end{gather}
which is the MGO counterpart of the GO equations \eq{eq:act0}. An advantage of this approach is the simplicity of the ray Hamiltonian, \m{\Lambda}. A disadvantage is that the term \m{\spinhall} causes the phase of \m{a} to evolve. Over large time, this can make \m{a} a rapidly oscillating function (across the rays in multi-dimensional problems), potentially undermining the assumed ordering and thus rendering \eq{eq:Pvw} inaccurate.

Alternatively, one can choose the reference ray such that \m{\Lambda - \spinhall = 0}, so \eq{eq:pen2} becomes
\begin{gather}\label{eq:Pvw2}
\pd_\tau a = \Gamma a,
\qquad
\pd_\tau |a|^2 = 2\Gamma |a|^2.
\end{gather}
The advantage of this approach is that the phase of \m{a} remains fixed, so \m{a} is the true envelope and is more likely to remain smooth. However, since the ray Hamiltonian is now \m{\Lambda - \spinhall}, the ray equations are now more complicated. The resulting reference ray deviates from the conventional GO trajectory governed by the Hamiltonian \m{\Lambda}. In GO, this deviation is called the spin-Hall effect. For further discussion, see \citep{my:spinfu, my:qdiel, my:quasiop1, my:covar, ref:littlejohn91} and references therein.

Solutions of the vector equations \eq{eq:Pvw} and \eq{eq:Pvw2} can be mapped to the \m{x}-representation just like described in \Sec{sec:mapping}. (Importantly, the fact that \m{\vec{\psi}_{\oper{\tau}}} is a vector has nothing to do with \Mwaves, which are constructed based just on the scalar ray Hamiltonian \m{\Lambda}, and therefore remain scalar.)

\subsection{Mode conversion}
\label{sec:mc}

Suppose now that \m{\matr{H}_\sr} is degenerate, so that more than one of its eigenvalues are \m{\bigO(\epsilon)}. This means that the wave can consist of multiple modes that can resonantly interact with each other, \ie experience linear mode conversion. In this case, one can represent \m{\vec{\psi}_{\oper{\tau}}} as 
\begin{gather}\label{eq:psiXib}
\vec{\psi}_{\oper{\tau}} = \vec{\Xi} \vec{a} + \bar{\vec{\Psi}},
\qquad
\bar{\vec{\Psi}} = \bigO(\epsilon),
\end{gather}
where the (generally non-square) matrix \m{\vec{\Xi}} has the orthonormal polarization vectors \m{\vec{\eta}_i} of the resonant modes (eigenvectors of \m{\matr{H}_\sr}) as its columns, \m{\vec{a}} is the vector that characterises the amplitudes of these modes, and \m{\bar{\vec{\Psi}}} is the nonresonant part of \m{\vec{\psi}_{\oper{\tau}}}. For details, see \citep{my:quasiop1}, which discusses a similar parametrization within GO.

The derivation of the envelope equation in this case is identical to that in \Sec{sec:vwaves} up to replacing \m{\vec{\eta}} with \m{\matr{\Xi}} (and \m{a} with \m{\vec{a}}), and one obtains
\begin{gather}\label{eq:pen3}
(\matr{\Lambda} - \matr{\spinhall} + \ii \matr{\Gamma})\vec{a}
- \ii \matr{\mc{V}} \pd_\tau \vec{a} - \frac{\ii}{2}\,\matr{\mc{V}}' \vec{a} = 0,
\qquad
\matr{\spinhall} \doteq ((\matr{\Xi}^\dag)' \matr{V} \matr{\Xi})_{\rm A}.
\end{gather}
Here, the index A denotes the anti-Hermitian part, and \m{\matr{\Lambda} = \matr{\Xi}^\dag\matr{H}_\sr \matr{\Xi}} is a diagonal matrix, \m{\matr{\Lambda} = \diag\{\Lambda_1, \Lambda_2, \ldots\}}, where \m{\Lambda_i} are the small eigenvalues of \m{\matr{H}_\sr}. Also, \m{\matr{\Gamma} = \matr{\Xi}^\dag\matr{\Gamma}_\sr \matr{\Xi}}, and \m{\matr{\mc{V}} = \matr{\Xi}^\dag\matr{V} \matr{\Xi}}, so the elements of the latter, \m{\mc{V}_{ij} = \vec{\eta}_i^\dag (\pd_h\matr{H}_\sr) \vec{\eta}_j}, can be expressed as
\begin{align}
\mc{V}_{ij} 
& = \frac{\pd}{\pd h}\, (\vec{\eta}_i^\dag \matr{H}_\sr \vec{\eta}_j)
- \frac{\pd \vec{\eta}_i^\dag}{\pd h}\, \matr{H}_\sr\vec{\eta}_j
- \vec{\eta}_i^\dag \matr{H}_\sr\,\frac{\pd \vec{\eta}_j}{\pd h}
\notag\\
& = \frac{\pd \Lambda_{ij}}{\pd h} 
- \frac{\pd \vec{\eta}_i^\dag}{\pd h}\,\Lambda_j\vec{\eta}_j
- \vec{\eta}_i^\dag \Lambda_i\,\frac{\pd \vec{\eta}_j}{\pd h}.
\label{eq:mcVij}
\end{align}
In the envelope equation, the coefficients \m{\mc{V}_{ij}} appear only in combination with \m{\pd_\tau = \bigO(\epsilon)}, so they are of interest only to the zeroth order in \m{\epsilon}. Then, the last two terms in \eq{eq:mcVij}, which are proportional to small eigenvalues of \m{\matr{H}_\sr}, are negligible; \ie \m{\vec{\mc{V}} \approx \pd_h \matr{\Lambda}}. In other words, \m{\vec{\mc{V}}} is approximately diagonal, so mode coupling is determined entirely by \m{\matr{U}} and \m{\matr{\Gamma}}. Also, if \m{\det\matr{\mc{V}} \approx \prod_i \pd_h \Lambda_i} is nonzero, \ie all \m{\pd_h \Lambda_i} are nonzero, then \m{\vec{\mc{V}}} is invertible and one can rewrite \eq{eq:pen3} as
\begin{gather}\label{eq:b}
\ii \pd_\tau \vec{b} 
= (\matr{\mc{N}}^{-1} (\matr{\Lambda} - \matr{\spinhall} + \ii \matr{\Gamma}) \matr{\mc{N}}^{-1} + \matr{\mcc{H}})\vec{b},
\end{gather}
where \m{\vec{b} \doteq \matr{\mc{N}}\vec{a}}, \m{\matr{\mc{N}} = \matr{\mc{V}}^{1/2}}, and\footnote{The fact that \m{\vec{\mc{V}}} is \textit{approximately} diagonal is enough to prove that it is invertible, \ie that \m{\matr{\mc{N}}^{-1}} exists, but not necessarily enough for \m{\matr{\mcc{H}}} to be negligible.}
\begin{align}
\matr{\mcc{H}} 
& = - \ii \matr{\mc{N}} (\matr{\mc{N}}^{-1})'
- \frac{\ii}{2}\,\matr{\mc{N}}^{-1}(\matr{\mc{N}}\matr{\mc{N}})'\matr{\mc{N}}^{-1}
\notag\\
& = \ii \matr{\mc{N}}' \matr{\mc{N}}^{-1}
- \frac{\ii}{2}\,\matr{\mc{N}}^{-1}\matr{\mc{N}}'\matr{\mc{N}}\matr{\mc{N}}^{-1}
- \frac{\ii}{2}\,\matr{\mc{N}}^{-1}\matr{\mc{N}}\matr{\mc{N}}'\matr{\mc{N}}^{-1}
\notag\\
& = \ii \matr{\mc{N}}' \matr{\mc{N}}^{-1}
- \frac{\ii}{2}\,\matr{\mc{N}}^{-1}\matr{\mc{N}}'
- \frac{\ii}{2}\,\matr{\mc{N}}'\matr{\mc{N}}^{-1}
\notag\\
& = \frac{\ii}{2}\,(\matr{\mc{N}}' \matr{\mc{N}}^{-1}
- \matr{\mc{N}}^{-1}\matr{\mc{N}}').
\end{align}
In particular, at zero \m{\matr{\Gamma}}, the operator on the right-hand side of \eq{eq:b} is Hermitian, so \m{|\vec{b}|^2} is conserved. This reflects conservation of the total action of resonant waves.

\begin{figure*}
\centering
\includegraphics[width=.99\textwidth]{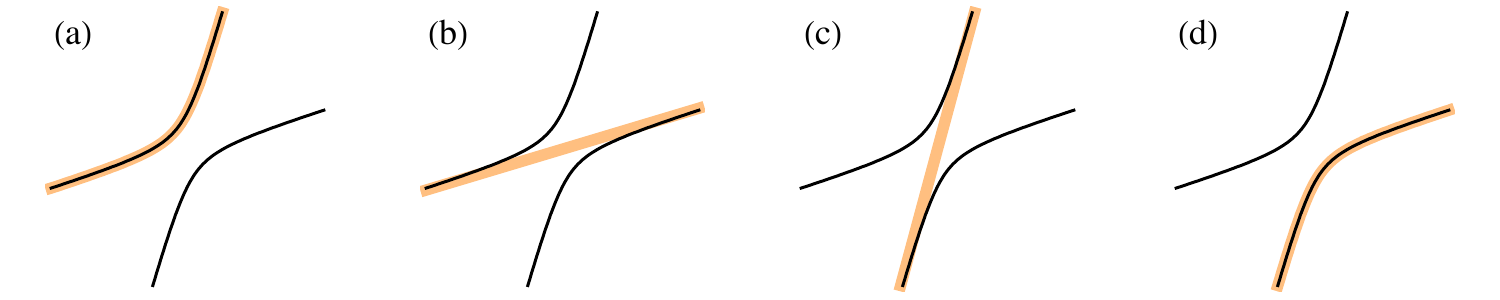}
\caption{\label{fig:mc1}
A typical structure of the dispersion curves (black) of two resonant waves in the mode-conversion region in phase space \m{\sz}. The orange curves indicate possible choices of the reference ray, depending on a problem that one needs to solve.
}
\end{figure*}

In a typical scenario involving two resonant modes \citep{book:tracy, ref:tracy03}, the corresponding dispersion curves are roughly parallel to each other in the mode-conversion region due to the `level repulsion' (\Fig{fig:mc1}). If the reference ray is chosen such that it propagates at least somewhat along one dispersion curve, then it also propagates somewhat along the other one as well. Then both \m{\pd_h \Lambda_1} and \m{\pd_h \Lambda_2} are nonzero, so \m{\matr{\mc{V}}} is invertible. Then, both \eq{eq:pen3} and \eq{eq:b} predict non-singular \m{\vec{\psi}_{\oper{\tau}}}, regardless of the orientation of the dispersion curves in the original \m{\sz}-space (\ie MGO can handle mode conversion near cutoffs). The optimal choice of the Hamiltonian for the reference ray in this case depends on a problem that one needs to solve (\Fig{fig:mc1}). For example, if one is interested in how much energy is retained on the initial branch after the mode conversion, one might want to choose \m{h = \Lambda_1} or \m{h = \Lambda_2}, depending on the initial conditions, so the reference ray simply follows one of the branches. Alternatively, if one is interested in how much energy is transmitted through the gap, one can choose \m{h = c_1\Lambda_1 + c_2\Lambda_2}, where the coefficients \m{c_1} and \m{c_2} are such that one of them vanishes at \m{\tau \to -\infty} and the other one vanishes at \m{\tau \to +\infty}.

\begin{figure*}
\centering
\includegraphics[width=.9\textwidth]{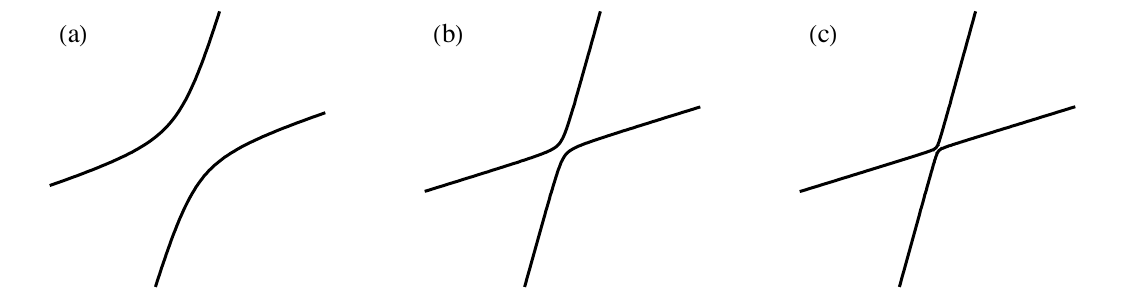}
\caption{\label{fig:mc2}
A schematic of how the dispersion curves of two resonance waves in the mode-conversion region in the \m{\sz}-space transition from smooth (left) to non-smooth (right) as the exact resonance is approached.
}
\end{figure*}

Keep in mind, however, that the above approach to mode conversion may not be optimal when the gap between the dispersion curves is particularly small, resulting in large \m{\matr{\Lambda}'} and \m{\matr{X}'} (\Fig{fig:mc2}). In this case, representing \m{\vec{\psi}_{\oper{\tau}}} in the eigenvector basis, as in \eq{eq:psiXib}, may be inconvenient. Instead, one might want to solve the original \eq{eq:Leqm}, where the coefficients are smooth, unlike in \eq{eq:pen3} and \eq{eq:b}. We will elaborate on this subject in a separate publication.

Solutions of \eq{eq:b} (or, if needed, \eq{eq:Leqm}) can be mapped to the \m{x}-representation mostly like described in \Sec{sec:mapping}. The only difference is that, in this case, the Wigner function of the wavefield becomes a Wigner \textit{matrix}, which is the symbol of \m{(2\upi)^{-1} \ket{\vec{\psi}}\bra{\vec{\psi}}}.

\section{Wave--particle interactions in MGO}
\label{sec:wp}

\subsection{Generalised Cherenkov resonance}

The MT formalism also suggests a way to rethink wave--particle interactions. Consider the Cherenkov condition of the wave--particle resonance:
\begin{gather}\label{eq:ch0}
\omega = kv_{\text{p}},
\end{gather}
where \m{\omega} is the wave frequency, \m{k} is the local wavenumber, and \m{v_{\text{p}}} is the particle velocity. This condition is not symplectically invariant but can be generalised within MGO, if one considers particles semiclassically.\footnote{See \citep{my:qponder, my:lens} for how the Cherenkov condition emerges in generic wave--wave interactions and semiclassical interactions in particular.} Like any other waves, a semiclassical particle can be assigned phase-space variables \m{(\tau, h)}. Suppose such a particle interacts with another (\eg classical electromagnetic) wave that is eikonal in these variables too, even though this wave might not propagate along the curve \m{h = 0}. We assume this wave in the form
\begin{gather}
\psi_{\oper{\tau}}(t, \tau) = \Psi(t, \tau)\ee^{\ii\theta(t, \tau)}.
\end{gather}
Here, \m{\Psi} is the envelope that is slow compared to the (real) phase \m{\theta}, and the time \m{t} can be considered as a parameter within our formalism. A particle can interact resonantly with such a wave if \m{\theta} changes slowly, or not at all, in the particle frame:
\begin{gather}\label{eq:ch1}
0 \approx \dot{\theta} 
= \pd_t \theta + \dot{\tau} \pd_\tau \theta,
\end{gather}
where the dot stands for \m{\dd/\dd t}.\footnote{Equation \eq{eq:ch1} subsumes \eq{eq:ch0} as a special case. Indeed, if the \m{\tau}-axis is parallel to the \m{x}-axis, one has \m{\dot{\tau} \pd_\tau \theta = \dot{x} \pd_x \theta = k v_{\text{p}}}, where we used \m{k \doteq \pd_x \theta} and \m{v_{\text{p}} \doteq \dot{x}}.} Provided that the particle Hamiltonian and the wave Hamiltonian are normalized consistently, one has \m{\dot{\tau} = 1} (cf.~\eq{eq:thre}). Then, assuming the standard definition \m{\omega \doteq - \pd_t \theta}, \eq{eq:ch1} leads to a \text{metaplectic resonance condition}:
\begin{gather}\label{eq:ch2}
\omega = \pd_\tau \theta.
\end{gather}
(An alternative derivation and interpretation of this resonance condition is given in \Sec{sec:heat}.) Since \m{\dot{\tau} = 1}, all quantities here are entirely classical, \ie \m{\hbar} is not involved. Also notice that the \m{\tau} axis can be aligned arbitrarily relative to the physical-space axis \m{x}. For example, a wave can resonantly interact with a particle even near a cutoff in the \m{x}-space. This unifies Cherenkov acceleration with Fermi acceleration, which we define here broadly as particle acceleration caused by interaction with moving walls.

To better understand the condition \eq{eq:ch2}, recall that \m{\dd\tau = \dd\phi/\Omega}. As a reminder, \m{\phi} and \m{\Omega} are the canonical phase and frequency of particle's OHO (\Sec{sec:iy}),
\begin{gather}\label{eq:ch6}
\Omega = \left|\det \big(\pd^2_\sz H_\sz\big)\right|^{1/2}\sgn\ki,
\end{gather}
where \m{H_\sz} is the particle Hamiltonian, \m{\sz} is the particle phase-space variables, and \m{\ki \doteq \ddot{\sz} \wedge \dot{\sz}}.\footnote{Since we treat the particle as a quantum wave, its classical Hamiltonian is \m{\hbar H_\sz} and its classical momentum is \m{\hbar k}. However, simultaneously rescaling the energy and momentum by \m{\hbar} (or any other constant factor) does not affect \eq{eq:ch6}, so one can as well interpret \m{H_\sz} in \eq{eq:ch6} as the classical Hamiltonian and \m{\sz} as the classical phase space.} Then, \eq{eq:ch2} can be written as
\begin{gather}\label{eq:ch3}
\frac{\pd \theta}{\pd \phi} = \frac{\omega}{\Omega}.
\end{gather}
For illustration, let us consider a special case when \m{\omega} is a constant and the particle oscillates (nonrelativistically) in a parabolic potential, so \m{\Omega} is a constant too. Then, assuming \eq{eq:ch3} is satisfied globally, integrating this equation over \m{\phi} from \m{0} to \m{2\upi} gives
\begin{gather}\label{eq:ch4}
\frac{\Delta \theta_\circlearrowright}{2\upi} = \frac{\omega}{\Omega},
\end{gather}
where \m{\Delta \theta_\circlearrowright} is the total change of \m{\theta} on said interval. Since \m{m \doteq \Delta \theta_\circlearrowright/2\upi} must be integer, \eq{eq:ch4} leads to the following condition for global wave--particle resonances:
\begin{gather}\label{eq:ch5}
\omega = m \Omega.
\end{gather}
This result is, of course, not unexpected. However, note that \eq{eq:ch5} is only a special case that illustrates a more general \textit{local} resonance condition \eq{eq:ch3}, which does not require the wave and particle oscillations to be bounded. 

\subsection{Metaplectic resonance heating}
\label{sec:heat}

The metaplectic resonance condition \eq{eq:ch2} also emerges naturally in plasma kinetic theory. Consider particles with a probability distribution function \m{f} governed by the Vlasov equation 
\begin{gather}
\pd_t f + \poisson{H, f} = 0,
\end{gather}
where \m{H} is the particle Hamiltonian. Suppose \m{H = H_0 + H_1}, where \m{H_1} is a perturbation small compared to \m{H_0}. Then, within linear theory, one has \m{f = f_0 + f_1}, where \m{f_0} is the background distribution and \m{f_1} is of order \m{H_1}. These functions satisfy
\begin{gather}
\pd_t f_0 + \poisson{H_0, f_0} = 0,
\qquad
\pd_t f_1 + \poisson{H_0, f_1} = - \poisson{H_1, f_0}.
\end{gather}
Let us assume that \m{H_0} is time-independent and \m{f_0} is an equilibrium distribution \m{f_0 = f_0(H_0)}. Then, in the canonical variables \m{(\tau, H_0)}, the Poisson brackets in the equation for \m{f_1} can be expressed as follows:
\begin{gather}
\poisson{H_0, f_1} = \pd_\tau f_1,
\qquad
\poisson{H_1, f_0(H_0)} = - (\pd_\tau H_1)f_0'(H_0).
\end{gather}
This leads to
\begin{gather}
(\pd_t + \pd_\tau) f_1 = (\pd_\tau H_1) f_0'(H_0).
\end{gather}
Let us assume that the perturbation is an eikonal wave in \m{\tau}, such that \m{f_1 = \re \tilde{f}}, \m{H_1 = \re \tilde{H}},
\begin{gather}
\pd_t \tilde{f} = - \ii \omega \tilde{f},
\qquad
\pd_\tau \tilde{f} = \ii (\pd_\tau \theta)\tilde{f},
\qquad
\pd_\tau \tilde{H} = \ii (\pd_\tau \theta)\tilde{H}.
\end{gather}
Hence,
\begin{gather}
\tilde{f} = - \frac{\pd_\tau\theta}{\omega - \pd_\tau\theta}\,\tilde{H} f_0'(H_0),
\end{gather}
so particles will be resonant with the wave under the condition \eq{eq:ch2}. 

As usual, the presence of a linear resonance implies a heating mechanism, which we call \textit{metaplectic resonance heating}. A quantitative theory of such heating for specific waves is beyond the scope of this paper, but note that the concept of power dissipation is naturally extendable from the physical \m{x}-space to metaplectically transformed fields. For example, as shown in \App{app:P}, there is a remarkably concise formula that connects the energy dissipated by an electromagnetic wave, per volume \m{\Gamma} of wave's extended phase space, in any linear dielectric medium and plasma in particular:
\begin{gather}
\frac{\dd\mcc{E}}{\dd\Gamma} = \tr(\matr{\sigma}_{\rm H}\matr{\mcc{W}}).
\end{gather}
Here, \m{\matr{\sigma}_{\rm H}} is the symbol of the Hermitian part of the conductivity operator, \m{\matr{\mcc{W}}} is the average Wigner matrix of the electric field, and \m{\tr} denotes the matrix trace. Traditionally, \m{\matr{\sigma}} is calculated analytically or semi-analytically using GO ordering. Using MGO, it should be possible to extend those calculations to regions where GO fails, cutoffs in particular.

\section{Conclusions}

Let us summarize. In this paper, we introduce a formulation of geometrical optics (GO) that is not limited to small \m{\lambda/L} in the physical space. How do we do it? The intrepid readers who hoped to find a short answer here are referred to the synopsis in \Sec{sec:spoiler}, which is as short as a meaningful answer can be. But here are some highlights. 

As commonly known by now \citep{book:tracy}, the natural language of GO is the Weyl symbol calculus and metaplectic transforms (MTs).\footnote{In fact, one can think of the envelope dynamics itself as a metaplectic image of the dynamics of the complete wavefield, as discussed in \Sec{sec:go}.} But neither Weyl symbols nor MTs are inherently tied to the physical space \m{x} or the spectral space \m{k} for that matter. Thus, there is no fundamental reason, other than tradition, why the envelope dynamics should be calculated in the \m{x}-space or the \m{k}-space. Furthermore, instead of wavefields, what really matters for practical use are their Wigner functions (or even integrals thereof). As we show, those can be easily recalculated between any two representations using the Airy transform. Then, it makes sense to work in the phase-space variables in which the GO applicability conditions are satisfied best and map the solution to the \m{x}-space only occasionally, when needed or convenient.

The most natural variables are the ray time \m{\tau} and the ray energy \m{h}. We explicitly derive the envelope equation in these variables; see \Sec{sec:mgogeneral} for the general case. As usual, the coefficients therein are determined by the Weyl symbol of the dispersion operator, which is the same (within the assumed accuracy) as the corresponding symbol in \m{(x, k)}. This means that integrating the envelope equation in MGO is just as easy as integrating the one in GO. Yet, because the derivatives in the MGO equation are taken along cutoff-free directions, the field remains non-singular. As we show, the standard Airy patterns that form in regions where conventional GO fails are successfully reproduced within MGO simply by remapping the field from the \m{\tau}-space to the \m{x}-space.

MGO will likely be useful, for example, for reduced modeling of the O--X conversion in inhomogeneous plasma near the critical density (work in progress). A generalization to quasioptical wave beams, which can be done mostly like in \citep{my:quasiop1, my:quasiop2, my:quasiop3}, is left to future work. Overall, MGO can replace GO for any practical purposes, because it better handles cutoffs and is a similar framework otherwise.

\section*{Funding}

This research was supported by the U.S.\ Department of Energy through contract No.\ DE-AC02-09CH11466, by an A*STAR SERC Central Research Fund, and by grant NSF PHY-2309135 to the Kavli Institute for Theoretical Physics (KITP).

The United States Government retains a non-exclusive, paid-up, irrevocable, world-wide license to publish or reproduce the published form of this manuscript, or allow others to do so, for United States Government purposes.

\section*{Declaration of interests}

The authors report no conflict of interest.

\appendix
\section{Ray equations}
\label{app:req}

Here, we rederive the ray equations, both for completeness and also to introduce some terminology. In doing so, we generally follow \citep{my:ql}, but see also \citep{book:tracy}. Note that the notation in this appendix differs from that in the main text.

Consider an eikonal wave with a rapid phase \m{\theta(t, \vec{x})}, where \m{t} is the physical time and \m{\vec{x}} is the coordinate in the physical space, which we now allow to be multi-dimensional for generality. The phase determines the wave frequency and wavevector:
\begin{gather}\label{eq:wk}
\bar{\omega}(t, \vec{x}) \doteq - \pd_t \theta,
\qquad
\bar{\vec{k}}(t, \vec{x}) \doteq \nabla \theta.
\end{gather}
(We use bars to distinguish these functions from generic coordinates in the spectral space.) Suppose that they satisfy some dispersion relation 
\begin{gather}\label{eq:HJ}
\Lambda(t, \vec{x}, \bar{\omega}, \bar{\vec{k}}) = 0,
\end{gather}
where \m{\Lambda} is a real function. Let us introduce \m{w(t, \vec{x}, \vec{k})} as the (relevant) solution of
\begin{gather}\label{eq:wdef}
\Lambda(t, \vec{x}, w(t, \vec{x}, \vec{k}), \vec{k}) = 0
\end{gather}
for a given \m{\vec{k}}. Differentiating \eq{eq:wdef} with respect to \m{t}, \m{\vec{x}}, and \m{\vec{k}} gives
\begin{subequations}\label{eq:Lambdaw}
\begin{align}
& \pd_{t} \Lambda + (\pd_\omega \Lambda)\pd_t w = 0,
\\
& \pd_{\vec{x}} \Lambda + (\pd_\omega \Lambda)\pd_{\vec{x}}w = 0,
\\
& \pd_{\vec{k}} \Lambda + (\pd_\omega \Lambda)\pd_{\vec{k}}w = 0,
\label{eq:vvg1}
\end{align}
\end{subequations}
where the derivatives of \(\Lambda\) are evaluated at \m{(t, \vec{x}, w(t, \vec{x}, \vec{k}), \vec{k})}. In particular, \eq{eq:vvg1} gives
\begin{gather}\label{eq:vvg2}
\vvg \doteq \frac{\pd w}{\pd \vec{k}} = - \frac{\pd_{\vec{k}} \Lambda}{\pd_\omega \Lambda},
\end{gather}
for the group velocity \m{\vvg}, whose physical meaning is yet to be introduced. The notation \m{\avvg} used below denotes \m{\vvg} evaluated on \m{\vec{k} = \bar{\vec{k}}(t, \vec{x})}.

From \eq{eq:wk}, one readily obtains
\begin{gather}
\frac{\pd \bar{k}_i}{\pd t} + \frac{\pd \bar{\omega}}{\pd x^i} = 0,
\qquad
\frac{\pd \bar{k}_j}{\pd x^i} = \frac{\pd \bar{k}_i}{\pd x^j},
\end{gather}
which are known as consistency relations. They lead to
\begin{align}
\bigg(\frac{\pd}{\pd t} & + \avvg \cdot \frac{\pd}{\pd \vec{x}}\bigg) \bar{k}_i(t, \vec{x})
= - \frac{\pd w(t, \vec{x}, \bar{\vec{k}}(t, \vec{x}))}{\pd x^i} + \avvg \cdot \frac{\pd \bar{k}_i(t, \vec{x})}{\pd \vec{x}}
\notag\\
& = - \left(\frac{\pd w(t, \vec{x}, \vec{k})}{\pd x^i}
\right)_{\vec{k} = \bar{\vec{k}}(t, \vec{x})}
- \avg^j\,\frac{\pd \bar{k}_j(t, \vec{x})}{\pd x^i} + \avg^j\, \frac{\pd \bar{k}_i(t, \vec{x})}{\pd x^j}
\notag\\
& = - \left(\frac{\pd w(t, \vec{x}, \vec{k})}{\pd x^i}\right)_{\vec{k} = \bar{\vec{k}}(t, \vec{x})},
\label{eq:kdw}
\end{align}
and similarly,
\begin{align}
\bigg(\frac{\pd}{\pd t} & + \avvg \cdot \frac{\pd}{\pd \vec{x}}\bigg) \bar{\omega}(t, \vec{x})
= \bigg(\frac{\pd}{\pd t} + \avvg \cdot \frac{\pd}{\pd \vec{x}}\bigg) w(t, \vec{x}, \bar{\vec{k}}(t, \vec{x}))
\notag\\
& = \left(\frac{\pd w(t, \vec{x}, \vec{k})}{\pd t} + \avg^i\, \frac{\pd w(t, \vec{x}, \vec{k})}{\pd x^i}\right)_{\vec{k} = \bar{\vec{k}}(t, \vec{x})}
+  \avg^i \left(\frac{\pd}{\pd t} + \avvg \cdot \frac{\pd}{\pd \vec{x}}\right) \bar{k}_i(t, \vec{x})
\notag\\
& = \left(\frac{\pd w(t, \vec{x}, \vec{k})}{\pd t}\right)_{\vec{k} = \bar{\vec{k}}(t, \vec{x})},
\label{eq:wdw}
\end{align}
where we used \eq{eq:kdw}. Using the convective derivative associated with the group velocity,
\begin{gather}\label{eq:conv2}
\dd/\dd t \doteq \pd_t + (\avvg \cdot \pd_{\vec{x}}),
\end{gather}
one can rewrite these compactly as
\begin{gather}\label{eq:kwdw}
\frac{\dd\bar{k}_i(t, \vec{x})}{\dd t} = - \left(\frac{\pd w(t, \vec{x}, \vec{k})}{\pd x^i}\right)_{\vec{k} = \bar{\vec{k}}(t, \vec{x})},
\quad
\frac{\dd\bar{\omega}(t, \vec{x})}{\dd t} = \left(\frac{\pd w(t, \vec{x}, \vec{k})}{\pd t}\right)_{\vec{k} = \bar{\vec{k}}(t, \vec{x})}.
\end{gather}
One can represent \eq{eq:kwdw} as \textit{ordinary} differential equations for \m{\bar{\vec{k}}(t) \doteq \bar{\vec{k}}(t, \bar{\vec{x}}(t))} and \m{\bar{\omega}(t) \doteq \bar{\omega}(t, \bar{\vec{x}}(t))}, where \m{\bar{\vec{x}}(t)} are the `ray trajectories' governed by
\begin{gather}\label{eq:xi}
\frac{\dd\bar{x}^i(t)}{\dd t} = \vg^i(t, \bar{\vec{x}}(t), \bar{\vec{k}}(t))
\end{gather}
and \m{\vvg} serves as the ray velocity. Together with \eq{eq:xi}, equations \eq{eq:kwdw} become Hamilton's equations:
\begin{gather}\label{eq:rays}
\frac{\dd x^i}{\dd t} = \frac{\pd w(t, {\vec{x}}, {\vec{k}})}{\pd {k}_i},
\qquad
\frac{\dd {k}_i}{\dd t} =- \frac{\pd w(t, {\vec{x}}, {\vec{k}})}{\pd {x}^i},
\qquad
\frac{\dd {\omega}}{\dd t} = \frac{\pd w(t, {\vec{x}}, {\vec{k}})}{\pd t},
\end{gather}
where we have finally dropped the bars for clarity.

The ray equations can also be expressed directly through the dispersion function \m{\Lambda}. To do this, let us introduce a new ray variable \m{\sigma}~via
\begin{gather}\label{eq:si}
\frac{\dd \sigma}{\dd t} = -\frac{1}{\pd_\omega \Lambda},
\end{gather}
where \m{\Lambda \equiv \Lambda(t, \vec{x}, \omega, \vec{k})}. (The minus sign on the right-hand side is a matter of convention.) Then, using \eq{eq:Lambdaw}, one can write
\begin{gather}
\frac{\pd w}{\pd t} = \frac{\pd \Lambda}{\pd t}\, \frac{\dd \sigma}{\dd t},
\qquad
\frac{\pd w}{\pd x^i} = \frac{\pd \Lambda}{\pd x^i}\,\frac{\dd \sigma}{\dd t},
\qquad
\frac{\pd w}{\pd k_i} = \frac{\pd \Lambda}{\pd k_i}\,\frac{\dd \sigma}{\dd t}.
\end{gather}
Substituting these into \eq{eq:rays}, one finds
\begin{subequations}
\begin{alignat}{2}
\frac{\dd t}{\dd \sigma} & = - \frac{\pd \Lambda}{\pd \omega},
\qquad
\frac{\dd \omega}{\dd \sigma} && = + \frac{\pd \Lambda}{\pd t},
\\
\frac{\dd x^i}{\dd \sigma} & = + \frac{\pd \Lambda}{\pd k_i},
\qquad
\frac{\dd k_i}{\dd \sigma} &&=- \frac{\pd \Lambda}{\pd x_i},
\label{eq:raysa2}
\end{alignat}
\end{subequations}
where we have also included \eq{eq:si} for completeness. One can view these as Hamilton's equations with Hamiltonian \m{\Lambda}, canonical pairs \m{(t, -\omega)} and \m{(x^i, k_i)}, and `ray time' \m{\sigma}, which is also known as the ray orbit parameter. The ray time coincides with the physical time if the ray Hamiltonian can be represented as 
\begin{gather}
\Lambda(t, \vec{x}, \omega, \vec{k}) = \lambda(t, \vec{x}, \vec{k}) - \omega.
\end{gather}

In the main text, we assume that \m{\vec{x}} and \m{\vec{k}} are one-dimensional, \m{\pd_t \Lambda \equiv 0}, overdots represent derivatives with respect to \m{\sigma}, \m{\tau} stands for \m{\sigma} expressed as a function of phase-space coordinates, and the physical time \m{t} is not used until \Sec{sec:wp}. Then, for \m{\Lambda = H_\sz}, \eq{eq:raysa2} can be written as \eq{eq:raysz}, and \eq{eq:dc} plays the role of \eq{eq:HJ}.

\section{Angle--action variables}
\label{app:aac}

Here, we elaborate on the idea of adopting MGO variables in the form \m{(\phi, j)^\T \equiv \sa}, where \m{\phi} is the angle variable of the OHO, \m{j \doteq J - J_0}, \m{J} is OHO's action, and \m{J_0 = n} is OHO's action on the ray. As mentioned in \Sec{sec:aac2}, this approach is not optimal. Nevertheless, it is still worth considering in that it helps us introduce certain concepts that are needed in other parts of this paper.

\subsection{Operator transformation}

\begin{figure*}
\centering
\includegraphics[width=.6\textwidth]{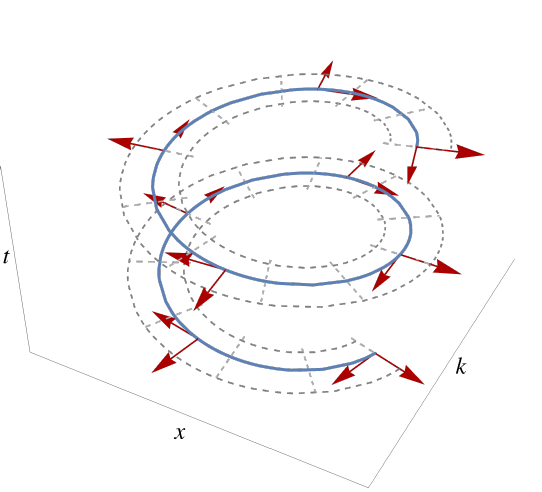}
\caption{\label{fig:spiral}
Schematic of the local \m{(\phi, j)} coordinate grid for an oscillator (same notation as in \Fig{fig:basis}). Near the ray on the \m{(\phi, j)}-plane, the isosurfaces of \m{j} are parallel to the ray and the isosurfaces of \m{\phi} are transverse to the ray. The mapping \m{(x, k) \mapsto (\phi, j)} is multi-valued, so \m{\phi} is not restricted to \m{[-\upi, \upi)}, as a canonical angle normally would, but can range from \m{-\infty} to \m{+\infty}.
}
\end{figure*}

Strictly speaking, we are looking for an \textit{operator} transformation, and the true angle variable cannot be promoted to an operator due to the discreteness of the oscillator's spectrum \citep{ref:susskind64}.\footnote{The normalised eigenvectors of \m{\oper{J}} with a discrete spectrum satisfy \m{\braket{\ev_{\oper{J}}(J_1) | \ev_{\oper{J}}(J_2)} = \delta_{J_1, J_2}}. This is incompatible with the canonical commutation relation \m{[\oper{\phi}, \oper{J}] = \ii}. Indeed, because \m{\braket{\ev_{\oper{J}}(J_1) | [\oper{\phi}, \oper{J}] | \ev_{\oper{J}}(J_2)} = (J_2  - J_1) \braket{\ev_{\oper{J}}(J_1) | \oper{\phi} | \ev_{\oper{J}}(J_2)}}, the commutation relation leads to \m{(J_2  - J_1) \braket{\ev_{\oper{J}}(J_1) | \oper{\phi} | \ev_{\oper{J}}(J_2)} = \ii \delta_{J_1, J_2}}, which cannot be satisfied at \m{J_1 = J_2}.} However, this is not a problem if the angle--action variables are defined only locally and no boundary conditions are imposed on them. By allowing \m{\phi} to be a multi-valued function of \m{\sz}, one can treat the \m{\phi}-axis as \m{\mathbb{R}} even for rays that actually undergo periodic oscillations (\Fig{fig:spiral}). The action variable can be handled similarly: although the domain of \m{J} is only half of the real axis, \m{j} is a signed quantity whose large absolute values do not matter, so the \m{j}-axis can be treated as \m{\mathbb{R}} too. Then, the Weyl symbol calculus can be introduced in the \m{(\phi, j)}-space as usual.

\begin{figure*}
\centering
\includegraphics[width=\textwidth]{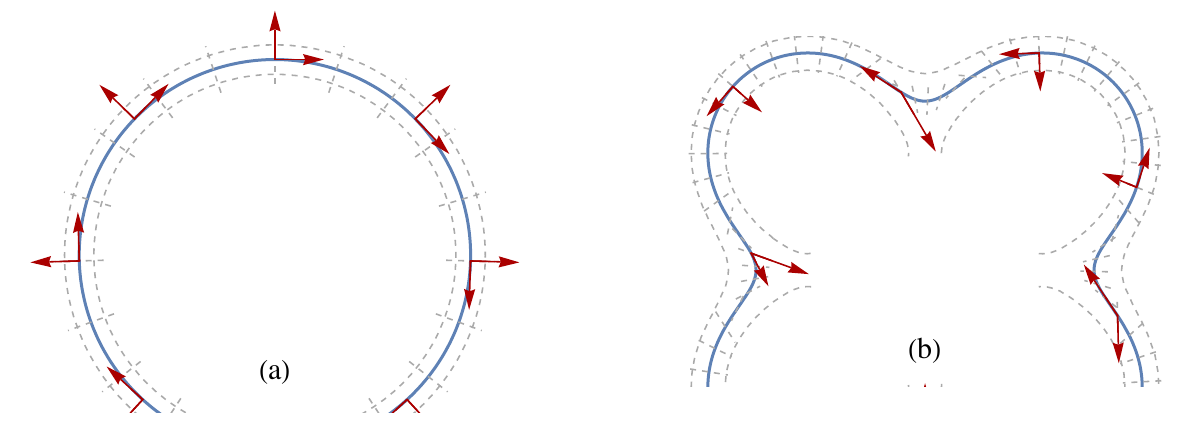}
\caption{\label{fig:basis}
Examples of the ray trajectories (blue) in the \m{\sz}-space with the corresponding local basis vectors \eq{eq:ss} (red arrows). Both figures are produced numerically for sample \m{H_\sz}. Dashed are the local isosurfaces of \m{\phi(\sz)} and \m{j(\sz)}. (a) constant \m{g > 0} and \m{\ki > 0}; (b) \m{g < 0} everywhere, while the sign of \m{\ki} varies. The directions of the arrows are consistent with those in \Fig{fig:signs}.
}
\end{figure*}

The angle--action variables of the OHO are introduced via the parametrization
\begin{subequations}\label{eq:qppj}
\begin{gather}
q = \sqrt{R^2 + 2j}\,\sin \phi,\label{eq:qpj2}
\\
p = - R + \sqrt{R^2 + 2j}\,\cos \phi.
\end{gather}
\end{subequations}
This leads to the following formulas for \m{(\phi, j)} as functions of \m{(\yq, \yp)} (for either sign of~\m{\ki}):
\begin{subequations}\label{eq:pj}
\begin{gather}
\phi = \arctan\left(\frac{\yq}{R + \yp}\right),
\label{eq:varphi}
\\
j = \frac{1}{2}\,(\yq^2 + (\yp + R)^2 - R^2).
\label{eq:j}
\end{gather}
\end{subequations}
We will now seek the corresponding operators such that
\begin{gather}\label{eq:spj}
[\oper{\phi}, \oper{j}] = \ii.
\end{gather}
This means that the corresponding symbols\footnote{Since the transformation \m{\sz \mapsto \sy} is linear, these can be equally considered as \m{\sy}-symbols or \m{\sz}-symbols.} \m{\phi(\yq, \yp)} and \m{j(\yq, \yp)} must satisfy \m{\moyal{\phi, j} = 1}, while \eq{eq:pj} satisfy \m{\poisson{\phi, j} = 1}. The Moyal bracket coincides with the Poisson bracket only to the extent that the third-order derivatives of \m{\phi} and \m{j} are negligible (\Sec{sec:weyl}), \ie in the limit of large \m{R}. Hence, one can consider \eq{eq:pj} as a large-\m{R} \textit{asymptotic} of the desired transformation, which helps to guess the operators that actually satisfy \eq{eq:spj} with an acceptable accuracy. Specifically, we proceed as follows. In \eq{eq:j}, we simply replace the symbols \m{\yq} and \m{\yp} with the corresponding operators without making additional changes, \ie adopt the standard action operator of a harmonic oscillator up to a shift:
\begin{subequations}\label{eq:phij}
\begin{gather}\label{eq:qpj}
\oper{j} = \frac{1}{2}\,(\oyq^2 + (\oyp + R)^2 - R^2). 
\end{gather}
For the angle variable, we consider the large-\m{R} asymptotic of \eq{eq:varphi} and then replace \m{\yq} and \m{\yp} with \m{\oyq} and \m{\oyp}, correspondingly, in such a way that the resulting operator remains Hermitian (or one can actually take the Weyl transform instead):
\begin{gather}\label{eq:pop}
\oper{\phi} = \frac{\oyq}{R} 
- \frac{\oyp\oyq + \oyq\oyp}{2R^2} 
+ \frac{3\oyp\oyq\oyp - \oyq^3}{3R^3} + \ldots
\end{gather}
\end{subequations}
Remember that \m{R} is the radius of curvature \textit{at a fixed point} \m{\sz_0}, so it is considered independent of \m{\phi} and \m{j} in this local approximation and thus commutes with \m{\oyq} and \m{\oyp}. By keeping \m{\bigO(R^{-m})} terms in \eq{eq:pop}, one satisfies \eq{eq:spj} up to \m{\bigO(R^{-m})}, as can be verified by a direct calculation. Hence, the above \m{\oper{\phi}} and \m{\oper{j}} can be used as base operators. Figure \ref{fig:basis} illustrates the construction of the corresponding coordinate mesh, reflecting the evolution of OHO's parameters along the ray. 

\subsection{\tMwaves}
\label{sec:mjphi}

For the specific transformation considered here, \Mwaves satisfy the equation of a quantum harmonic oscillator (QHO), \m{\ii \pd_\phi \inv{M} = j(q, -\ii \pd_q)\inv{M}}. The propagator for this equation is known exactly \citep[Section 2.6.1]{book:sakurai20}. Ironically, though, the initial conditions for said equation, \m{\inv{M}(q, 0) = \braket{\ev_{\oper{\phi}}(0) | \ev_{\oper{q}}(q)}}, are not, so the exact propagator is not immediately applicable. Because of this, below we use an alternative, asymptotic, approach that follows the general recipe from \Sec{sec:mwavego}. Let us adopt \mm{\inv{M}(q, \phi) \equiv M^*(\phi, q) = \ee^{\ii \Theta(q, \phi)}\mcc{M}(q, \phi)}. Then, by \eq{eq:kp}, one has\footnote{Remember that \Sec{sec:mwavego} describes a transformation \m{(x, k) \mapsto (q, p)}, where \m{(q, p)} serve the `new' variables. Here, in contrast, we consider a transformation \m{(q, p) \mapsto (\phi, h)}, so \m{(q, p)} serve as the `old' variables. The signs in \eq{eq:dTpj} have been adjusted accordingly.}
\begin{subequations}\label{eq:dTpj}
\begin{gather}
\pd_q \Theta(q, \phi) = p (q, \phi) = q \cot \phi - R,
\\
\pd_\phi \Theta(q, \phi) = - j(q, \phi) = \frac{R^2}{2} - \frac{q^2}{2 \sin^2\phi},
\label{eq:dThj}
\end{gather}
\end{subequations}
where the latter equality is obtained by expressing \m{j} from \eq{eq:qpj2}. This leads to
\begin{gather}\label{eq:TD1}
\Theta(q, \phi) = \Theta_0 + \frac{q^2}{2} \cot \phi - R q + \frac{R^2\phi}{2},
\end{gather}
where \m{\Theta_0} is an integration constant. Then, \m{\pd_{q\phi}\Theta = -q/\sin^2\phi}, so \eq{eq:Mampl2} gives
\begin{gather}\label{eq:Mosc}
\mcc{M}^2(q, \phi) = \frac{|q|}{2\upi \sin^2\phi}.
\end{gather}

The constant phase \m{\Theta_0} in \eq{eq:TD1} is convenient to choose such that the eigenvectors of \m{\oper{q}} and \m{\oper{\phi}}, \m{\ket{\ev_{\oper{q}}(q)}} and \m{\ket{\ev_{\oper{\phi}}(\phi)}}, coincide at small \m{q \approx R \phi} up to normalization. This corresponds to \m{M(\phi, q) \approx \sqrt{R}\,\delta(q - R\phi)}, where the normalization factor \m{\sqrt{R}} is introduced to satisfy \eq{eq:Mnorm}; cf.\ \Sec{sec:resc}. Hence, we adopt
\begin{gather}\label{eq:Mqho}
M(\phi, q) = \sqrt{\frac{\ii q}{2\upi \sin^2 \phi}}\,\exp\left(
-\frac{\ii q^2}{2} \cot \phi + \ii R q - \frac{\ii R^2\phi}{2}
\right).
\end{gather}
Equation \eq{eq:Mqho} is consistent with \eq{eq:TD1} and \eq{eq:Mosc} and, at small \m{\phi}, gives, as desired,
\begin{gather}\label{eq:Mosc2}
M(\phi, q) \approx \sqrt{\frac{\ii q}{2\upi \phi^2}}\,\exp\left(
- \frac{\ii (q - R\phi)^2}{2\phi}\right) \approx \sqrt{R}\,\delta(q - R\phi),
\end{gather}
where the latter equality holds due to the fact that one can replace \m{q} with \m{R\phi} in the pre-exponential factor and we interpret the phase of the square root as \m{(\upi/4)\sgn q}. Then, the phase changes by \m{\upi/2} across \m{q = 0}, that is, from \m{-\upi/4} to \m{\upi/4}.  

The corresponding pseudo-measure \eq{eq:Wdef} is given by
\begin{align}
\mu
& = \frac{1}{2\upi}\int \dd s\,\dd s'\,
M\left(\phi + \frac{s'}{2}, q + \frac{s}{2}\right) 
M^*\left(\phi - \frac{s'}{2}, q - \frac{s}{2}\right)
\ee^{-\ii j s' + \ii p s}
\notag\\
& = \frac{1}{(2\upi)^2}\int \dd s\,\dd s'\,
\sqrt{\frac{|(q + s/2)(q - s/2)|}{\sin^2(\phi + s'/2)\sin^2(\phi - s'/2)}}\,\ee^{\ii \varphi},
\notag\\
& = \frac{1}{(2\upi)^2}\int \dd s\,\dd s'\,
\frac{\sqrt{|q^2 - s^2/4|}}{|\sin^2\phi - \sin^2(s'/2)|}\,\ee^{\ii \varphi},
\label{eq:auxint}
\end{align}
where 
\begin{gather}
\varphi 
= -\Theta\left(q + \frac{s}{2}, \phi + \frac{s'}{2}\right) 
+ \Theta\left(q - \frac{s}{2}, \phi - \frac{s'}{2}\right) - j s' + p s.
\end{gather}
This can be used to recalculate the symbol representation as usual,
\begin{gather}\label{eq:Ozeta}
O_\sa(\phi, j) = \int \dd q\,\dd p\,\mu\, O_\sy(q, p).
\end{gather}
For any smooth symbol \m{O_\sy}, the integral \eq{eq:Ozeta} can be handled like the one in \eq{eq:auxenv}. That is, one can Taylor-expand \m{\Theta} in \m{s} and \m{s'}, keep the linear terms under the exponent, transform the rest and the pre-exponential factor into a polynomial in \m{s} and \m{s'}, and rewrite it using \m{s = -\ii \pd_p \ee^{\ii p s}} and \m{s' = \ii \pd_j \ee^{-\ii j s'}}. This leads to
\begin{align}
O_\sa(\phi, j)
& \approx \frac{1}{(2\upi)^2}\int \dd q\,\dd p\,\dd s\,\dd s'\left(1 + \bigO(s^2, s'^2)\right)
\frac{|q|}{\sin^2\phi}\,O_\sy(q, p)\,\ee^{\ii (p - p(q, \phi))s -\ii (j - j(q, \phi))s'}
\notag\\
& = \frac{1}{(2\upi)^2}\int \dd q\,\dd p\,\frac{|q|}{\sin^2\phi}\,O_\sy(q, p)
\left(1 + \bigO(\pd_p^2, \pd_j^2)\right) \int \dd s\,\dd s'
\ee^{\ii (p - p(q, \phi))s -\ii (j - j(q, \phi))s'}
\notag\\
& = \int \dd q\,\dd p\,
\frac{|q|}{\sin^2\phi}\,O_\sy(q, p)
 \left(1 + \bigO(\pd_p^2, \pd_j^2)\right)\delta(j - j(q, \phi))\,\delta(p - p(q, \phi)).
\end{align}
Due to \eq{eq:dThj}, one has
\begin{gather}
\frac{|q|}{\sin^2\phi}\,\delta(j - j(q, \phi)) 
= \left|\frac{\pd j(q, \phi)}{\pd q}\right|\delta(j - j(q, \phi)) 
= \delta(q - Q(\phi, j)),
\end{gather}
where \m{Q(\phi, j) \doteq \sqrt{R^2 + 2j}\,\sin \phi}.\footnote{Formally, the equation \m{j(q, \phi) = j} has two solutions for \m{q} as a function of \m{\phi} at given \m{j}, \m{q = \pm \sqrt{R^2 + 2j}\,\sin \phi}. The one with the minus sign is ignored here because, as we already mentioned earlier, the MT to the angle--action variables is defined only locally.} This leads to
\begin{gather}
O_\sa(\phi, j)
= \int \dd q\,\dd p\,(O_\sy(q, p) + \bigO(\pd_\sy^2 O))\,
\delta(q - Q(\phi, j))\,\delta(p - P(\phi, j)),
\end{gather}
where \m{P(\phi, j) \doteq -R + \sqrt{R^2 + 2j}\,\cos \phi}. The term \m{\bigO(\pd_\sy^2 O)} can be estimated as \m{O/R^2}, so it is negligible within the assumed accuracy. Then, unsurprisingly, the leading-order transformation of symbols is simple:
\begin{gather}\label{eq:Oho}
O_\sa(\phi, j) \approx O_\sy(Q(\phi, j), P(\phi, j)).
\end{gather}

However, remember that the approximation \eq{eq:Oho} is applicable only to smooth symbols. Also, the usability of \eq{eq:pop} as an asymptotic series is conditioned upon smallness of \m{\kappa}, which is due to the fact that higher-order derivatives of \m{H_\sz} have been neglected. Also, \m{\dot{\phi}} is not globally constant, so \m{(\phi, j)} are not true angle--action variables of the original system even approximately. That is why we prefer the alternative variable transformations discussed in \Secs{sec:aac2} and \ref{sec:mgogen}, which are not associated with such inconveniences.

\section{Auxiliary functions}
\label{app:ai}

\subsection{Properties of \mt{\Ai_\gamma}}

Here, we discuss some useful properties of the function \m{\Ai_\gamma}. By Taylor-expanding the integrand in \eq{eq:airyr} in \m{\gamma}, one obtains
\begin{align}
\Ai_\gamma(z) 
& = \frac{1}{2\upi}\int_{-\infty}^{\infty} \dd z \sum_{n = 0}^\infty
\frac{1}{n!}\left(\frac{\ii \gamma t^3}{24}\right)^n
\ee^{\ii z t}
\notag\\
& = \sum_{n = 0}^\infty
\frac{1}{n!}\left(\frac{\ii \gamma}{24}\, (-\ii \pd_z)^3\right)^n
\frac{1}{2\upi}\int_{-\infty}^{\infty} \dd z \, \ee^{\ii z t}
\notag\\
& = \sum_{n = 0}^\infty
\frac{1}{n!}\left(-\frac{\gamma}{24}\, \pd_z^3\right)^n
\delta(z)
\notag\\
& = \exp\left(-\frac{\gamma}{24}\, \pd_z^3\right)\delta(z).
\label{eq:ainorm60}
\end{align}
In other words, 
\begin{gather}\label{eq:ainorm6}
\Ai_\gamma(z) = \delta(z) - \frac{\gamma}{24}\,\delta'''(z) + \ldots.
\end{gather}
Notice that the first two derivatives of the delta function do not contribute to the right-hand side of \eq{eq:ainorm6}.

\subsection{Properties of \mt{\GAF}}

Let us also consider the function \m{\GAF} defined in \eq{eq:gaf2}. First of all, a direct calculation shows that
\begin{gather}
\Phi = (H_\sy(\sy) - h)s 
- \frac{\ii s^3}{24}\, V^3(q)\,p''(q)
- \frac{\ii s^5}{1920}\,V^5(q)\,p''''(q)  + \ldots
\end{gather}
Let us rewrite this as
\begin{gather}
\Phi \doteq s \Delta H
+ \frac{s^3}{24}\, \ki_2 
+ \frac{s^5}{1920}\, \ki_4 + \ldots,
\label{eq:Thetagaf}
\end{gather}
where \m{\Delta H \doteq H_\sy - h} and \m{\ki_n \doteq -V^{n+1} p^{(n)}}. Note that the characteristic scales of \m{q} and \m{p} are both of order \m{R}, and \m{V \sim v}, so
\begin{gather}\label{eq:PP}
\ki_2 \sim v^3/R \sim \ki,
\qquad
\ki_4 \sim v^5/R^3 \sim g\ki,
\end{gather}
where we used \m{g \sim \Omega^2 \sim v^2/R^2}. The characteristic values of \m{s} that contribute to the integral \eq{eq:gaf2} are of the order of those at which \m{\Phi} is stationary, \ie those that satisfy
\begin{gather}\label{eq:sp}
0 = \pd_s \Phi = \Delta H
+ \frac{s^2}{8}\, \ki_2
+ \frac{s^4}{384}\, \ki_4
+ \ldots
\end{gather}
Let us use this to estimate the characteristic scale of \m{\GAF} with respect to \m{\Delta H}. Assuming that the nonlinear, in \m{s}, terms in \eq{eq:Thetagaf} are not too large at values of \m{s} that satisfy \eq{eq:sp}, one can rewrite \eq{eq:gaf2} as follows:
\begin{subequations}
\begin{align}
\GAF(\Delta H, \ldots)
& =  \frac{1}{2\upi}\int \dd s\,\bigg(1 + \frac{s^3}{24}\, \ki_2
+ \frac{s^5}{1920}\,\ki_4 + \ldots \bigg)\ee^{\ii s \Delta H}
\label{eq:cat}\\
& =  \bigg(1 + \frac{\ii}{24}\, \ki_2 \pd_{\Delta H}^3
- \frac{\ii}{1920}\, \ki_4 \pd_{\Delta H}^5 + \ldots \bigg)
\frac{1}{2\upi}\int \dd s\,\ee^{\ii s \Delta H}
\\
& = \bigg(1 + \underbrace{\frac{\ii}{24}\,\ki_2 \pd_{\Delta H}^3}_{\sim \ki_2/\Delta H^3}
- \underbrace{\frac{\ii}{1920}\, \ki_4 \pd_{\Delta H}^5}_{\sim \ki_4/\Delta H^5}
+ \ldots \bigg) \delta(\Delta H).
\label{eq:gefest}
\end{align}
\end{subequations}
(The integral in \eq{eq:cat} can be recognised as a `catastrophe integral' like those that emerge in theory of caustics; see \cite[equation (22)]{my:mgoinv} and references therein.) Then, \m{\GAF}'s characteristic scale (roughly, the value of \m{\Delta H} at which \m{\GAF} has its first zero) is determined by
\begin{gather}
1 + \bigO(\ki_2/\Delta H^3) +  \bigO(\ki_4/\Delta H^5) + \ldots = 0,
\label{eq:OOO}
\end{gather}
and the values of \m{s} that matter satisfy \m{s \sim \Delta H^{-1}}, as seen from the derivation of \eq{eq:gefest}. The higher-order terms denoted by the ellipsis can be dealt with just like the third term, so they do not need to be considered separately and are ignored from now on. Then, there are two options.

First, suppose \m{\ki_2 \ll \ki_4/\Delta H^2}. Then, \eq{eq:OOO} yields \m{\Delta H \sim (g\ki)^{1/5}}, and \mm{s \sim \Delta H^{-1} \sim (g\ki)^{-1/5}}, which is also consistent with \eq{eq:sp}. The term \m{\propto\, \ki_2} is negligible when \m{s^3 \ki_2 \ll 1},~\ie 
\begin{gather}\label{eq:c1}
1 \gg (g\ki)^{-3/5} \ki \sim (\ki^{2/3}/g)^3 \sim R^{4/5}.
\end{gather}
Now, suppose \m{\ki_2 \gg \ki_4/\Delta H^2}. Then, \eq{eq:OOO} yields \m{\Delta H \sim \ki^{1/3}}, and \m{s \sim \Delta H^{-1} \sim \ki^{-1/3}}, which is also consistent with \eq{eq:sp}. The term \m{\propto\, \ki_4} is negligible when \m{s^5 \ki_4 \ll 1},~\ie 
\begin{gather}
1 \gg \ki^{-5/3} g\ki \sim g/\ki^{2/3} \sim R^{-4/3}.
\end{gather}
which is just the condition opposite to \eq{eq:c1}. In summary then, \m{\GAF} is approximately equal to RAF when \m{R \gg 1}.

\section{Quantization condition for closed orbits}
\label{app:qz}

For closed orbits, the operators \m{\oper{\tau}} and \m{\oper{h}} can be introduced much like the angle--action operators in \App{app:aac}, \ie with \m{\tau} as a multi-valued function of \m{q} and~\m{p}. Then, one can notice that, actually, multiple parts of the \m{\tau}-space represent the same parts of the \m{q}-space. By doing so, one effectively replaces the commonly used reflecting boundary conditions in the \m{q}-space with periodic boundary conditions in the \m{\tau}-space. This readily yields the familiar quantization rule for the discrete spectrum of closed orbits as follows. 

First of all, \m{M_{\oper{q} \mapsfrom \oper{\tau}}(q, \tau)} satisfies
\begin{gather}\label{eq:M1aux}
M_{\oper{q} \mapsfrom \oper{\tau}}(0, \tau) \approx M_0 \delta(\tau),
\end{gather}
where \m{M_0 \doteq [\tau'(0)]^{-1/2}}. Accordingly,
\begin{gather}\label{eq:psixt0}
\psi_{\oper{q}}(0) \approx M_0 \psi_{\oper{\tau}}(0).
\end{gather}
But one can also express \m{\psi_{\oper{q}}(0)} through \m{M_{\oper{q} \mapsfrom \oper{\tau}}} after a full rotation of the ray around the closed orbit, \m{\tau = T}. The corresponding MT is given by
\begin{gather}\label{eq:M2aux}
M_{\oper{q} \mapsfrom \oper{\tau}}(0, \tau) \approx -M_0 \delta(\tau)
\exp(\ii\Theta_\circlearrowright),
\end{gather}
where the sign change is due to the cumulative Maslov phase \m{\pm\upi} that originates from the square root in \eq{eq:semiMxt} (\Sec{sec:mwavego})\footnote{Near each zero of said square root, \m{M} can be approximated by the asymptotic form \eq{eq:Mosc2} (up to rescaling \m{\dd\tau = \dd\phi/\Omega}). Then, the phase of \m{M_{\oper{\tau} \mapsfrom \oper{q}}} increases by \m{\upi} over a closed orbit at clockwise direction, so the phase of \m{M_{\oper{q} \mapsfrom \oper{\tau}}} decreases by \m{\upi}. At counter-clockwise rotation, the phases have the opposite signs, but, in either case, the \Mwave changes sign.} and \m{\Theta_\circlearrowright} is the phase change of \m{M_{\oper{q} \mapsfrom \oper{\tau}}} around the orbit. Then,
\begin{gather}\label{eq:psixt3}
\psi_{\oper{q}}(0) \approx M_0 \psi_{\oper{\tau}}(T) \exp(-\ii \upi + \ii\Theta_\circlearrowright).
\end{gather}
Since \m{\psi_{\oper{\tau}}} is conserved on the ray, one has \m{\psi_{\oper{\tau}}(T) = \psi_{\oper{\tau}}(0)}. Then, by comparing \eq{eq:psixt0} with \eq{eq:psixt3}, one finds that \m{-\upi + \Theta_\circlearrowright} must be equal to \m{2\upi n}, where \m{n} is integer; \ie \m{\Theta_\circlearrowright = 2\upi n + \upi}. Also, from \eq{eq:ktTh}, one has \m{\Theta = \int (p\,\dd q - h\,\dd \tau)} and \m{h = 0} on the ray, so
\begin{gather}
\Theta_\circlearrowright = \oint p\,\dd q 
= \frac{1}{2} \oint \dd \sy\wedge \sy
= \frac{1}{2} \oint \dd \sz\wedge \sz
= \oint k\,\dd x.
\end{gather}
This leads to the well-known Einstein--Brillouin--Keller quantization rule \citep{ref:stone05},
\begin{gather}
\oint k\,\dd x = 2\upi \left(n + \frac{1}{2}\right),
\end{gather}
which is also known as the Bohr--Sommerfeld condition in the limit when \m{1/2} is negligible compared to \m{n}.

\section{Dissipation power}
\label{app:P}

Let us consider a wave with electric field \m{\tilde{\vec{E}}} in a linear medium with a conductivity operator \m{\boper{\sigma}}. Such a wave induces the current density
\begin{gather}
\tilde{\vec{\mc{J}}}(t, \vec{x}) = (\boper{\sigma}\tilde{\vec{E}})(t, \vec{x})
\equiv \int \dd t'\,\dd\vec{x}'\,\underline{\matr{\sigma}}(t, \vec{x}, t', \vec{x}')\,\tilde{\vec{E}}(t', \vec{x}')
\end{gather}
and dissipates the energy
\begin{gather}\label{eq:P}
\mcc{E} = \int \dd t\,\dd \vec{x}\,\tilde{\vec{\mc{J}}} \cdot \tilde{\vec{E}}
= \int \dd t\,\dd \vec{x}\,\tilde{\vec{E}}{}^\T \boper{\sigma} \tilde{\vec{E}}
= \int \dd t\,\dd \vec{x}\,\tilde{\vec{E}}{}^\dag \boper{\sigma}_{\rm H} \tilde{\vec{E}},
\end{gather}
where \m{_{\rm H}} denotes the Hermitian part. (We treat \m{\tilde{\vec{E}}} as a column vector, so \m{\tilde{\vec{E}}{}^\T} is a row vector, and \m{\tilde{\vec{E}}{}^\dag = \tilde{\vec{E}}{}^\T}, because \m{\tilde{\vec{E}}} is real.) Using \eq{eq:WE} together with (2.53a) from \citep{my:ql}, one can rewrite \eq{eq:P} as follows:
\begin{gather}\label{eq:EE0}
\mcc{E} = \int \dd t\,\dd \vec{x}\,\dd\omega\,\dd \vec{k}\,
\tr(\matr{\sigma}_{\rm H} \star \matr{W}_{\tilde{\vec{E}}}).
\end{gather}
Here, \m{\tr} denotes the matrix trace, \m{\star} is the Moyal product, and \m{\matr{\sigma} \equiv \matr{\sigma}(t, \vec{x}, \omega, \vec{k})} is the Weyl symbol of \m{\boper{\sigma}}:
\begin{gather}
\matr{\sigma}(t, \vec{x}, \omega, \vec{k}) \doteq \int \dd\tau \,\dd\vec{s}\,\ee^{\ii \omega \tau - \ii \vec{k} \cdot \vec{s}}\,
\underline{\matr{\sigma}}\left(t + \frac{\tau}{2}, \vec{x} + \frac{\vec{s}}{2}, t - \frac{\tau}{2}, \vec{x} - \frac{\vec{s}}{2}\right),
\end{gather}
which satisfies
\begin{gather}\label{es:sigsign}
\matr{\sigma}(t, \vec{x}, - \omega, - \vec{k}) 
= \matr{\sigma}^*(t, \vec{x}, \omega, \vec{k}),
\end{gather}
because \m{\underline{\matr{\sigma}}} is real by definition. Also, \m{\matr{W}_{\tilde{\vec{E}}} \equiv \matr{W}_{\tilde{\vec{E}}}(t, \vec{x}, \omega, \vec{k})} is the Wigner matrix of \m{\tilde{\vec{E}}}:
\begin{gather}\label{eq:WE}
\matr{W}_{\tilde{\vec{E}}} \doteq \frac{1}{(2\upi)^{1 + n}}\,\int \dd \tau \,\dd\vec{s}\,
\ee^{-\ii \vec{k} \cdot \vec{s} + \ii \omega \tau}
\tilde{\vec{E}}
\left(t + \frac{\tau}{2}, \vec{x} + \frac{\vec{s}}{2}\right)\,
\tilde{\vec{E}}{}^\dag\left(t - \frac{\tau}{2}, \vec{x} - \frac{\vec{s}}{2}\right),
\end{gather}
where \m{n \doteq \dim \vec{x}}. By a known theorem\footnote{See, for example, Corollary~7 in Appendix~B of \citep{phd:ruiz17}.}, \eq{eq:EE0} can also be written as
\begin{gather}\label{eq:EE1}
\mcc{E} 
= \int \dd t\,\dd \vec{x}\,\dd\omega\,\dd \vec{k}\,
\tr (\matr{\sigma}_{\rm H} \matr{W}_{\tilde{\vec{E}}}).
\end{gather}
Assuming that \m{\matr{\sigma}} satisfies MGO applicability conditions for the characteristic phase-space scales, \ie (cf.\ \eq{eq:mgopar})
\begin{gather}
\Delta t \Delta \omega \gg 1,
\qquad
\Delta x \Delta k \gg 1,
\end{gather}
one can replace \eq{eq:EE1} with
\begin{gather}\label{eq:EE2}
\mcc{E} 
= \int \underbrace{\dd t\,\dd \vec{x}\,\dd\omega\,\dd \vec{k}\vphantom{\big\lbrace}}_{\dd\Gamma}\,
\tr \left(\matr{\sigma}_{\rm H} \matr{\mcc{W}}\right),
\end{gather}
where \m{\matr{\mcc{W}}} is the local average of \m{\matr{W}_{\tilde{\vec{E}}}} over a phase-space volume that is large compared to unity yet small compared to the scale of \m{\matr{\sigma}_{\rm H}}. As shown in \citep{my:ql}, this averaging makes \m{\mcc{W}} positive-semidefinite \textit{and} a local property of a wave (as opposed to the original \m{\matr{W}_{\tilde{\vec{E}}}}, which can have oscillating tails that can extend over a whole system). Then, one obtains the following phase-space density of \m{\mcc{E}}:
\begin{gather}\label{eq:Pd}
\frac{\dd\mcc{E}}{\dd\Gamma} = \tr(\matr{\sigma}_{\rm H}\matr{\mcc{W}}).
\end{gather}

To illustrate the connection of \eq{eq:Pd} with familiar results, let us consider a medium that is weakly inhomogeneous (if at all) in time and space, so that standard GO applies. Then, the wavefield can be locally represented as \m{\tilde{\vec{E}} = \re(\ee^{-\ii \bar{\omega} t + \ii \bar{\vec{k}} \cdot \vec{x}}\, \Eenv)}, with constant \m{\bar{\omega}}, \m{\bar{\vec{k}}}, and \m{\Eenv}. This corresponds to \citep[Appendix A.2]{my:ql}
\begin{gather}\label{eq:W0}
\matr{\mcc{W}} = 
\frac{\Eenv\Eenv^\dag}{4}\, \delta(\omega - \bar{\omega})\, \delta(\vec{k} - \bar{\vec{k}})
+ \frac{(\Eenv\Eenv^\dag)^*}{4}\, \delta(\omega + \bar{\omega})\, \delta(\vec{k} + \bar{\vec{k}}).
\end{gather}
Then, the energy dissipated per unit spacetime volume \m{\dd t\,\dd\vec{x}} (`dissipation power density') can be expressed as
\begin{gather}\label{eq:aux5}
\mcc{P} \doteq \int \frac{\dd\mcc{E}}{\dd\Gamma}\,\dd\omega\,\dd\vec{k} 
= \frac{1}{2}\,\Eenv^\dag \matr{\sigma}_{\rm H}(t, \vec{x}, \bar{\omega}, \bar{\vec{k}}) \Eenv.
\end{gather}
To the leading (zeroth) order in \m{(\Delta t\Delta \omega)^{-1}}, one also has \m{\matr{\sigma}_{\rm H} = (\omega/4\upi)\,\matr{\varepsilon}_{\rm A}}, where \m{\matr{\varepsilon}_{\rm A}} is the anti-Hermitian part of the symbol of the dielectric operator, or the dispersion tensor. Thus, \eq{eq:aux5} leads to the familiar formula \citep[Sec.~4-2]{book:stix}
\begin{gather}
\mcc{P} = \frac{\bar{\omega}}{8\upi}\,\Eenv^\dag \matr{\varepsilon}_{\rm A}(t, \vec{x}, \bar{\omega}, \bar{\vec{k}}) \Eenv.
\label{eq:GOP}
\end{gather}

\section{Selected notation}
\label{app:notation}

\begin{spacing}{1.2}
\begin{longtable}{l@{\qquad}l@{\quad}l}
\hline Symbol & Definition & Properties/Explanation\\
\hline\\
\m{\doteq} & definition & \\
\m{\mapsto} & variable transformation & \\
\m{\Leftrightarrow} & Wigner--Weyl isomorphism & \Sec{sec:weyl} \\
\m{\placeholder} & placeholder & \\
\m{\dot{\placeholder}} & time derivative on a ray & \\
\m{\inv{\placeholder}} & refers to the inverse transform &\\
\m{\oper{\placeholder}} & operator & \\
\m{\placeholder^*} & complex conjugate & \\
\m{\placeholder^\dag} & Hermitian adjoint & \\
\m{\placeholder^\T} & transposition & \\
\m{\placeholder_0} & evaluated at \m{\sz = \sz_0} & \\
\m{\placeholder_{\oper{x}}} & \m{x}-representation & \Sec{sec:notation}\\
\m{\placeholder_{\oper{q}}} & \m{q}-representation & \Sec{sec:notation}\\
\m{\placeholder_\sr} & \m{\sr}-symbol & \Sec{sec:mgo}\\
\m{\placeholder_\sy} & \m{\sy}-symbol & \Sec{sec:transform}\\
\m{\placeholder_\sz} & \m{\sz}-symbol & \Sec{sec:transform}\\
\m{\ket{\placeholder}} & Dirac ket & vector in \m{\mathbb{H}}\\ 
\m{\bra{\placeholder}} & Dirac bra & covector in \m{\mathbb{H}}\\ 
\m{\placeholder\cdot\placeholder} 
    & \m{\symp{a} \cdot \symp{b} \doteq \symp{a}^\T \symp{b}} 
    & \m{\symp{a} \cdot \symp{b} = \symp{b} \cdot \symp{a} = \symp{a}_\alpha \symp{b}^\alpha}\\
\m{\placeholder\wedge\placeholder}
    & \m{\symp{a} \wedge \symp{b} \doteq \symp{a} \cdot \sJ \symp{b}} 
    & \m{\symp{a} \wedge \symp{b}= - \symp{b} \wedge \symp{a}}, \m{\symp{a} \wedge \symp{a} = 0}\\
\m{\placeholder \star \placeholder} & Moyal star & \Sec{sec:weyl}\\
\m{[\placeholder, \placeholder]} & commutator & \\
\m{\poisson{\placeholder, \placeholder}}
    & Poisson bracket
    & \m{\poisson{a, b} = (\pd_\sz a) \cdot \sJ (\pd_\sz b) = - \poisson{b, a}}\\
\m{\moyal{\placeholder, \placeholder}} & Moyal bracket & \Sec{sec:weyl}\\
&&\\
\m{\oper{1}} & unit operator & \\
\m{\pd} & partial derivative & \\
&&\\
\m{\oper{\Gamma}}, \m{\boper{\Gamma}} & anti-Hermitian part of \m{\oper{D}} & \Sec{sec:trad}\\
\m{\Gamma} & symbol of \m{\oper{\Gamma}}, & \Sec{sec:trad}\\
& also phase-space volume & \eq{eq:EE2} \\
\m{\Omega} & OHO's angular frequency & \m{\sigma_\ki\sqrt{|g|}}, \Sec{sec:iy} \\
&&\\
\m{\delta} & Dirac delta function & \\
\m{\epsilon} & MGO parameter & \eq{eq:mgopar} \\
\m{\ki} & \m{(\ddot{\sz} \wedge \dot{\sz})_0} & \m{\ki = - uv = \sv \cdot \sg\sv \equiv ||\sv||^2_\sg}\\
\m{\theta} & eikonal & rapid phase of a GO wave\\
\m{\kappa} & symplectic curvature & \eq{eq:kapep} \\
\m{\MWf} & \m{2\upi W^{(M)}} & \eq{eq:Wdef}\\
\m{\sigma_\placeholder} & \m{\sgn \placeholder} & \\
&&\\
\m{\ket{\ev_\placeholder(\placeholder)}}
    & normalised eigenvector
    & \m{\oper{A}\ket{\ev_{\oper{A}}(\lambda)} = \lambda\ket{\ev_{\oper{A}}(\lambda)}}\\
\m{\kc} & reference wavevector & \m{\pd_\sz\theta} \\
&&\\
\m{\mathbb{H}} & Hilbert space & \\
\m{\mathbb{R}} & real line & \m{(-\infty, \infty)}\\
&&\\
\m{\sI} & unit matrix & \\
\m{\sJ} & canonical symplectic form & \eq{eq:JI} \\
\m{\sS} & symplectic matrix & \m{\sS^\T \sJ \sS = \sJ, \,\, \det\sS = 1} \\
\m{\inv{\sS}} & \m{\sS^{-1}} & \m{\sSi^\T \sJ \sSi = \sJ,\,\, \det\inv{\sS} = 1} \\
\m{\sg} & \m{(\pd^2_{\sz\sz} H)_0} & \Sec{sec:mgo} \\
\m{\sr} & phase-space variables & \m{(\tau, h)^\T} \\
\m{\seta_q, \seta_p} & columns of \m{\sSi} & \Sec{sec:tang} \\
\m{\sv} & local phase-space velocity & \m{\sv = \dot{\sz}_0 = \sJ (\pd_\sz H_\sz)_0}\\
\m{\su} & local phase-space acceleration & \m{\su = \ddot{\sz}_0 = \sJ \sg\sv}\\
\m{\sy} & phase-space variables & \m{(q, p)^\T} \\
\m{\sz} & phase-space variables & \m{(x, k)^\T} \\
\m{\sz_0} & selected point in the \m{\sz}-space & \\
&&\\
\m{\bigO} & big O & \\
&&\\
\m{\Ai_\gamma} & rescaled Airy function & \eq{eq:airyr}\\
\m{\dd} & differential & \\
\m{\ee} & Euler's number & \m{\exp(1)} \\
\m{\ii} & imaginary unit & \m{\sqrt{-1}} \\
\m{\symb\placeholder} & Weyl symbol & \Sec{sec:weyl} \\
&&\\
\m{\oper{D}}, \m{\boper{D}} & dispersion operator & \\
\m{\oper{H}}, \m{\boper{H}} & Hermitian part of \m{\oper{D}}, \m{\boper{D}} & \Sec{sec:trad} \\
\m{H}, \m{\matr{H}} & symbol of \m{\oper{H}}, \m{\boper{H}} & \\
\m{M(q, x)} & same as \m{M_{\oper{q} \mapsfrom \oper{x}}(q, x)} 
 & \Sec{sec:Mdef}\\
\m{M_{\oper{q} \mapsfrom \oper{x}}(q, x)}
    & MT kernel & \m{\braket{\ev_{\oper{q}}(q) | \ev_{\oper{x}}(x)}}\\
\m{\inv{M}(x, q)} & same as \m{M_{\oper{x} \mapsfrom \oper{q}}(x, q)} 
& \Sec{sec:Mdef}\\
\m{M_{\oper{x} \mapsfrom \oper{q}}(x, q)}
    & inverse-MT kernel & \m{\braket{\ev_{\oper{x}}(x) | \ev_{\oper{q}}(q)}}\\
\m{\oper{O}} & generic operator & \\
\m{O} & symbol of \m{\oper{O}} & \\
\m{R} & symplectic radius (scale) & \Secs{sec:iy} and \ref{sec:opt}\\
\m{W^{(\placeholder)}} & Wigner function & \Sec{sec:wigner} \\
\m{g} & \m{\det \sg} & \Sec{sec:iy} \\
\m{u} & \m{\sv \wedge \su/v} & \Secs{eq:etap} and \ref{sec:iy} \\
\m{v} & symplectic speed & \m{(\ki^2/|g|)^{1/4}}, \Sec{sec:iy} \\
&&\\
\hline
\end{longtable}
\end{spacing}

\begin{spacing}{1.2}
\begin{longtable}{l@{\qquad}l}
\hline Abbreviation & Meaning\\
\hline\\
GAF & generalised Airy function\\
GO & geometrical optics\\
LST & linear symplectic transformation\\
MGO & metaplectic geometrical optics\\
MT & metaplectic transform\\
OHO & osculating harmonic oscillator\\
QHO & quantum harmonic oscillator\\
RAF & rescaled Airy function\\
SE & Schr\"odinger equation\\
WKB & Wentzel--Kramers--Brillouin\\
&\\
\hline
\end{longtable}
\end{spacing}


\end{document}